\documentclass[PhD]{iitcsthesis}
\usepackage{array}
\usepackage{amsmath}
\usepackage{amssymb}
\usepackage{upref}
\usepackage{amsfonts}
\usepackage{graphicx}
\usepackage{epsfig}
\usepackage{wasysym}

\usepackage{slashbox}
\usepackage{tikz}

\usepackage{mathrsfs}

\newcommand{\deff}{\mbox{$\stackrel{\rm def}{=}$}}

\newcommand{\floorenv}[1]{\left\lfloor #1 \right\rfloor}

\newcommand{\sbinom}[2]{\left[ \begin{array}{c} #1 \\ #2 \end{array} \right] }
\newcommand{\sbinomq}[2]{\sbinom{#1}{#2}_q }

\newcommand{\dS}{\field{S}}
\newcommand{\T}{\field{T}}

\newcommand{\field}[1]{\mathbb{#1}}
\newcommand{\C}{\field{C}}
\newcommand{\F}{\field{F}}
\newcommand{\Z}{\field{Z}}

\newcommand{\cF}{{\cal F}}
\newcommand{\cH}{{\cal H}}
\newcommand{\cA}{{\cal A}}
\newcommand{\cB}{{\cal B}}
\newcommand{\cC}{{\cal C}}
\newcommand{\cG}{{\cal G}}
\newcommand{\cL}{{\cal L}}

\newcommand{\cS}{{\cal S}}

\newcommand{\cP}{{\cal P}}

\newcommand{\cX}{{\cal X}}

\newcommand{\sP}{\cP}
\newcommand{\sG}{\cG}

\newcommand{\Gr}{\smash{{\sG\kern-1.5pt}_q\kern-0.5pt(n,k)}}
\newcommand{\Grtwo}{\smash{{\sG\kern-1.5pt}_2\kern-0.5pt(n,k)}}
\newcommand{\Gkone}{\smash{{\sG\kern-1.5pt}_q\kern-0.5pt(n,k_1)}}
\newcommand{\Gktwo}{\smash{{\sG\kern-1.5pt}_q\kern-0.5pt(n,k_2)}}
\newcommand{\Ps}{\smash{{\sP\kern-2.0pt}_q\kern-0.5pt(n)}}
\newcommand{\Span}[1]{{\left\langle {#1} \right\rangle}}

\newcommand{\Gk}{\smash{{\sG\kern-1.5pt}_q\kern-0.5pt(n,k_1)}}
\newcommand{\Gkk}{\smash{{\sG\kern-1.5pt}_q\kern-0.5pt(n,k_2)}}
\newcommand{\Grr}{\smash{{\sG\kern-1.5pt}_q\kern-0.5pt(n,r)}}

\newcommand{\CMRD}{\C^{\textmd{MRD}}}

\newcommand{\Gauss}[2]{\begin{footnotesize}\left[\begin{array}
{c}#1\\#2\end{array}\right]_{q}\end{footnotesize}}

\newcommand{\GaussLarge}[2]{\left[\begin{array}
{c}#1\\#2\end{array}\right]_{q}}

\newcommand{\GaussBin}[2]{\begin{footnotesize}\left[\begin{array}
{c}#1\\#2\end{array}\right]_{2}\end{footnotesize}}

\newtheorem{theorem}{Theorem}

\newtheorem{lemma}{Lemma}
\newtheorem{remark}{Remark}
\newtheorem{corollary}{Corollary}
\newtheorem{example}{Example}
\newtheorem{conjecture}{Conjecture}

\DeclareMathOperator{\rank}{rank}

\newenvironment{algorithm}{%
       \begin{minipage}{\columnwidth}\vspace{0.5ex}%
       \makebox[0ex]{}\hrulefill\makebox[0ex]{}\\*\normalsize}{%
             \makebox[0ex]{}\hrulefill\makebox[0ex]{}\end{minipage}}

\begin{document}

\authorEnglish{Natalia Silberstein}

\titleEnglish{Coding Theory and \\Projective Spaces}

\supervisorEnglish{This Research Thesis was done under the
supervision of  Prof.~Tuvi Etzion  in the Department of
Computer Science.}

\GregorianDateEnglish{September 2011} \JewishDateEnglish{Elul 5771}

\personalThanksEnglish{}

\financialThanksEnglish{The Generous Financial Help Of The Technion, Israeli Science Foundation, and
 Neaman Foundation Is Gratefully Acknowledged}

\publicationListJournals{
\begin{enumerate}
\item T. Etzion and N. Silberstein, \emph{{}``Error-Correcting Codes in
Projective Spaces Via Rank-Metric Codes and Ferrers Diagrams}'',
IEEE Transactions on Information Theory, Vol. 55, No. 7, pp. 2909--2919,
July 2009.
\item N. Silberstein and T. Etzion, \emph{{}``Enumerative Coding for Grassmannian
Space}'', IEEE Transactions on Information Theory, Vol. 57, No. 1,
pp. 365 - 374, January 2011.
\item N. Silberstein and T. Etzion,\emph{ {}``Large Constant Dimension
Codes and Lexicodes}'', Advances in Mathematics of Communications
(AMC), vol. 5, No. 2, pp. 177 - 189, 2011.
\item T. Etzion and N. Silberstein, \emph{{}``Codes and Designs Related
to Lifted MRD Codes}'', submitted to IEEE Transactions on Information
Theory.
\end{enumerate}
}

\publicationListConferences{
\begin{enumerate}
\item T. Etzion and N. Silberstein, \emph{{}``Construction of Error-Correcting
Codes For Random Network Coding}'', in IEEE 25th Convention of Electrical
\& Electronics Engineers in Israel (IEEEI 2008), pp. 70 - 74, Eilat, Israel, December
2008.
\item N. Silberstein and T. Etzion, \emph{{}``Enumerative Encoding in the
Grassmannian Space}'', in 2009 IEEE Information Theory Workshop (ITW
2009), pp. 544 - 548, Taormina, Sicily, October 2009.
\item N. Silberstein and T. Etzion,\emph{ {}``Large Constant Dimension
Codes and Lexicodes}'', in Algebraic Combinatorics and Applications
(ALCOMA 10), Thurnau, Germany, April 2010.
\item N. Silberstein and T. Etzion, {}``\emph{Codes and Designs Related
to Lifted MRD Codes}'', in IEEE International Symposium on Information
Theory (ISIT 2011), pp. 2199 - 2203,  Saint Petersburg, Russia, July-August 2011.
\end{enumerate}
}

\maketitleEnglish


\abstractEnglish
The projective space of order $n$ over a finite field $\F_q$, denoted
by $\mathcal{P}_{q}(n)$, is a set of all subspaces of the vector space
$\F_q^{n}$. The projective space is a metric space with the distance
function $d_{s}(X,Y)=\mbox{dim}(X)+\mbox{dim}(Y)-2\mbox{dim}(X\cap Y)$,
for all $X, Y\in\mathcal{P}_{q}(n)$. A code in the projective space
is a subset of $\mathcal{P}_{q}(n)$.
Coding in the projective space has received recently a lot of
attention due to its application in random network coding.

If the dimension of each codeword
is restricted to a fixed nonnegative integer $k\leq n$, then the code forms a subset of a
Grassmannian, which is the set of all $k$-dimensional subspaces of $\F_q^{n}$, denoted by $\Gr$.
Such a code is called a constant dimension code. Constant dimension
codes in the projective space are analogous to constant weight codes in the Hamming space.

In this work, we consider error-correcting codes in the projective space, focusing mainly on
constant dimension codes.

We start with the different representations of subspaces in $\Ps$.
These representations involve matrices in reduced row echelon form,
associated binary vectors, and Ferrers diagrams. Based on these representations, we
provide a new formula for the computation of the distance between any two subspaces
in the projective space.

We examine lifted maximum rank distance (MRD) codes, which are
nearly optimal constant dimension codes.
We prove that a lifted MRD
code can be represented in such a way that it forms a block
design known as a transversal design. A slightly different
representation of this design makes it similar to a $q$-analog of
transversal design. The incidence matrix of the transversal design derived from a
lifted MRD code can be viewed as a parity-check matrix of a linear code in
the Hamming space.
We find the properties of these codes which can be viewed also as
LDPC codes.

We present new bounds and constructions
for constant dimension codes. First, we
present a  multilevel construction for constant dimension codes, which
can be viewed as a generalization of a lifted MRD codes construction.
This construction
is based on a new type of rank-metric codes, called Ferrers diagram rank-metric codes.
We provide an upper bound on the size of Ferrers diagram rank-metric codes and present
a construction of codes that attain this bound.
Then we derive upper bounds
on the size of constant dimension codes which contain the
lifted MRD code, and provide a construction for two families of codes,
that attain these upper bounds. Most of the codes obtained by these constructions are
the largest known constant dimension codes.
We generalize the
well-known concept of a punctured code for a code in the
projective space to obtain large codes which are not constant dimension.

We present efficient enumerative
encoding and decoding techniques for the Grassmannian. These
coding techniques are based on two different lexicographic orders for the
Grassmannian induced by different representations of
$k$-dimensional subspaces of $\F_q^n$. Finally we describe a search method
for constant dimension lexicodes. Some of the codes obtained by
this search are the largest known constant dimension codes with
their parameters.

\abbreviationsAndNotationsEnglish
\begin{tabular}{lcl}
$\F_q$& --- & a finite field of size $q$\\
$\Ps$& --- & the projective space of order $n$\\
$\Gr$& --- & the Grassmannian\\
$d_S(\cdot,\cdot)$& --- & the subspace distance\\
$d_R(\cdot,\cdot)$& --- &the rank distance\\
$d_H(\cdot,\cdot)$& --- &the Hamming distance\\
$\C$& --- &a code in the projective space\\
$\CMRD$& --- &the lifted MRD code\\
$\cC$& --- &a rank-metric code\\
$\bf{C}$& --- &a code in the Hamming space\\
RREF& --- &reduced row echelon form\\
$\mbox{RE}(X)$& --- &a subspace $X$ in RREF\\
$v(X)$& --- &the identifying vector of a subspace $X$\\
$\mbox{FE}(X)$& --- &the Ferrers echelon form of a subspace $X$ \\
$\cF$& --- & Ferrers diagram\\
$\cF_X$& --- &the Ferrers diagram of a subspace $X$\\
$\cF(X)$& --- &the Ferrers taubleux form of a subspace $X$\\
$\mbox{EXT}(X)$& --- &the extended representation of a subspace $X$\\
$\Gauss{n}{k}$ & --- &the $q$-ary Gaussian coefficient\\
$\textmd{TD}_{\lambda}(t,k,m)$& --- &a transversal design of blocksize $k$, groupsize $m$, \\
&  &strength $t$ and index $\lambda$\\
$\textmd{TD}_{\lambda}(k,m)$& --- &a transversal design  $\textmd{TD}_{\lambda}(2,k,m)$\\
$\textmd{STD}_q(t,k,m)$& --- &a subspace transversal design of block dimension $k$,\\
&  &groupsize $q^m$ and strength $t$\\
$\textmd{OA}_{\lambda}(N,k,s,t)$& --- &an $N\times k$ orthogonal array with $s$ levels, strength $t$, and index $\lambda$\\
\end{tabular}

\cleardoublepage


\chapter{Introduction}
\label{ch:Introduction}
\section{Codes in Projective Space}
Let $(M,d)$ be a metric space, where $M$ is a finite set, and $d$
is a metric defined on $M$. A code $C$ in $M$ is a collection of
elements of $M$; it has minimum distance $d$,
if for each two different elements $A,B\in M$, $d(A,B)\geq d$.

Let $\F_q$ be the finite field of size $q$. The {\it projective space}
of order $n$ over~\smash{$\F_q$}, denoted by $\Ps$, is the set of all subspaces
of the vector space~\smash{$\F_q^n$}.  Given a nonnegative integer
$k \leq n$, the set of all $k$-dimensional subspaces of \smash{$\F_q^n$}
forms the \emph{Grassmannian space} (Grassmannian in short) over $\F_q$,
which is denoted by $\Gr$. Thus, $ \Ps = \bigcup_{0 \le k \le n} \Gr $.
It is well known that
$$|\Gr|=\Gauss{n}{k}=\prod_{i=0}^{k-1}\frac{q^{n-i}-1}{q^{k-i}-1},$$
where $\Gauss{n}{k}$ is the $q$-ary Gaussian coefficient.
The projective space and the Grassmannian are metric spaces with
the distance function, called \emph{subspace distance}, defined by
\begin{equation}
\label{def_subspace_distance} d_S (X,\!Y) \deff  \dim X + \dim Y
-2 \dim\bigl( X\, {\cap}Y\bigr),
\end{equation}
for any two subspaces $X$ and $Y$ in $\Ps$.

A subset $\C$ of the projective space is called an  $(n,M,d_S)_q$
{\it code in projective space} if it has size $M$ and minimum
distance $d_S$.
If an $(n,M,d_S)_q$ code $\C$ is contained in $\Gr$
for some~$k$,~we say that $\C$ is an $(n,M,d_S,k)_q$  {\it constant
dimension code}.
The $(n,M,d)_q$, respectively $(n,M,d,k)_q$,
codes in projective space are akin to the familiar codes in the
Hamming space, respectively constant-weight codes in the Johnson
space, where the Hamming distance serves as the metric.

Koetter and Kschischang~\cite{KK} showed that codes in $\Ps$ are
precisely what is needed for error-correction in random network
coding~\cite{YeCa06,CaYe06}. This is the motivation to
explore error-correcting codes in $\Ps$.

\section{Random Network Coding}

A network is a directed graph, where the edges represent pathways
for information. Using the max-flow min-cut theorem, one can calculate
the maximum amount of information that can be pushed through this
network between any two graph nodes. It was shown that simple forwarding
of information between the nodes is not capable of attaining the max-flow
value. Rather, by allowing mixing of data at intermediate network
nodes this value can be achieved. Such encoding is referred to as
\emph{network coding}~\cite{ACLY00,HKMKE, HMKKESL}.

\begin{figure}
    \begin{center}
\includegraphics[scale=0.15]{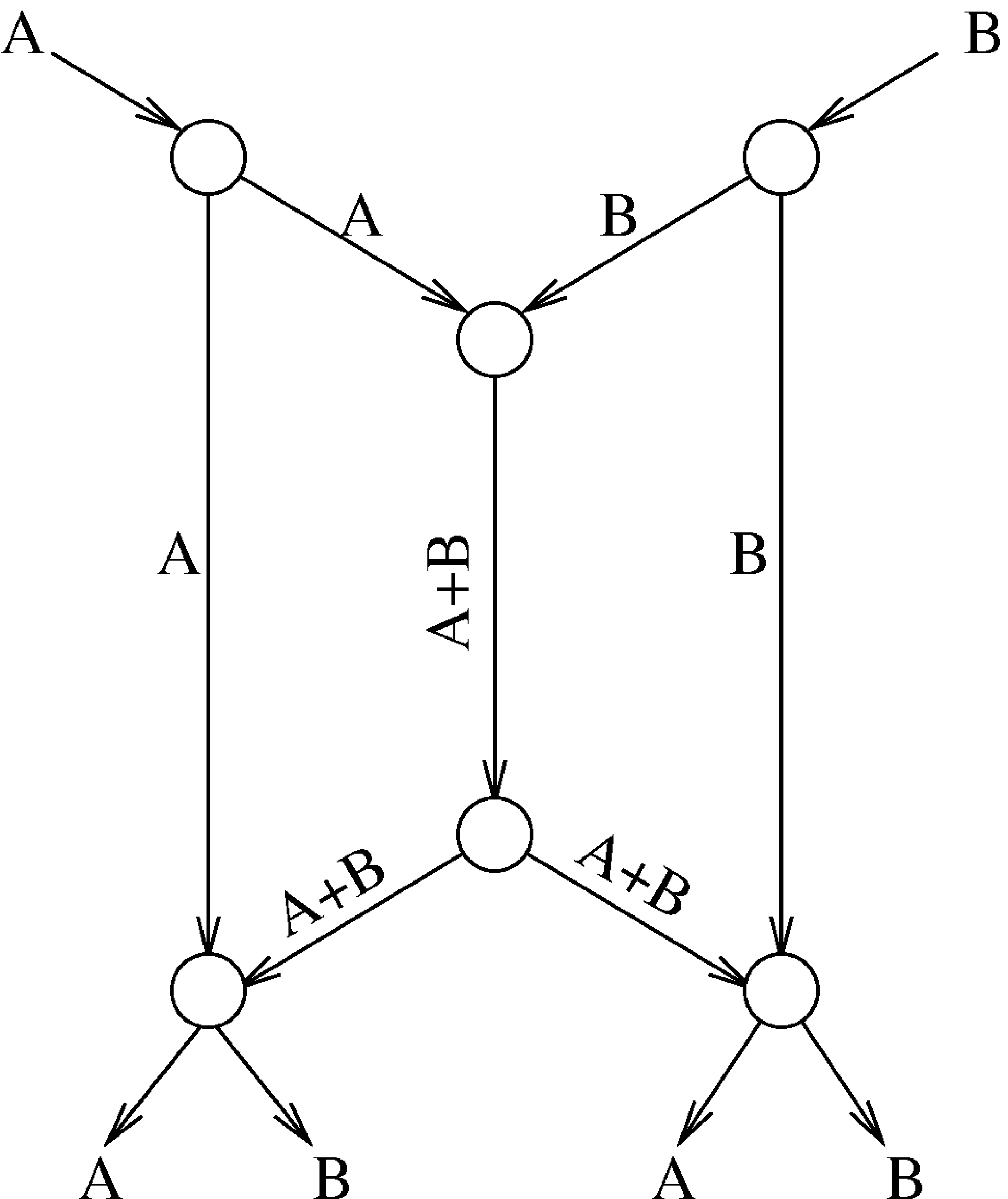}
\end{center}
\caption{\label{fig:Butterfly-Network}Network coding example. Max-flow is
attainable only through the mixing of information at intermediate
nodes.}
\end{figure}

In the example in Figure~\ref{fig:Butterfly-Network}, two sources
having access to bits A and B at a rate of one bit per unit time, have
to communicate these bits to two sinks, so that both sinks receive
both bits per unit time. All links have a capacity of one bit per
unit time. The network problem can be satisfied with the transmissions
specified in the example but cannot be satisfied with only forwarding
of bits at intermediate packet nodes.

\subsection{Errors and Erasures Correction in Random Network Coding}

Now we describe the network coding model proposed by Koetter and Kschischang~\cite{KK}.
Consider a communication between a single source and a single destination node. During each
generation, the source node injects $m$ packets $x_1,x_2,\ldots,x_m\in \F_q^n$ into the network.
When an intermediate node has a transmission opportunity,
it creates an outgoing packet as a random $\F_q$-linear combination of the incoming packets.
The destination node collects such randomly generated packets $y_1,y_2,\ldots,y_N\in \F_q^n$,
and tries to recover the injected packets into the network.
The matrix form representation of the transmission model is
$$
Y = HX,
$$
where $H$ is a random $N\times m$ matrix, corresponding to the overall linear
transformation applied to the network, $X$ is the $m\times n$ matrix
whose rows are the transmitted
packets, and $Y$ is the $N\times n$ matrix whose rows are the received packets.
Note, that there is no assumption here that the network operates synchronously
or without delay or that the network is acyclic.

If we consider the extension of this model by incorporation of $T$ packet errors
$e_1,e_2,\ldots, e_T$ then the matrix form representation of the transmission model is given by
$$
Y = HX + GE,
$$
where $X, Y$, and $E$ are $m\times n$, $N\times n$, and $T\times n$ matrices, respectively,
whose rows represent the transmitted,
received, and erroneous packets, respectively, and $H$ and $G$ are
corresponding random $N\times m$ and $N\times T$ matrices induced by linear network coding.

Note, that the only property of the matrix $X$ that is preserved
under the unknown linear transformation applied by random network coding,
is its row space. Therefore, the information can be encoded by the
choice of the vector space spanned by the rows of $X$, and not by the choice of $X$.
Thus, the input and output alphabet for the underlying channel,
called \emph{operator channel}, is $\Ps$. In other words, an operator channel
takes in a vector space and outputs another vector space, possibly
with errors, which can be of two types: \emph{erasures }(deletion of vectors from the
transmitted space), and \emph{errors} (addition of vectors to the transmitted space).

It was proved in~\cite{KK}, that an $(n,M,d)_q$ code in the projective space can
correct any $t$ packet errors and
any $\rho$ packet erasures introduced (adversatively) anywhere in
the network as long as $2t + 2\rho < d$.
\section{Rank-Metric Codes}
\label{sec:rank-metric codes}
Rank-metric codes were
introduced by Delsarte~\cite{Del78} and rediscovered
in~\cite{Gab85,Rot91}.
These
codes have
found a new application in the construction of error-correcting
codes for random network coding~\cite{SKK08}.

For two $m \times \eta$ matrices $A$ and $B$ over $\F_q$ the {\it
rank distance} is defined by
$$
d_R (A,B) \deff \text{rank}(A-B)~.
$$
An  $[m \times \eta,\varrho,\delta]$ {\it rank-metric code} $\cC$
is a linear code, whose codewords are $m \times \eta$ matrices
over $\F_q$; they form a linear subspace with dimension $\varrho$ of
$\F_q^{m \times \eta}$, and for each two distinct codewords $A$
and $B$, $d_R (A,B) \geq \delta$.
For an $[m \times \eta,\varrho,\delta]$ rank-metric code $\cC$ we have $\varrho \leq
\text{min}\{m(\eta-\delta+1),\eta(m-\delta+1)\}$
~\cite{Del78,Gab85,Rot91}. This bound, called Singleton bound for rank metric,
is attained for all
possible parameters. The codes which attain this bound are called {\it
maximum rank distance} codes (or MRD codes in short).
This definition is
generalized for a nonlinear rank-metric code,
which is a subset of $\F_q^{m \times \eta}$ with
minimum distance $\delta$ and size $q^\varrho$.
If $\varrho = \text{min}\{m(\eta-\delta+1),\eta(m-\delta+1)\}$,
then such a code  will be also called an MRD code.

An important family of MRD linear codes is presented by Gabidulin~\cite{Gab85}.
These codes can be seen as the analogs of Reed-Solomon codes for rank metric.
Without loss of generality, assume $\eta\leq m$ (otherwise we consider the transpose
of all the codewords).
A codeword $c$ in an $[m \times \eta, \varrho , \delta]$
rank-metric code $\cC$, can be represented by a
vector $c=(c_1 , c_2 , \ldots , c_{\eta})$, where $c_i \in \F_{q^m}$, since $\F_{q^m}$
can be viewed as an $m$-dimensional vector space over $\F_q$.
Let $g_i\in \F_{q^m}$, $1\leq i\leq \eta$,  be linearly independent over $\F_q$.
The generator matrix $G$ of an  $[m \times \eta,\varrho,\delta]$
Gabidulin MRD code is given by
$$
G=\left(\begin{array}{cccc}
g_{1} & g_{2} & \ldots & g_{\eta}\\
g_{1}^{[1]} & g_{2}^{[1]} & \ldots & g_{\eta}^{[1]}\\
g_{1}^{[2]} & g_{2}^{[2]} & \ldots & g_{\eta}^{[2]}\\
\ldots & \ldots & \ldots & \ldots\\
g_{1}^{[K-1]} & g_{2}^{[K-1]} & \ldots & g_{\eta}^{[K-1]}\end{array}\right),
$$
where $K=\eta-\delta+1$, $\varrho=mK$, and  $[i]=q^{i \textmd{~mod}~m}$.

\section {Related Work}

\subsection{Bounds}

Let $\mathcal{A}_{q}(n,d)$  denotes the maximum number of codewords
in an $(n,M,d)$ code in projective space, and let $\mathcal{A}_{q}(n,2\delta,k)$
denotes the maximum number of codewords in an $(n,M,2\delta,k)$ constant dimension
code. (Note that the distance between any two elements in $\Gr$ is always even).

Without loss of generality we will
assume that $k \leq n-k$. This assumption
can be justified  as a consequence of the following
lemma~\cite{EV08}.

\begin{lemma}

If $\C$ is an $(n,M,d ,k)_q$ constant dimension code then
$\C^\perp = \{ X^\perp : X \in\C\}$, where $X^\perp$ is the
orthogonal subspace of $X$, is an $(n,M,d ,n-k)_q$ constant
dimension code.
\end{lemma}


Let $S_{n,k}(X,t)$ denotes a sphere of radius $t$ in $\Gr$
centered at a subspace ${X\in \Gr}$. It was proved~\cite{KK} that the volume of
$S_{n,k}(X,t)$ is independent on $X$, since the Grassmann graph, corresponding
to $\mathcal{G}_{q}(n,k)$, is distance regular. Then we denote the volume
of a sphere of radius $t$ in $\Gr$ by $|S_{n,k}(t)|$.
\begin{lemma}~\cite{KK} Let $t\leq k$. Then
$$
|S_{n,k}(t)|=\sum_{i=0}^tq^{i^2}\Gauss{k}{i}\Gauss{n-k}{i}.
$$
\end{lemma}

Koetter and Kschischang~\cite{KK} established the following
sphere-packing and sphere-covering bounds for $\mathcal{A}_{q}(n,2\delta,k)$:

\begin{theorem}(Sphere-packing bound)
Let $t=\left\lfloor \frac{\delta-1}{2}\right\rfloor $. Then
\begin{equation}
\mathcal{A}_{q}(n,2\delta,k)\leq\frac{|\mathcal{G}_{q}(n,k)|}
{|S_{n,k}(t)|}=\frac{\left[\begin{array}{c}
n\\
k\end{array}\right]_{q}}{\overset{t}{\underset{i=0}
{\sum}}q^{i^{2}}\left[\begin{array}{c}
k\\
i\end{array}\right]_{q}\left[\begin{array}{c}
n-k\\
i\end{array}\right]_{q}}.
\label{eq:sphere-pack}
\end{equation}
\end{theorem}

\begin{theorem}(Sphere-covering bound)
\begin{equation}
\mathcal{A}_{q}(n,2\delta,k)\geq\frac{|\mathcal{G}_{q}(n,k)|}{|S_{n,k}
(\delta-1)|}=\frac{\left[\begin{array}{c}
n\\
k\end{array}\right]_{q}}{\overset{\delta-1}{\underset{i=0}{\sum}
}q^{i^{2}}\left[\begin{array}{c}k\\
i\end{array}\right]_{q}\left[\begin{array}{c}n-k\\
i\end{array}\right]_{q}}.
\label{eq:sphere-cover}
\end{equation}
\end{theorem}

Koetter and Kschischang \cite{KK} also developed
the Singleton-type bound, which is always stronger than the sphere-packing bound~(\ref{eq:sphere-pack}):
\begin{theorem}(Singleton bound)
\label{trm:Singleton}
\begin{equation}
\mathcal{A}_{q}(n,2\delta,k)\leq\left[\begin{array}{c}
n-\delta+1\\
k-\delta+1\end{array}\right]_{q}.
\label{eq:Singleton}
\end{equation}
\end{theorem}

Xia in~\cite{Xia08} showed a Graham-Sloane type lower bound:
\begin{theorem}
$$
\mathcal{A}_{q}(n,2\delta,k)\geq\frac{(q-1)\left[\begin{array}{c}
n\\
k\end{array}\right]_{q}}{(q^{n}-1)q^{n(\delta-2)}}.
$$
\end{theorem}
However, this bound is weaker than the bound~(\ref{eq:sphere-cover}).

Wang, Xing and Safavi-Naini~\cite{WXS-N03} introduced
the linear authentication codes. They showed that an $(n,M,2\delta,k)_q$
constant dimension code is exactly an $\left[n,M,n-k,\delta\right]$
linear authentication code over $GF(q)$. They also established an
upper bound on linear authentication codes, which is equivalent to
the following bound on constant dimension codes:

\begin{theorem}
\label{trm:Steiner}
\begin{equation}
\mathcal{A}_{q}(n,2\delta,k)\leq\frac{\left[\begin{array}{c}
n\\
k-\delta+1\end{array}\right]_{q}}{\left[\begin{array}{c}
k\\
k-\delta+1\end{array}\right]_{q}}.
\label{eq:Steiner}
\end{equation}
\end{theorem}

This bound was proved by using a different method by Etzion and
Vardy in~\cite{EV,EV08}. This method based on
bounds on anticodes in the Grassmannian. In~\cite{XiFu09} was shown
that the bound~(\ref{eq:Steiner}) is always stronger than the Singleton
bound~(\ref{eq:Singleton}). Furthermore, it was proved~\cite{EV, EV08} that the
codes known as Steiner structures attain the bound~(\ref{eq:Steiner}).

The following Johnson-type
bounds were presented in~\cite{EV,EV08,XiFu09}:

\begin{theorem}(Johnson bounds)
\begin{equation}
\mathcal{A}_{q}(n,2\delta,k)\leq\frac{q^{n}-1}{q^{k}-1}\mathcal{A}_{q}(n-1,2\delta,k-1),
\label{eq:Johnson1}
\end{equation}
\begin{equation}
\mathcal{A}_{q}(n,2\delta,k)\leq\frac{q^{n}-1}{q^{n-k}-1}\mathcal{A}_{q}(n-1,2\delta,k).
\label{eq:Johnson2}
\end{equation}
\end{theorem}

Using bounds~(\ref{eq:Johnson1}),
and~(\ref{eq:Johnson2}) recursively,
and combining with the observation
that $\mathcal{A}_{q}(n,2\delta,k)=1$ for all $k<2\delta$, the following
bound is obtained~\cite{EV,EV08,XiFu09}:
\begin{theorem}
$$
\mathcal{A}_{q}(n,2\delta,k)\leq\left\lfloor \frac{q^{n}-1}{q^{k}-1}\left\lfloor \frac{q^{n-1}-1}{q^{k-1}-1}\cdots\left\lfloor \frac{q^{n-k+\delta}-1}{q^{\delta}-1}
\right\rfloor \cdots\right\rfloor \right\rfloor .
$$
\end{theorem}

The upper and lower bounds on $\mathcal{A}_{q}(n,2\delta,k)$ when
$\delta=k$ were considered in~\cite{EV, EV08}:
\begin{theorem}
\begin{equation}
\mathcal{A}_{q}(n,2k,k)\leq\left\lfloor \frac{q^{n}-1}{q^{k}-1}\right
\rfloor -1,\;\mbox{if }k\nmid n,
\label{eq:up bound k not div n}\end{equation}
\begin{equation}
\mathcal{A}_{q}(n,2k,k)=\frac{q^{n}-1}{q^{k}-1},\mbox{ if }k\mid n,
\label{eq:up bound k div n}\end{equation}
\begin{equation}
\mathcal{A}_{q}(n,2k,k)\geq\frac{q^{n}-q^{k}(q^{r}-1)-1}{q^{k}-1},
\mbox{ where }n\equiv r\:(\mbox{mod }k).
\label{eq:lower bound EV}
\end{equation}
\end{theorem}

\vspace{0.5cm}
The following two bounds on $\mathcal{A}_q(n,d)$ are presented in~\cite{EV,EV08}.

\begin{theorem}(Gilbert-Varshamov bound)
\[
\mathcal{A}_{q}(n,d)\geq\frac{|\mathcal{P}_{q}(n)|^{2}}{\underset{k=0}
{\overset{n}{\sum}}\underset{j=0}{\overset{d-1}{\sum}}\underset{i=0}
{\overset{j}{\sum}}\left[\begin{array}{c}
n-k\\
j-i\end{array}\right]_{q}\left[\begin{array}{c}
k\\
i\end{array}\right]_{q}\left[\begin{array}{c}
n\\
k\end{array}\right]_{q}q^{i(j-i)}}.\]
\end{theorem}

This lower bound generalize the Gilbert-Varshamov bound for graphs that are not
necessarily distance-regular.

The following upper bound on $\mathcal{A}_q(n,d)$~\cite{EV,EV08} is obtained
by using a linear programming (LP) method.
\begin{theorem}(LP bound)
\[
\mathcal{A}_{q}(n,2e+1)\leq f^{\star},\]
where $f^{\star}=\max\left\{ D_{0}+D_{1}+\cdots D_{n}\right\} $,
subject to the following $2n+2$ linear constraints:\[
\sum_{j=-e}^{e}c(i+j,i,e)D_{i+j}\leq\left[\begin{array}{c}
n\\
i\end{array}\right]_{q}\]
\[
\mbox{and }D_{i}\leq\mathcal{A}_{q}(n,2e+2,i),\]
for all $0\leq i\leq n$, where $D_{i}$ denote the number of codewords
with dimension $i$ and $c(k,i,e)$ denote the size of the set
$\left\{ X\::\: d_{s}(X,Y)\leq e,\:\dim X=i\right\} $
for a $k$-dimensional subspace $Y$.

\end{theorem}

\subsection{Constructions of Codes}
\label{subsec:known constructions}

Koetter and Kschischang~\cite{KK} presented a construction
of Reed-Solomon like constant dimension codes.
They showed that these codes attain the Singleton bound asymptotically.

Silva, Koetter, and
Kschischang~\cite{SKK08} showed that this construction can be
described in terms of rank-metric codes.

Let $A$ be an $m \times \eta$ matrix over $\F_q$, and let $I_m$ be an
$m \times m$ identity matrix. The matrix $[ I_m ~ A ]$ can be
viewed as a generator matrix of an $m$-dimensional subspace of $\F_q^{m+\eta}$.
This subspace is called the \emph{lifting} of $A$ ~\cite{SKK08}.

\begin{example}
Let $A$ and $[I_3 ~A]$ be the following matrices over $\F_2$
$$A=\left( \begin{array}{ccc}
1& 1 & 0\\
0& 1 & 1\\
0& 0 & 1
\end{array}
\right) ~,
[I_3 ~ A] =\left( \begin{array}{cccccc}
1&0&0&1& 1 & 0\\
0&1&0&0& 1 & 1\\
0&0&1&0& 0 & 1
\end{array}
\right),
$$
then the $3$-dimensional subspace $X$, the lifting of $A$, is given by the following $8$ vectors:
$$X = (\{1 0 0 1 1 0),
(0 1 0 0 1 1), (0 0 1 0 0 1),(1 1 0 1 0 1),$$
$$(1 0 1 1 1 1),(0 1 1 0 1 0),
(1 1 1 1 0 0),(0 0 0 0 0 0)\}.
$$
\end{example}

 A constant dimension
code $\mathbb{C}\subseteq\Gr$ such that all its codewords are lifted codewords
of a rank-metric code $\cC\subseteq\F_q^{k\times(n-k)}$, i.e., $\C = \{\textmd{row space} [I_k ~ A] : A \in \cC\}$,
is called the \emph{lifting of $\cC$}~\cite{SKK08}.

\begin{theorem}~\cite{SKK08}
\label{trm:CDC from rank-metric}
If $\cC$ is a $[k \times (n-k),\varrho,\delta]$ rank-metric code,
then the constant dimension code $\C$ obtained by the lifting of
$\cC$
is an $(n,q^\varrho,2\delta,k)_q$ code.
\end{theorem}

A constant dimension
code $\mathbb{C}$ such that all its codewords are lifted codewords
of an  MRD code
is called a \emph{lifted MRD code}~\cite{SKK08}.
This code will be denoted by $\C^{\textmd{MRD}}$.
\vspace{0.3cm}

Manganiello, Gorla and Rosenthal~\cite{MGR08} showed the
construction of spread codes, i.e. codes that have the maximal possible
distance in the Grassmannian. This construction can be viewed as a generalization of the lifted
MRD code construction.

Skachek~\cite{Ska10} provided a recursive construction for constant dimension codes, which
can be viewed as a generalization of the construction in~\cite{MGR08}.

Gadouleau and Yan~\cite{GaYa10} proposed a construction of constant dimension codes based on
constant rank codes.

Etzion and Vardy~\cite{EV,EV08} introduced a construction
of codes in $\mathcal{G}_{q}(n,k)$ based on a Steiner structure, that
attain the bound (\ref{eq:Steiner}). They proved that any Steiner structure
$S_{q}(t,k,n)$ is an $(n,M,2\delta,k)$ code in $\mathcal{G}_{q}(n,k)$
with $M=\left[\begin{array}{c}
n\\
t\end{array}\right]_{q}/\left[\begin{array}{c}
k\\
t\end{array}\right]_{q}$ and ${\delta=k-t+1}$.
They also
developed computational methods to search for the codes with a certain
structure, such as cyclic codes, in $\mathcal{P}_{q}(n)$.

Kohnert and Kurz~\cite{KoKu08} described a construction of constant dimension codes
in terms of $0-1$ integer programming. However, the dimensions of such an optimization problem
are very large in this context. It was shown in~\cite{KoKu08} that by prescribing a group
of automorphisms of a code, it is possible significantly
reduce the size of the problem. Large codes with constant dimension $k=3$ and $n\leq 14$ were
constructed by using this method.

\begin{remark} Silva and Kschischang~\cite{SK09} proposed a new subspace metric, called the injection metric,
for error correction in network coding, given by
$$d_I(X, Y)=\max\{\dim(X),\dim(Y)\}-\dim(X\cap Y),
$$
for any two subspaces  $X,Y\in\Ps$.
It was shown~\cite{SK09} that codes in $\Ps$ designed for $d_I$ may have higher rates
than those designed for $d_S$.
The injection distance and the subspace distance are closely related~\cite{SK09}:
$$
d_I(X, Y)=\frac{1}{2}d_S(X, Y)+\frac{1}{2}|\dim(X)-\dim(Y)|,
$$
therefore, these two metrics are equivalent for the Grassmannian.
The bounds and constructions of codes in $\Ps$ for the injection metric are presented in~\cite{GaYa10a},
~\cite{KhKs09}, and ~\cite{KhSK09}.
\end{remark}

\section{Organization of This Work}
The rest of this thesis is organized as follows.
In Chapter~\ref{ch:representation and distance} we discuss different
representations of subspaces in the projective space and present a new
formula for the computation of the
distance between any two different subspaces in $\Ps$.
In  Section~\ref{sec:representations}
we consider the
representations of subspaces in $\Ps$.
We define the reduced row echelon
form of a $k$-dimensional subspace and its Ferrers diagram. These
two concepts combined with the identifying vector of a
subspace will be our main tools for the
representation of subspaces. In Section~\ref{sec:distance} we present a formula
for an efficient computation of the distance between two subspaces in the projective space.


In Chapter~\ref{ch:designs} we consider lifted MRD codes.
In Section~\ref{sec:lifted MRD} we discuss properties of these codes related to block
designs. We prove that the codewords of a lifted MRD code form a design
called a transversal design, a structure which is known to be
equivalent to the well known orthogonal array. We also prove that
the same codewords form a subspace transversal design, which is
akin to the transversal design, but not its $q$-analog.
In Section~\ref{sec:Linear codes} we show that these designs can be used to derive a new
family of linear codes in the Hamming space, and in particular,
LDPC codes. We provide upper and lower bounds on the minimum distance,
the stopping distance and the dimension of such codes. We prove that there
are no small trapping sets in such codes.  We prove that some of these codes are
quasi-cyclic and attain the Griesmer bound.

In Chapter~\ref{ch:bounds and constr} we present new bounds and constrictions
for constant dimension codes. In Section~\ref{sec:multilevel construction} we
present the multilevel construction. This construction
requires rank-metric codes in which some of the entries are forced
to be zeroes due to constraints given by the Ferrers diagram. We
first present an upper bound on the size of such codes. We show
how to construct some rank-metric codes which attain this bound.
Next, we describe the
multilevel construction of the constant dimension codes. First, we
select a constant weight code $\bf{C}$. Each codeword of $\bf{C}$
defines a skeleton of a basis for a subspace in reduced row
echelon form. This skeleton contains a Ferrers diagram on which we
design a rank-metric code. Each such rank-metric code is lifted to
a constant dimension code. The union of these codes is our final
constant dimension code. We discuss the parameters of these codes
and also their decoding algorithms.
In Section~\ref{sec:bounds contructions from designs}
we derive upper bounds on codes that contain
lifted MRD codes, based on their combinatorial structure, and
provide  constructions for two families of codes
that attain these upper bounds. The first construction can be considered
as a generalization of the multilevel method presented in
Section~\ref{sec:multilevel construction}.
This construction based also
on an one-factorization of a complete graph. The second construction is based
on the existence of a $2$-parallelism in $\mathcal{G}_q(4,2)$.
In Section~\ref{sec:punctured}
we generalize the well-known concept of a punctured code for a
code in the projective space. Puncturing in the projective space
is more complicated than its counterpart in the Hamming space. The
punctured codes of our constant dimension codes have larger size
than the codes obtained by using the multilevel approach described
in Section~\ref{sec:multilevel construction}.
We discuss the parameters of the
punctured code and also its decoding algorithm.

The main goal of Chapter~\ref{ch:Enum and Lexi} is to present efficient enumerative
encoding and decoding techniques for the Grassmannian and to describe
a general search method for constant dimension
lexicodes. In Section~\ref{sec:lexi order} we present two lexicographic orders for the
Grassmannian, based on different representations of subspaces in the Grassmannian.
In Section~\ref{sec:enum coding} we describe the enumerative coding methods, based
on different lexicographic orders, and discuss their computation complexity.
Section~\ref{sec:lexicodes} deals with constant dimension lexicodes.

Finally, we conclude with Chapter~\ref{ch:conclusion}, where we summarize our results and present a list
of open problems for further research.


\chapter[Representations of Subspaces and Distance Computation]{Representations of Subspaces and Distance Computation\let\thefootnote\relax\footnotetext{The results
presented in this chapter were published in~\cite{SiEt09}
and~\cite{SiEt10}.}}

\label{ch:representation and distance}

In this chapter we first consider different representations of a subspace in $\Ps$.
The constructions for
codes in $\Ps$ and $\Gr$, the enumerative coding methods, and the search for  lexicodes,
presented in the following chapters, are based on these
representations.
Next, we present a new formula for the computation of the
distance of two different subspaces in $\Ps$. This formula enables to simplify
the computations that lead to the next subspace in the search for a constant dimension
lexicode which will be described in the sequel.

\section{Representations of Subspaces}
\label{sec:representations}

In this section we define the reduced row echelon
form of a $k$-dimensional subspace and its Ferrers diagram. These
two concepts combined with the identifying vector of a
subspace will be our main tools for the
representation of subspaces. We also define and discuss some types
of integer partitions which have an important role in our exposition.

\subsection{Reduced Row Echelon Form Representation}
\label{subsec:reduced row echelon form representation}

A matrix is said to be in {\it row echelon form} if each nonzero
row has more leading zeroes than the previous row.

A $k \times n$ matrix with rank $k$ is in {\it reduced row echelon
form} (RREF) if the following conditions are satisfied.
\begin{itemize}
\item The leading coefficient (pivot) of a row is always to the right of
the leading coefficient of the previous row.
\item All leading coefficients are {\it ones}.
\item Each leading coefficient is the only nonzero entry in its
column.
\end{itemize}

A $k$-dimensional subspace $X$ of $\F_q^n$ can be represented by a
$k \times n$ {\it generator matrix} whose rows form a basis for
$X$.  There is exactly one such matrix in RREF and it will be denoted by
$\mbox{RE} (X)$. For simplicity, we will
assume that the entries in $\mbox{RE} (X)$ are taken from $\Z_q$
instead of $\F_q$, using an appropriate bijection.

\begin{example}
\label{exm:running} We consider the $3$-dimensional subspace  $X$ of
$\F_2^7$ with the following eight elements.
\begin{align*}
\begin{array}{cccccccc}
\text{1)} & (0 & 0 & 0 & 0 & 0 & 0 & 0) \\
\text{2)} & (1 & 0 & 1 & 1 & 0 & 0 & 0) \\
\text{3)} & (1 & 0 & 0 & 1 & 1 & 0 & 1) \\
\text{4)} & (1 & 0 & 1 & 0 & 0 & 1 & 1) \\
\text{5)} & (0 & 0 & 1 & 0 & 1 & 0 & 1) \\
\text{6)} & (0 & 0 & 0 & 1 & 0 & 1 & 1) \\
\text{7)} & (0 & 0 & 1 & 1 & 1 & 1 & 0) \\
\text{8)} & (1 & 0 & 0 & 0 & 1 & 1 & 0)
\end{array} .
\end{align*}
The subspace $X$ can be represented by a $3 \times 7$ generator matrix whose
rows form a basis for the subspace. There are 168 different
matrices for the 28 different bases. Many of these matrices are in
row echelon form. One of them is
\begin{align*}
\left( \begin{array}{ccccccc}
1 & 0 & 1 & 0 & 0 & 1 & 1 \\
0 & 0 & 1 & 1 & 1 & 1 & 0 \\
0 & 0 & 0 & 1 & 0 & 1 & 1
\end{array}
\right) .
\end{align*}
Exactly one of these 168 matrices is in reduced row echelon form:
\begin{align*}
\emph{RE}(X) =\left( \begin{array}{ccccccc}
1 & 0 & 0 & 0 & 1 & 1 & 0 \\
0 & 0 & 1 & 0 & 1 & 0 & 1 \\
0 & 0 & 0 & 1 & 0 & 1 & 1
\end{array}
\right).
\end{align*}
\end{example}
\vspace{0.6cm}

\subsection{Ferrers Tableaux Form Representation}
\label{subsec:Ferrers Tableaux Form Representation}

\vspace{0.3cm}
\subsubsection{Partitions}
\label{subsubsec:partitions}

A \textit{partition} of a positive integer $t$ is a representation
of $t$ as  a sum of positive integers, not necessarily distinct.
We order this collection of integers in a decreasing order.

A {\it Ferrers diagram} $\cF$ represents a partition as a pattern
of dots with the $i$-th row having the same number of dots as the
$i$-th term in the partition~\cite{And84,vLWi92,Sta86}. In the
sequel, a {\it dot} will be denoted by a $"\bullet"$. A Ferrers
diagram satisfies the following conditions.
\begin{itemize}
\item The number of dots in a row is at most the number of dots in
the previous row.
\item All the dots are shifted to the right of the diagram.
\end{itemize}

\begin{remark}
Our definition of Ferrers diagram is slightly different from the
usual definition~\cite{And84,vLWi92,Sta86}, where the dots in
each row are shifted to the left of the diagram.
\end{remark}

\vspace{0.3cm}
Let $| \cF |$ denotes the {\it size} of a Ferrers diagram $\cF$,
i.e., the number of dots in $\cF$.
The {\it number of rows (columns)} of the Ferrers diagram $\cF$ is
the number of dots in the rightmost column (top row) of $\cF$. If
the number of rows in the Ferrers diagram is $m$ and the number of
columns is $\eta$ we say that it is an $m \times \eta$ Ferrers
diagram.

If we read the Ferrers diagram by columns we get another partition
which is called the {\it conjugate} of the first one. If the
partition forms an $m \times \eta$ Ferrers diagram then the
conjugate partition forms an $\eta \times m$ Ferrers diagram.

\begin{example}
\label{ex:Ferrers} Assume we have the partition $6+5+5+3+2$ of 21.
The $5 \times 6$ Ferrers diagram $\cF$ of this partition is given
by
\begin{align*}
\begin{array}{cccccc}
\bullet & \bullet & \bullet & \bullet & \bullet & \bullet \\
& \bullet & \bullet & \bullet & \bullet & \bullet \\
& \bullet & \bullet & \bullet & \bullet & \bullet \\
& & & \bullet & \bullet & \bullet \\
& & & & \bullet & \bullet
\end{array}.
\end{align*}
The number of rows in $\cF$ is 5 and the number of columns is 6.
The conjugate partition is the partition $5+5+4+3+3+1$ of 21 and
its $6 \times 5$ Ferrers diagram is given by
\begin{align*}
\begin{array}{ccccc}
\bullet & \bullet & \bullet & \bullet &  \bullet \\
\bullet & \bullet & \bullet & \bullet & \bullet \\
& \bullet & \bullet & \bullet & \bullet \\
& & \bullet & \bullet & \bullet \\
& & \bullet & \bullet & \bullet  \\
& & & & \bullet
\end{array} .
\end{align*}
\end{example}
\vspace{0.3cm}

The \emph{partition function} $p(t)$ is the number of different partitions of
$t$~\cite{And84,vLWi92,Sta86}.
The following lemma presented in~\cite[p. 160]{vLWi92} provides an upper bound on this function.

\begin{lemma}
\label{lm:upper bound partitions}
 $p(t)< e^{\pi \sqrt{\frac{2}{3}t}}$.
\end{lemma}

Let $ \cF $ be an $m\times \eta$ Ferrers diagram. If $m\leq \alpha$ and $\eta\leq \beta$,
we say that $\cF$ is \emph{embedded} into an $\alpha\times \beta$  box.
Let $p(\alpha,\beta,t)$ be the
number of partitions of $t$ whose Ferrers diagrams can  be embedded into an $\alpha\times \beta$ box.
The following result was given in~\cite[pp. 33-34]{And84}.
\begin{lemma}
\label{lem: recursion} $p(\alpha, \beta,t)$ satisfies the following
recurrence relation:
\begin{equation}
\label{eq:rec_part} p(\alpha, \beta,t)=p(\alpha, \beta-1,t-\alpha)+p(\alpha-1, \beta, t),
\end{equation}
with the initial conditions
\begin{equation}
\label{eq:initial_cond}
 p(\alpha, \beta,t)=0 ~~ \textrm{if} ~~ t < 0~
\textrm{or} ~ t > \beta \cdot \alpha ,
~\textrm{ and }~ p(\alpha, \beta,0)=1.
\end{equation}
\end{lemma}
\vspace{0.2cm}

\begin{lemma}
\label{lem:bound_p} For any given $\alpha$, $\beta$, and $t$, we have
${p(\alpha, \beta,t)< e^{\pi \sqrt{\frac{2}{3}t}}}$.
\end{lemma}
\begin{proof}
Clearly, $p(\alpha, \beta,t)\leq p(t)$, where $p(t)$ is the number of
unrestricted partitions of $t$. Then by Lemma~\ref{lm:upper bound partitions}
we have
that $p(t)< e^{\pi \sqrt{\frac{2}{3}t}}$ and thus ${p(\alpha, \beta,t)< e^{\pi \sqrt{\frac{2}{3}t}}}$.
\end{proof}

\vspace{0.3cm}
The following theorem~\cite[p. 327]{vLWi92} provides a connection
between the $q$-ary Gaussian coefficients and
partitions.
\begin{theorem}
\label{thm:vanLint} For any given integers $k$ and $n$, $0 < k
\leq n$, \[ \left[\begin{array}{c}
n\\
k\end{array}\right]_{q}=\sum_{t=0}^{k(n-k)}\alpha_t q^t,\] where
$\alpha_t=p(k,n-k,t)$.
\end{theorem}

\vspace{0.2cm}

\subsubsection{Ferrers Tableaux Form Representation}
\label{subsubsec:Ferrers Tableaux Form Representation}

For each $X\in \Gr$  we associate a binary vector of length $n$
and weight $k$, denoted by $v(X)$, called the \emph{identifying vector} of
$X$, where the \textit{ones} in $v(X)$ are exactly in the
positions where $\mbox{RE}(X)$ has the leading \textit{ones}.

\begin{example}
Consider the $3$-dimensional subspace $X$ of
Example~\ref{exm:running}. Its identifying vector is
$v(X)=1011000$.
\end{example}

\begin{remark}
We can consider an identifying vector $v(X)$ for some
$k$-dimensional subspace $X$ as a characteristic vector of a
$k$-subset. This coincides with the definition of rank- and
order-preserving map $\phi$ from $\Gr$ onto the lattice of subsets
of an $n$-set, given by Knuth~\cite{Knu71} and discussed by
Milne~\cite{Milne82}.
\end{remark}

The {\it echelon Ferrers form} of a binary vector $v$ of length
$n$ and weight $k$, denoted by $\mbox{EF}(v)$, is the $k\times n$ matrix in RREF
with leading entries (of rows) in the columns indexed by the
nonzero entries of $v$ and ``$\bullet$''  in all entries which do
not have terminal {\it zeroes} or {\it ones}. This
notation is also given in~\cite{vLWi92,Sta86}.
The dots of this matrix form the Ferrers diagram~$\cF$ of $\mbox{EF}(v)$.
Let $v(X)$ be the identifying vector of  a subspace $X\in \Gr$.
Its echelon Ferrers form $\mbox{EF}(v(X))$ and the  corresponding Ferrers diagram,
denoted by $\cF_X$, will be called the echelon Ferrers
form  and the Ferrers diagram of the subspace $X$, respectively.

\vspace{0.3cm}
\begin{example}
For the vector $v=1011000$, the echelon Ferrers form $\emph{EF}(v)$ is
the following $3 \times 7$ matrix:
\begin{align*}
\emph{EF}(v)=\left[ \begin{array}{ccccccc}
1 & \bullet & 0 & 0 & \bullet & \bullet & \bullet \\
0 & 0 & 1 & 0 & \bullet & \bullet & \bullet \\
0 & 0 & 0 & 1 & \bullet & \bullet & \bullet
\end{array}
\right]~.
\end{align*}
The Ferrers diagram of $\emph{EF}(v)$ is given by
\begin{align*}
\begin{array}{cccc}
\bullet & \bullet & \bullet & \bullet \\
&\bullet & \bullet & \bullet\\
&\bullet & \bullet & \bullet
\end{array}.
\end{align*}
\end{example}

\begin{remark}
All the  binary vectors of the length $n$ and weight $k$
can be considered as  the identifying vectors of all the subspaces
in  $\Gr$. These  $\binom{n}{k}$ vectors partition  $\Gr$
into the $\binom{n}{k}$ different classes, where each class
consists of all the subspaces in  $\Gr$ with the same identifying
vector. These classes are called \emph{Schubert cells}~\cite[p. 147]{Ful97}.
Note that each Schubert cell contains all the subspaces with the same
given echelon Ferrers form.
\end{remark}

The {\it Ferrers tableaux form} of a subspace $X$, denoted by
$\cF(X)$, is obtained by assigning the values of $\mbox{RE}(X)$ in the
Ferrers diagram $\cF_X$ of $X$. In other words, $\cF(X)$ is obtained
from $\mbox{RE}(X)$ first by removing from
each row of $\mbox{RE}(X)$ the {\it zeroes} to the left of the
leading coefficient; and after that removing the columns which
contain the leading coefficients. All the remaining entries are
shifted to the right. Each Ferrers tableaux form
represents a unique subspace in $\Gr$.

\begin{example}  Let $X$ be a subspace in $\mathcal G_2(7,3)$ from Example~\ref{exm:running}.
Its
echelon Ferrers form, Ferrers diagram, and Ferrers tableaux form
are given by
$$\left[ \begin{array}{ccccccc}
1 & \bullet & 0 & 0 & \bullet & \bullet & \bullet \\
0 & 0 & 1 & 0 & \bullet & \bullet & \bullet \\
0 & 0 & 0 & 1 & \bullet & \bullet & \bullet
\end{array}\right],~~~
\begin{array}{cccc}
 \bullet & \bullet & \bullet & \bullet \\
  & \bullet & \bullet & \bullet   \\
  & \bullet & \bullet & \bullet  \\
\end{array},~~~
\; \textrm{and }\;
\begin{array}{cccc}
0 & 1 & 1 & 0 \\
&1 & 0 & 1  \\
&0 & 1 & 1
\end{array},\; \textrm{respectively }.$$
\end{example}

\subsection{Extended Representation}
\label{subsec:Extended Representation}
Let $X\in\Gr$ be a $k$-dimensional subspace. The {\it extended
representation}, $\mbox{EXT}(X)$, of $X$ is a $(k+1)\times n$
matrix obtained by combining the identifying vector
$v(X)=(v(X)_n,\ldots,v(X)_1)$ and the RREF
$\mbox{RE}(X)=(X_n,\ldots,X_1)$, as follows
\begin{align*}
\mbox{EXT}(X)=\left( \begin{array}{cccc}
 v(X)_n & \ldots & v(X)_2 & v(X)_1 \\
 X_n & \ldots & X_2 & X_1
\end{array}
\right).
\end{align*}
Note, that $v(X)_n$ is the most significant bit of $v(X)$. Also,
$X_i$ is a column vector and $v(X)_i$ is the most significant bit
of the column vector $\begin{footnotesize} \left(
\begin{array}{c} v(X)_i
\\X_i
\end{array}\right) \end{footnotesize}$.

\vspace{0.1cm}

\begin{example} Consider the $3$-dimensional subspace $X$ of
Example~\ref{exm:running}. Its extended representation is given by
\begin{align*}
\emph{EXT}(X)= \left( \begin{array}{ccccccc}
1 & 0 & 1 & 1 & 0 & 0 & 0\\
1 & 0 & 0 & 0 & 1 & 1 & 0 \\
0 & 0 & 1 & 0 & 1 & 0 & 1 \\
0 & 0 & 0 & 1 & 0 & 1 & 1
\end{array}
\right) ~.
\end{align*}
\end{example}

\vspace{0.1cm}

The extended representation is redundant since the RREF defines a
unique subspace. Nevertheless, we will see in the sequel
that this representation will lead to
more efficient enumerative coding.
 Some insight for this will be
the following well known equality given in~\cite[p. 329]{vLWi92}.
\begin{lemma}
\label{lem:pascal} For all integers $q$, $k$, and $n$, such that
$k\leq n$ we have
\begin{equation}
\label{eq:pascal} \begin{small}\left[\begin{array}{c}
n\\
k\end{array}\right]\end{small}_{q}=q^k\begin{small}\left[\begin{array}{c}
n-1\\
k\end{array}\right]\end{small}_{q}+\begin{small}\left[\begin{array}{c}
n-1\\
k-1\end{array}\right]\end{small}_{q}~.
\end{equation}
\end{lemma}
\vspace{0.4cm} The lexicographic order of the Grassmannian that will be
discussed in Section~\ref{sec:lexi order} is
based on Lemma~\ref{lem:pascal} (applied recursively). Note that
the number of subspaces in which $v(X)_1=1$ is
$\begin{footnotesize} \left[\begin{array}{c}
n-1\\
k-1\end{array}\right]_{q}
\end{footnotesize}$ and the number of subspaces in which $v(X)_1=0$
is $\begin{footnotesize} q^k\left[\begin{array}{c}
n-1\\
k\end{array}\right]_{q}
\end{footnotesize}$.

\section{Distance Computation}
\label{sec:distance}

The research on error-correcting codes in the projective space in
general and on the search for lexicodes in the Grassmannian (which will
be considered in the sequel) in particular,
requires many computations of the distance between two subspaces
in $\Ps$.
The motivation is to simplify the computations that lead to the next
subspace which will be joined to a lexicode.

Let $A*B$ denotes the concatenation  $\left(\begin{footnotesize}
\begin{array}{c}
A\\B\end{array}\end{footnotesize}\right)$ of two matrices $A$ and $B$
with the same number of columns.
By the definition of the subspace distance (\ref{def_subspace_distance}),
it follows that
\begin{align}
d_S(X,Y) & =  2\rank(\textmd{RE}(X)*\textmd{RE}(Y))
 -  \rank(\textmd{RE}(X)) - \rank(\textmd{RE}(Y)).\label{distance-rank}
\end{align}

Therefore, the computation of $d_S(X,Y)$ can be done by using
Gauss elimination. In this section we present an improvement on
this computation by using the  representation of subspaces by
Ferrers tableaux forms, from which their identifying vectors  and
their RREF are easily determined. We will present an alternative
formula for the computation of the distance between two subspaces
$X$ and $Y$ in $\Ps$.

For $X \in \Gk$ and $Y \in \Gkk$, let  $\rho(X,Y)$ [$\mu(X,Y)$] be
a set of  coordinates with common \textit{zeroes}
[\textit{ones}] in $v(X)$  and $v(Y)$, i.e.,
\[
 \rho(X,Y)=\left\{  i |\;v(X)_i=0\mbox{ and }v(Y)_i
=0\right\} \]
 and \[
\mu(X,Y) =\left\{  i|\;v(X)_i=1\mbox{ and }v(Y)_i
=1\right\} .\]

Note that $|\rho(X,Y)|+|\mu(X,Y)|+d_{H}(v(X),v(Y))=n$, where $d_H(\cdot ,\cdot)$
denotes the Hamming distance, and
\begin{equation}
|\mu(X,Y)|=\frac{k_1+k_2-d_{H}(v(X),v(Y))}{2}.\label{commom_ones}
\end{equation}

Let $X_\mu$ be the $|\mu(X,Y)| \times n$ sub-matrix of $\textmd{RE}(X)$
which consists of the rows with leading \textit{ones} in the
columns related to (indexed by) $\mu(X,Y)$. Let $X_{\mu^C}$  be
the $(k_1-|\mu(X,Y)|)\times n$ sub-matrix of $\textmd{RE}(X)$  which
consists of all the rows  of  $\textmd{RE}(X)$ which are not contained in
$X_\mu$. Similarly, let $Y_{\mu}$ be the $|\mu(X,Y)| \times n$
sub-matrix of $\textmd{RE}(Y)$ which consists of the rows with leading
\textit{ones} in the columns related to  $\mu(X,Y)$. Let
$Y_{\mu^C}$ be the $(k_2-|\mu(X,Y)|)\times n$ sub-matrix of
$\textmd{RE}(Y)$ which consists of all the rows  of $\textmd{RE}(Y)$ which are not
contained in $Y_{\mu}$.

Let $\widetilde{X}_\mu$  be the $|\mu(X,Y)|\times n$ sub-matrix of
$\textmd{RE}(\textmd{RE}(X) * Y_{\mu^C})$ which consists of the rows with
leading\textit{ ones} in the columns indexed by  $\mu(X,Y)$.
Intuitively, $\widetilde{X}_\mu$ obtained by concatenation of the
two matrices, $\textmd{RE}(X)$ and $Y_{\mu^C}$,  and "cleaning" (by adding
the corresponding rows of $Y_{\mu^C}$) all the nonzero entries in
columns of $\textmd{RE}(X)$ indexed by leading \textit{ones} in
$Y_{\mu^C}$. Finally, $\widetilde{X}_\mu$ is obtained by taking
only the rows which are indexed by $\mu(X,Y)$. Thus,
$\widetilde{X}_\mu$ has all-zero columns indexed by
\textit{ones} of $v(Y)$  and  $v(X)$ which are not in $\mu(X,Y)$.
Hence $\widetilde{X}_\mu$ has nonzero elements only in columns
indexed by $\rho(X,Y)\cup\mu(X,Y)$.

Let $\widetilde{Y}_\mu$ be the $|\mu(X,Y)|\times n$ sub-matrix of
$\textmd{RE}(\textmd{RE}(Y) * X_{\mu^C})$ which consists of the rows with
leading\textit{ ones} in the columns indexed by  $\mu(X,Y)$.
Similarly to $\widetilde{X}_\mu$, it can be verified that
$\widetilde{Y}_\mu$ has nonzero elements only in columns indexed
by $\rho(X,Y)\cup\mu(X,Y)$.

\begin{corollary}
\label{cor:size of X-Y}
Nonzero entries in  $\widetilde{X}_\mu - \widetilde{Y}_\mu$
can appear only in columns indexed by $\rho(X,Y)$.
\end{corollary}
\begin{proof}
An immediate consequence from the definition of $\widetilde{X}_\mu$
and $\widetilde{Y}_\mu$, since the columns of $\widetilde{X}_\mu$
and $\widetilde{Y}_\mu$ indexed by $\mu(X,Y)$ form a
$|\mu(X,Y)|\times |\mu(X,Y)|$ identity matrix.
\end{proof}

\begin{theorem}
\label{thm:distance}
\begin{equation}
d_{S}(X,Y)=d_H (v(X),v(Y))+2d_R (\widetilde{X}_\mu , \widetilde{Y}_\mu)
\label{subspace_distance}.
\end{equation}
\end{theorem}

\vspace{0.2cm}

\begin{proof}
By (\ref{distance-rank}) it is sufficient to proof that
\begin{align}
2\rank (\textmd{RE}(X)*\textmd{RE}(Y))& = k_1+k_2+d_H(v(X),v(Y))
+2d_R(\widetilde{X}_\mu , \widetilde{Y}_\mu).\label{eq.thm}
\end{align}

It is easy to verify that
\begin{align}
\rank
\left(\begin{array}{c}
\textmd{RE}(X)\\\textmd{RE}(Y)\end{array}\right)= \rank \left(\begin{array}{c}
\textmd{RE}(X)\\Y_{\mu^C}\\Y_\mu\end{array}\right)
=\rank \left(\begin{array}{c}
\textmd{RE}(X)\\Y_{\mu^C}\\\widetilde{Y}_\mu\end{array}\right)\nonumber\\
=\rank\left(\begin{array}{c}
\textmd{RE}(\textmd{RE}(X)*Y_{\mu^C})\\
\widetilde{Y}_\mu\end{array}\right)
=\rank \left(\begin{array}{c}
\textmd{RE}(\textmd{RE}(X)*Y_{\mu^C})\\
\widetilde{Y}_\mu-\widetilde{X}_\mu\end{array}\right)\label{eq:rank}.
\end{align}
We note that the positions of the leading \textit{ones} in all the
rows of $\textmd{RE}(X)*Y_{\mu^C}$ are in $\{1,2,\ldots,n\}\setminus
\rho(X,Y)$. By Corollary \ref{cor:size of X-Y}, the positions of
the leading \textit{ones} of all the rows of
$\textmd{RE}(\widetilde{Y}_\mu-\widetilde{X}_\mu)$ are in $\rho(X,Y)$.
Thus, by (\ref{eq:rank}) we have
\begin{align}
\rank (\textmd{RE}(X)*\textmd{RE}(Y)) =&\rank(\textmd{RE}(\textmd{RE}(X)*Y_{\mu^C})+\rank
(\widetilde{Y}_\mu-\widetilde{X}_\mu).\label{eq:1}
\end{align}
Since the sets of positions of the leading \textit{ones} of
$\textmd{RE}(X)$ and $Y_{\mu^C}$ are disjoint, we have that $\rank
(\textmd{RE}(X)*Y_{\mu^C})= k_1+(k_2-|\mu(X,Y)|)$, and thus, by (\ref{eq:1}) we have
\begin{align}
\rank (\textmd{RE}(X)*\textmd{RE}(Y))
=&k_1+k_2-|\mu(X,Y)|+\rank(\widetilde{Y}_\mu-\widetilde{X}_\mu).
\label{eq:2}
\end{align}
Combining (\ref{eq:2}) and (\ref{commom_ones}) we obtain
\begin{align*}
2\rank (\textmd{RE}(X)*\textmd{RE}(Y))
= k_{1}+k_{2}+d_{H}(v(X),v(Y))+2d_R(\widetilde{Y}_\mu,\widetilde{X}_\mu),
\end{align*}
and by (\ref{eq.thm}) this proves the theorem.

\end{proof}

The following two results will play an important role in our
constructions for error-correcting codes in the projective
space and in our search for constant dimension lexicodes.

\begin{corollary}
\label{cor:distance}
For any two subspaces $X,Y\in \Ps$,
$$d_{S}(X,Y)\geq d_{H}(v(X),v(Y)).$$
\end{corollary}

\begin{corollary}
\label{cor:distanceSameID}
Let $X$  and $Y$ be two subspaces in $\Ps$ such that $v(X)=v(Y)$. Then
$$d_{S}(X,Y)= 2\rank(\emph{RE}(X)-\emph{RE}(Y)).$$
\end{corollary}

\chapter[Codes and Designs Related to Lifted MRD Codes]{Codes and Designs
Related to Lifted MRD Codes\let\thefootnote\relax\footnotetext{The material
in this chapter was presented in part in~\cite{SiEt11is}.}}
\label{ch:designs}

There is a close connection between error-correcting codes in the
Hamming space and combinatorial designs. For example, the
codewords of weight $3$ in the Hamming code form a Steiner triple
system, MDS codes are equivalent to orthogonal arrays, Steiner
systems (if exist) form optimal constant weight codes~\cite{AAK01}.

The well-known concept of $q$-analogs replaces subsets by
subspaces of a vector space over a finite field and their orders
by the dimensions of the subspaces. In particular, the $q$-analog
of a constant weight code in the Hamming space is a constant
dimension code in the projective space. Related to constant
dimension codes are $q$-analogs of block designs. $q$-analogs of
designs were studied in~\cite{AAK01,Bra05,EV08,EV10,ScEt02,Tho87}. For
example, in~\cite{AAK01} it was shown that Steiner structures (the
$q$-analog of Steiner system), if exist, yield optimal codes in
the Grassmannian. Another connection is the constructions of
constant dimension codes from spreads which are given
in~\cite{EV08} and~\cite{MGR08}.

In this chapter we consider the lifted MRD codes.
We prove that the codewords of such a code form a design
called a transversal design, a structure which is known to be
equivalent to the well known orthogonal array.
We also prove that the same codewords form a
subspace transversal design, which is akin to the transversal
design, but not its $~~~q$-analog.
The incidence matrix of the transversal design derived from
a lifted MRD code can be viewed as a parity-check matrix of a linear code in
the Hamming space. This way to construct linear codes from designs
is well-known~\cite{AHKXL04,JoWe01,KV03,KLF01,LaMi07, LTLMH08,VKK02,VaMi04}.
We find the properties of these codes
which can be viewed also as LDPC codes.
\section{Lifted MRD Codes and Transversal Designs}
\label{sec:lifted MRD}
 MRD codes can be viewed as
maximum distance separable (MDS) codes~\cite{Gab85}, and as such
they form combinatorial designs known as orthogonal arrays and
transversal designs~\cite{HSS99}. We consider some properties of
lifted MRD codes which are derived from their combinatorial
structure. These properties imply that lifted MRD codes yield
transversal designs and orthogonal arrays with other parameters.
Moreover, the codewords of these codes form the blocks of a new
type of transversal designs, called subspace transversal designs.

\subsection{Properties of Lifted MRD Codes}
Recall, that a lifted MRD code $\CMRD$
(defined in Subsection~\ref{subsec:known constructions})
is a constant dimension code such that all its
codewords are the lifted codewords of an MRD code.

For simplicity, in the sequel we will consider only the linear MRD
codes constructed by Gabidulin~\cite{Gab85}, which are presented
in Section~\ref{sec:rank-metric codes}. It does not restrict
our discussion as such codes exist for all parameters. However,
even lifted nonlinear MRD codes also have all the properties and
results which we  consider (with a possible exception of Lemma~\ref{lm:resolv}).

\begin{theorem}\cite{SKK08}
\label{trm:param lifted MRD}
If $\cC$ is a $[k \times (n-k), (n-k)(k-\delta +1),\delta ]$ MRD
code, then its lifted code $\C^{\emph{MRD}}$ is an $(n,q^{(n-k)(k-\delta+1)},
2\delta, k)_{q}$ code.
\end{theorem}

The parameters of the $[k \times (n-k), (n-k)(k-\delta +1),\delta
]$ MRD code $\cC$ in Theorem~\ref{trm:param lifted MRD} implies by the
definition of an MRD code that $k \leq n-k$. Hence, all our
results are only for $k \leq n-k$. The results cannot be
generalized for $k > n-k$ (for example Lemma~\ref{lm:k-delta+1}
does not hold for $k > n-k$ unless $\delta=1$ which is a trivial
case). We will also assume~that~$k > 1$.
%

Let $\cL$ be the set of $q^{n}-q^{n-k}$ vectors of length $n$ over
$\F_q$ in which not all the first $k$ entries are {\it zeroes}.
The following lemma is a simple observation.
\begin{lemma}
\label{lem:CMRD=L} All the nonzero vectors which are contained in
codewords of $\C^{\emph{MRD}}$ belong to~$\cL$.
\end{lemma}
For a set $\cS \subseteq \F_q^n$, let $\Span{\cS}$ denotes the
subspace of $\F_q^n$ spanned by the elements of $\cS$. If
$\cS=\{v\}$ is of size one, then we denote $\Span{\cS}$ by
$\Span{v}$. Let $\mathbb{V}=\{\Span{v}:v\in \cL\}$ be the set of
$\frac{q^n-q^{n-k}}{q-1}$ one-dimensional subspaces of $\F_q^n$
whose nonzero vectors are contained in $\cL$. We identify each
one-dimensional subspace  $A$ in
$\mathcal{G}_q(\omega,1)$, for any given $\omega$, with
the vector $v_A \in A$ (of length $\omega$)
in which the first nonzero entry is an {\it one}.

For each $A\in \mathcal{G}_q(k,1)$ we define
$$
\mathbb{V}_A \deff \{X \;|\;X= \Span {  v },\;v\;=\;v_Az ,\; z \in
\F_q^{n-k} \}.
$$
$\{ \mathbb{V}_A : A \in \mathcal{G}_q(k,1)\}$ contains
$\frac{q^k-1}{q-1}$ sets, each one of the size $q^{n-k}$. These
sets partition the set $\mathbb{V}$, i.e., these sets are disjoint
and $\mathbb{V}=\bigcup_{A\in\mathcal{G}_q(k,1)}\mathbb{V}_A$. We
say that a vector $v\in \F_q^n$ is in $\mathbb{V}_A$  if $v\in X$
for $X\in \mathbb{V}_A$. Clearly, $\Span { \{ v_Az',\;v_Az'' \}
}$, for $A \in \mathcal{G}_q(k,1)$ and $z' \neq z''$, contains a vector
with $k$ leading \emph{zeroes}, which does not belong to
$\cL$. Hence, by Lemma~\ref{lem:CMRD=L} we have
\begin{lemma}
\label{lem:CMRD=VA} For each $A\in \mathcal{G}_q(k,1)$, a codeword
of $\C^{\emph{MRD}}$ contains at most one element from~$\mathbb{V}_A$.
\end{lemma}
Note that each $k$-dimensional subspace of $\F_q^n$ contains
$\Gauss{k}{1}=\frac{q^k-1}{q-1}$ one-dimensional subspaces.
Therefore,  by Lemma~\ref{lem:CMRD=L}, each codeword of
$\C^{\text{MRD}}$ contains $\frac{q^k-1}{q-1}$ elements of
$\mathbb{V}$. Hence, by Lemma~\ref{lem:CMRD=VA} and since
$|\mathcal{G}_q(k,1)|=\frac{q^k-1}{q-1}$ we have

\begin{corollary}
\label{cor:exactlyone} For each $A\in \mathcal{G}_q(k,1)$, a
codeword of $\C^{\emph{MRD}}$ contains exactly one element from
$\mathbb{V}_A$.
\end{corollary}

\begin{lemma}\label{lm:k-delta+1}
Each  $(k-\delta+1)$-dimensional subspace $Y$ of $\F_q^n$, whose
nonzero vectors are contained in $\cL$, is contained in exactly
one codeword of $\C^{\emph{MRD}}$.
\end{lemma}

\begin{proof}
Let $\dS \deff \{Y\in \mathcal{G}_q(n,k-\delta+1):~ |Y \cap \cL | =
q^{k-\delta+1}-1 \}$, i.e. $\dS$ consists of all $(k-\delta+1)$-dimensional subspaces
of $\mathcal{G}_q(n,k-\delta+1)$ in which all the nonzero vectors
are contained in $\cL$.

Since the minimum distance of $\C^{\text{MRD}}$ is $2\delta$ and
its codewords are $k$-dimensional subspaces, it follows that the
intersection of any two codewords is at most of dimension
$k-\delta$. Hence, each $(k-\delta+1)$-dimensional subspace of
$\F_q^n$ is contained in at most one codeword. The size of
$\C^{\text{MRD}}$ is $q^{(n-k)(k-\delta+1)}$, and the number of
$(k-\delta+1)$-dimensional subspaces in a codeword is exactly
$\Gauss{k}{k-\delta+1}$. By Lemma~\ref{lem:CMRD=L},
each $(k-\delta+1)$-dimensional subspace,
of a codeword, is contained in $\dS$. Hence, the
codewords of $\C^{\text{MRD}}$ contain exactly
$\Gauss{k}{k-\delta+1}q^{(n-k)(k-\delta+1)}$ distinct
$(k-\delta +1)$-dimensional subspaces of $\dS$.

To complete the proof we only have to show that $\dS$ does not
contain more $(k-\delta+1)$-dimensional subspaces. Hence, we will
compute the size of $\dS$.
Each element of $\dS$ intersects with each
$\mathbb{V}_{A}$, $A\in\mathcal{G}_q(k,1)$ in at most one 1-dimensional subspace.
There are $\Gauss{k}{k-\delta+1}$ ways to choose an arbitrary $(k
- \delta +1)$-dimensional subspace of $\F_q^k$. For each such
subspace $Y$ we choose an arbitrary basis $\{ x_1 , x_2 , \ldots ,
x_{k - \delta +1} \}$ and denote $A_i=\Span{x_i}$, $1\leq i\leq
k-\delta+1$. A basis for a $(k-\delta+1)$-dimensional subspace
of $\dS$ will be generated by concatenation of $x_i$ with a vector $z\in\F_q^{n-k}$ for each $i$,
$1 \leq i \leq k-\delta+1$.
Therefore, there are $q^{(k-\delta+1)(n-k)}$ ways to choose a basis for
an element of $\dS$. Hence,
$|\dS|=\Gauss{k}{k-\delta+1}q^{(n-k)(k-\delta+1)}$.

Thus, the lemma follows.
\end{proof}

\begin{corollary}
\label{cor:k-delta-i} Each $(k-\delta-i)$-dimensional subspace of
$\F_q^n$, whose nonzero vectors are contained in $\cL$, is
contained in exactly $q^{(n-k)(i+1)}$ codewords of
$\C^{\emph{MRD}}$.
\end{corollary}
\begin{proof}
The size of $\C^{\text{MRD}}$ is $q^{(n-k)(k-\delta +1)}$.
The number of $(k-\delta-i)$-dimensional subspaces in a codeword
is exactly $\Gauss{k}{k-\delta-i}$. Hence, the total number
of $(k-\delta -i)$-dimensional subspaces in $\C^{\text{MRD}}$ is
$\Gauss{k}{k-\delta -i} q^{(n-k)(k-\delta+1)}$. Similarly to the
proof of Lemma~\ref{lm:k-delta+1}, we can prove that the total
number of $(k-\delta-i)$-dimensional subspaces which contain
nonzero vectors only from $\cL$ is $\Gauss{k}{k-\delta -i}
q^{(n-k)(k-\delta-i)}$. Thus, each $(k-\delta-i)$-dimensional
subspace of $\F_q^n$, whose nonzero vectors are contained in
$\cL$, is contained in exactly
$$
\frac{\Gauss{k}{k-\delta -i} q^{(n-k)(k-\delta
+1)}}{\Gauss{k}{k-\delta -i} q^{(n-k)(k-\delta-i)}}=
q^{(n-k)(i+1)}
$$
codewords of $\C^{\text{MRD}}$.
\end{proof}

\begin{corollary}
\label{cor:2} Any one-dimensional subspace $X\in \mathbb{V}$ is
contained in exactly $q^{(n-k)(k-\delta)}$ codewords of
$\C^{\emph{MRD}}$.
\end{corollary}

\begin{corollary}
\label{cor:1} Any two elements $X_1,X_2\in \mathbb{V}$, such that
$X_1\in \mathbb{V}_A$ and  $X_2\in \mathbb{V}_B$, $A \neq B$, are
contained in exactly $q^{(n-k)(k-\delta-1)}$ codewords of
$\C^{\emph{MRD}}$.
\end{corollary}

\begin{proof}
Apply Corollary~\ref{cor:k-delta-i} with $k-\delta-i=2$.
\end{proof}

\begin{lemma}\label{lm:resolv}
$\C^{\emph{MRD}}$ can be partitioned into $q^{(n-k)(k-\delta)}$
sets, called parallel classes, each one of size $q^{n-k}$, such
that in each parallel
class each element of $\mathbb{V}$ is contained in
exactly one codeword.
\end{lemma}

\begin{proof}
First we prove that a lifted MRD code contains a lifted MRD
subcode with disjoint codewords (subspaces). Let $G$ be the
generator matrix of a $[k \times (n-k),(n-k)(k-\delta +1),
\delta]$ MRD code $\cC$~\cite{Gab85}, $n-k \geq k$. Then $G$ has the
following form
\[G=\left(
      \begin{array}{cccc}
        g_1 & g_2 & \ldots & g_k \\
        g_1^{q} & g_2^{q} & \ldots & g_k^{q} \\
        \vdots & \vdots & \cdots & \vdots \\
        g_1^{q^{k-\delta}} &  g_2^{q^{k-\delta}} & \ldots  &  g_k^{q^{k-\delta}} \\
      \end{array}
    \right) ~,
\]
where $g_i\in \mathbb{F}_{q^{n-k}}$ are linearly independent over
$\mathbb{F}_q$.
If the last $k - \delta$ rows  are removed
from $G$, the result is an MRD subcode of $\cC$  with the minimum
distance $k$. In other words, an
$[k \times (n-k),n-k, k]$ MRD  subcode $\tilde{\cC}$
of $\cC$ is obtained. The corresponding lifted code is an $(n,q^{n-k},2k,k)_q$
lifted MRD subcode of $\C^{\text{MRD}}$.

Let $\tilde{\cC}_1 = \tilde{\cC},~\tilde{\cC}_2 , \ldots ,
~\tilde{\cC}_{q^{(n-k)(k - \delta )}}$ be the $q^{(n-k)(k - \delta
)}$ cosets of $\tilde{\cC}$ in $\cC$. All these $q^{(n-k)(k - \delta
)}$ cosets are nonlinear rank-metric codes with the same
parameters as the $[k \times (n-k),n-k, k]$ MRD code. Therefore,
their lifted codes form a partition of $\C^{\text{MRD}}$ into
$q^{(n-k)(k-\delta)}$ parallel classes each one of size $q^{n-k}$,
such that each element of $\mathbb{V}$ is contained in exactly one
codeword of each parallel class.
\end{proof}

\subsection{Transversal Designs from Lifted MRD Codes}

A \emph{transversal design} of groupsize $m$, blocksize $k$,
\emph{strength} $t$ and \emph{index} $\lambda$,  denoted by
$\text{TD}_{\lambda}(t, k, m)$ is a triple
$(V,\mathcal{G},\mathcal{B})$, where
\begin{enumerate}
\item $V$ is a set of $km$ elements (called \emph{points});
\item $\mathcal{G}$ is a partition of $V$ into $k$ classes
(called \emph{groups}), each one of size $m$;
\item $\mathcal{B}$ is a collection of $k$-subsets of $V$
(called \emph{blocks});
\item each block meets each group in exactly one point;
\item every $t$-subset of points that meets each group in at most
one point is contained in exactly $\lambda$ blocks.
\end{enumerate}
When $t=2$, the strength is usually not mentioned, and the design
is denoted by $\text{TD}_{\lambda}(k,m)$.
A $\text{TD}_{\lambda}(t,k,m)$ is \emph{resolvable} if the set $\mathcal{B}$
can be partitioned into sets $\mathcal{B}_1,...,\mathcal{B}_s$,
where each element of $V$ is contained in exactly one block of
each $\mathcal{B}_i$. The sets $\mathcal{B}_1,...,\mathcal{B}_s$ are called
\emph{parallel classes}.

\begin{example}
\label{ex:TD(3,4)}
Let $V=\{1,2,\ldots,12\}$; $\mathcal{G}=\{G_1,G_2,G_3\}$, where
$G_1=\{1,2,3,4\}$, $G_2=\{5,6,7,8\}$, and $G=\{9,10,11,12\}$;
$\mathcal{B}=\{B_1,B_2,\ldots, B_{16}\}$, where
$B_1=\{1,5,9\}$, $B_2=\{2,8,11\}$, $B_3=\{3,6,12\}$, $B_4=\{4,7,10\}$,
$B_5=\{1,6,10\}$, $B_6=\{2,7,12\}$, $B_7=\{3,5,11\}$, $B_8=\{4,8,9\}$,
$B_9=\{1,7,11\}$, $B_{10}=\{2,6,9\}$, $B_{11}=\{3,8,10\}$, $B_{12}=\{4,5,12\}$,
$B_{13}=\{1,8,12\}$, $B_{14}=\{2,5,10\}$, $B_{15}=\{3,7,9\}$, and $B_{16}=\{4,6,11\}$.
These form a resolvable $\emph{TD}_1(3,4)$ with  four parallel classes
$\mathcal{B}_1=\{B_1,B_2,B_3,B_4\}$, $\mathcal{B}_2=\{B_5,B_6,B_7,B_8\}$,
$\mathcal{B}_3=\{B_9,B_{10},B_{11},B_{12}\}$, and $\mathcal{B}_4=\{B_{13},B_{14},B_{15},B_{16}\}$.
\end{example}

\begin{theorem}\label{trm:CDC=TD}
The codewords of an
$(n,q^{(n-k)(k-\delta+1)},2\delta,k)_q$ code $\C^{\emph{MRD}}$
form the blocks of a resolvable transversal design
$\emph{TD}_{\lambda}(\frac{q^k-1}{q-1},\;q^{n-k})$,
$\lambda=q^{(n-k)(k-\delta-1)}$, with $q^{(n-k)(k-\delta)}$
parallel classes, each one of size $q^{n-k}$.
\end{theorem}

\begin{proof}
Let $\mathbb{V}$ be the set of $\frac{q^n - q^{n-k}}{q-1}$ points
for the design. Each set $\mathbb{V}_A$, $A \in \mathcal{G}_q(k,1)$, is
defined to be a group, i.e., there are $\frac{q^k-1}{q-1}$
groups, each one of size $q^{n-k}$. The {$k$-dimensional} subspaces
(codewords) of $\C^{\text{MRD}}$ are the blocks of the design. By
Corollary~\ref{cor:exactlyone}, each block meets each group in
exactly one point. By Corollary~\ref{cor:1}, each $2$-subset which
meets each group in at most one point is contained in exactly
$q^{(n-k)(k-\delta-1)}$ blocks. Finally, by Lemma~\ref{lm:resolv},
the design is resolvable with $q^{(n-k)(k-\delta)}$ parallel
classes, each one of size $q^{n-k}$.

\end{proof}

An $N\times k$ array $\cA$ with entries from a set of $s$ elements
is an \emph{orthogonal array} with $s$ \emph{levels},
\emph{strength} $t$ and \emph{index} $\lambda$, denoted by
$\text{OA}_{\lambda}(N,k,s,t)$, if every $N\times t$ subarray of $\cA$ contains
each $t$-tuple exactly $\lambda$  times as a row. It is
known~\cite{HSS99} that a $\text{TD}_{\lambda}(k,m)$ is
equivalent to an orthogonal array $\text{OA}_{\lambda}(\lambda\cdot
m^2,k,m,2)$.

\begin{remark}
By the equivalence of transversal designs and orthogonal arrays, we have that
an $(n, q^{(n-k)(k-\delta+1)}, 2\delta,k)_q$ code $\C^{\emph{MRD}}$
induces an
$\emph{OA}_{\lambda}(q^{(n-k)(k-\delta+1)},\frac{q^k-1}{q-1}, q^{n-k}, 2)$
with $\lambda=q^{(n-k)(k-\delta-1)}$.
\end{remark}

\begin{remark}A $[k\times (n-k),(n-k)(k-\delta+1),\delta]$ MRD code
$\cC$ is an MDS code if it is viewed  as a code of length $k$  over
$GF(q^{n-k})$. Thus its codewords form an orthogonal array
$\emph{OA}_{\lambda}(q^{(n-k)(k-\delta+1)},k,q^{n-k},k-\delta+1)$
with $\lambda=1$~\cite{HSS99}, which is also an orthogonal array
$\emph{OA}_{\lambda}(q^{(n-k)(k-\delta+1)},k,q^{n-k},2)$
with $\lambda=q^{(n-k)(k-\delta-1)}$.
\end{remark}

Now we define a new type of transversal designs in terms of subspaces, which
will be called a subspace transversal design. We will show that
such a design is induced by the codewords of a lifted MRD code. Moreover,
in the following chapter
we will show that this design is useful to obtain upper bounds
on the codes that contain the lifted MRD codes, and in a construction
of large constant dimension codes.

Let $\mathbb{V}_0$ be a set of one-dimensional subspaces in
$\mathcal{G}_q(n,1)$, that contains only  vectors  starting with
$k$ \emph{zeroes}. Note that $\mathbb{V}_0$ is isomorphic to
$\mathcal{G}_q(n-k,1)$.

A \emph{subspace transversal  design} of  groupsize $q^m$,
$m=n-k$, block dimension $k$, and \emph{strength} $t$, denoted by
$\text{STD}_{q}(t, k, m)$, is a triple
$(\mathbb{V},\mathbb{G},\mathbb{B})$, where

\begin{enumerate}
\item $\mathbb{V}$ is the subset of all elements of
$\mathcal{G}_q(n,1)\setminus \mathbb{V}_0$,
$|\mathbb{V}|=\frac{(q^{k}-1)}{q-1} q^{m}$ (the \emph{points});

\item $\mathbb{G}$ is a partition of $\mathbb{V}$ into
$\frac{q^{k}-1}{q-1}$ classes of size $q^m$ (the \emph{groups});
\item $\mathbb{B}$ is a collection of $k$-dimensional
subspaces which contain only points from $\mathbb{V}$ (the
\emph{blocks});
\item each block meets each group in exactly one point;
\item every $t$-dimensional subspace (with points from $\mathbb{V}$) which
meets each group in at most one point is contained in exactly
one block.
\end{enumerate}
As a direct consequence form Lemma~\ref{lm:k-delta+1}
and Theorem~\ref{trm:CDC=TD} we have the following theorem.

\begin{theorem}\label{trm:MRD=STD}
The codewords of an $(n,q^{(n-k)(k-\delta+1)},2\delta,k)_q$ code
$\C^{\emph{MRD}}$ form the blocks of a resolvable
$\emph{STD}_{q}(k-\delta+1,k, n-k)$, with the set of points
$\mathbb{V}$ and the set of groups $\mathbb{V}_A$,
$A\in\mathcal{G}_q(k,1)$, defined  previously in this section.
\end{theorem}
\begin{remark}
There is no known nontrivial $q$-analog of a block design with
$\lambda =1$ and $t >1$. An $\emph{STD}_{q}(t, k, m)$ is very
close to such a design.
\end{remark}
\begin{remark}
An $\emph{STD}_{q}(t, k, n-k)$ cannot exist if $k > n-k$, unless
$t=k$.  Recall, that the case $k >
n-k$ was not considered in this section (see
Theorem~\ref{trm:param lifted MRD}).
\end{remark}
\section{Linear Codes Derived from Lifted MRD Codes}
\label{sec:Linear codes}
In this section we study the properties of linear codes in the Hamming space whose parity-check
matrix is an incidence matrix of a transversal design derived from a lifted MRD code.
These codes may also be of interest as LDPC codes.

For each codeword $X$ of a constant dimension code
$\C^{\textmd{MRD}}$ we define its binary \emph{incidence} vector
$x$ of length $|\mathbb{V}|=\frac{q^{n}-q^{n-k}}{q-1}$ as follows:
$x_z=1$ if and only if the point $z \in \mathbb{V}$ is contained
in $X$.

Let $H$ be the $|\C^{\textmd{MRD}}|\times|\mathbb{V}|$ binary matrix whose
rows are the incidence vectors of the codewords of $\C^{\textmd{MRD}}$.
By Theorem~\ref{trm:CDC=TD}, this matrix $H$ is the
\emph{incidence matrix} of $TD_{\lambda}(\frac{q^k-1}{q-1},\;q^{n-k})$, with
$\lambda=q^{(n-k)(k-\delta-1)}$. Note that
the rows of the incidence matrix~$H$ correspond to the blocks of the transversal design,
and the columns of $H$ correspond to the points of the transversal design.
If $\lambda=1$ in such a design (or, equivalently,
$\delta=k-1$ for $\C^{\textmd{MRD}}$), then $H^T$ is an
incidence matrix of a \emph{net}, the dual structure
to the transversal  design~\cite[p. 243]{vLWi92}.

An $[N,K,d]$ linear code is a linear subspace of dimension $K$
of $\F_2^N$ with the minimum Hamming distance $d$.

Let $C$ be the linear code with the parity-check matrix $H$, and let
$C^T$ be the linear code with the parity-check matrix $H^T$.
This approach for construction of linear
codes is widely used for LDPC codes. For example, codes whose parity-check
matrix is an incidence matrix of a block design are considered in
~\cite{AHKXL04,JoWe01,KV03,KLF01,LaMi07, LTLMH08,VaMi04,VKK02}.
Codes obtained from nets and
transversal designs are considered in \cite{Dou93},
\cite{JoWe04}.

The parity-check matrix $H$ corresponds to a bipartite graph, called the \emph{Tanner graph}
of the code. The rows and the columns of $H$ correspond to the two
parts of the vertex set of the graph, and the nonzero entries of $H$
correspond the the edges of the graph.

Given $TD_{\lambda}(\frac{q^k-1}{q-1},\;q^{n-k})$, if $\lambda=1$,
then the corresponding Tanner graph has girth 6 (girth is the length of the shortest cycle).
If $\lambda \geq 1$, then the girth of the Tanner graph is~4.

\subsection{Parameters of Linear Codes Derived from $\C^{\textmd{MRD}}$}

The code $C$ has length $\frac{q^{n}-q^{n-k}}{q-1}$ and the code $C^T$ has length
$q^{(n-k)(k-\delta +1)}$.
By Corollary~\ref{cor:2}, each column of $H$ has $q^{(n-k)(k-\delta)}$
ones; since each $k$-dimensional subspace contains $\frac{q^{k}-1}{q-1}$
one-dimensional subspaces, each row has $\frac{q^{k}-1}{q-1}$ ones.

\begin{remark}
Note that if  $\delta=k$, then the column weight of $H$ is one.
Hence, the minimum distance of $C$ is $2$. Moreover, $C^T$ consists
only of the all-zero codeword. Thus, these codes are not
interesting  and hence in the sequel we assume that $\delta\leq
k-1$.
\end{remark}
\begin{lemma}\label{lm:partition to permutation blocks}
The matrix $H$ obtained from an $(n,q^{(n-k)(k-\delta+1)},2\delta,k)_q$
$\C^{\emph{MRD}}$ code can be decomposed into blocks, where each
block is a $q^{n-k}\times q^{n-k}$ permutation matrix.
\end{lemma}

\begin{proof} It follows from Lemma~\ref{lm:resolv}
that the related transversal design is resolvable.
In each parallel class each element of $\mathbb{V}$ is contained
in exactly one codeword of $\C^{\textmd{MRD}}$. Each class has
$q^{n-k}$ codewords, each group has $q^{n-k}$ points, and each
codeword meets each group in exactly one point. This implies that
each $q^{n-k}$ rows of $H$ related to such a class can be
decomposed into $\frac{q^k-1}{q-1}$ $\;q^{n-k}\times q^{n-k}$
permutation matrices.
\end{proof}

\begin{example}
The $[12,4,6]$ code $C$ and the $[16,8,4]$ code $C^T$ are obtained
from the $(4,16,2,2)_2$ lifted MRD code $\C^\emph{{MRD}}$. The
incidence matrix for corresponding transversal design $\emph{TD}_1(3,4)$
(see Example~\ref{ex:TD(3,4)})
is given by the following $16\times 12$ matrix. The four rows
above this matrix represent the column vectors for the points of the
design.

\begin{footnotesize}
$$\begin{array}{c}
\left.\begin{array}{cccc|cccc|cccc}
  \textbf{0} & \textbf{0} & \textbf{0} & \textbf{0} &
  \textbf{1} & \textbf{1} & \textbf{1} & \textbf{1} &
  \textbf{1} & \textbf{1} & \textbf{1} & \textbf{1} \\
  \textbf{1} & \textbf{1} & \textbf{1} & \textbf{1} &
  \textbf{0} & \textbf{0} & \textbf{0} & \textbf{0} &
  \textbf{1} & \textbf{1} & \textbf{1} & \textbf{1} \\
  \textbf{0} & \textbf{0} & \textbf{1} & \textbf{1} &
  \textbf{0} & \textbf{0} & \textbf{1} & \textbf{1} &
  \textbf{0} & \textbf{0} & \textbf{1} & \textbf{1} \\
  \textbf{0} & \textbf{1} & \textbf{0} & \textbf{1} &
  \textbf{0} & \textbf{1} & \textbf{0} & \textbf{1} &
  \textbf{0} & \textbf{1} & \textbf{0} & \textbf{1}
\end{array}\right.\\
\left(\begin{array}{cccc|cccc|cccc}\hline\hline
  1 & 0 & 0 & 0 & 1 & 0 & 0 & 0 & 1 & 0 & 0 & 0 \\
   0 & 1 & 0 & 0 & 0 & 0 & 0 & 1 & 0 & 0 & 1 & 0 \\
   0 & 0 & 1 & 0 & 0 & 1 & 0 & 0 & 0 & 0 & 0 & 1 \\
   0 & 0 & 0 & 1 & 0 & 0 & 1 & 0 & 0 & 1 & 0 & 0 \\\hline
   1 & 0 & 0 & 0 & 0 & 1 & 0 & 0 & 0 & 1 & 0 & 0 \\
   0 & 1 & 0 & 0 & 0 & 0 & 1 & 0 & 0 & 0 & 0 & 1 \\
   0 & 0 & 1 & 0 & 1 & 0 & 0 & 0 & 0 & 0 & 1 & 0 \\
   0 & 0 & 0 & 1 & 0 & 0 & 0 & 1 & 1 & 0 & 0 & 0 \\\hline
   1 & 0 & 0 & 0 & 0 & 0 & 1 & 0 & 0 & 0 & 1 & 0 \\
   0 & 1 & 0 & 0 & 0 & 1 & 0 & 0 & 1 & 0 & 0 & 0 \\
   0 & 0 & 1 & 0 & 0 & 0 & 0 & 1 & 0 & 1 & 0 & 0 \\
   0 & 0 & 0 & 1 & 1 & 0 & 0 & 0 & 0 & 0 & 0 & 1 \\\hline
   1 & 0 & 0 & 0 & 0 & 0 & 0 & 1 & 0 & 0 & 0 & 1 \\
   0 & 1 & 0 & 0 & 1 & 0 & 0 & 0 & 0 & 1 & 0 & 0 \\
   0 & 0 & 1 & 0 & 0 & 0 & 1 & 0 & 1 & 0 & 0 & 0 \\
   0 & 0 & 0 & 1 & 0 & 1 & 0 & 0 & 0 & 0 & 1 & 0
\end{array}\right)
\end{array}
$$
\end{footnotesize}

\end{example}

\begin{corollary}\label{cor:even weight}
All the codewords of the code $C$, associated with the parity-check matrix
$H$, and of the code $C^T$, associated with the parity-check matrix
$H^T$, have  even weights.
\end{corollary}

\begin{proof} Let $c$ be a codeword of $C$ ($C^T$).
Then $Hc=\textbf{0}$ ($H^Tc=\textbf{0}$),
where $\textbf{0}$ denotes the  all-zero column vector. Assume that $c$ has an odd weight.
Then there is an odd number of columns of $H$ ($H^T$) that can be added to
obtain the all-zero column vector, and that is a contradiction, since by
Lemma \ref{lm:partition to permutation blocks}, $H$ ($H^T$) is
an array consisting of permutation matrices.
\end{proof}

\begin{corollary} \label{cor:upperbound on d}
The minimum Hamming distance $d$ of $C$  and the minimum Hamming distance $d^T$
of $C^T$ are upper bounded by $2q^{n-k}$.
\end{corollary}

\begin{proof}We take all the columns of $H$ ($H^T$) corresponding to any two
blocks of permutation matrices mentioned in
Lemma \ref{lm:partition to permutation blocks}. These columns
sums to all-zero column vector, and hence we found $2q^{n-k}$ depended
columns in $H$ ($H^T$). Thus $d\leq 2q^{n-k}$ ( $d^T\leq 2q^{n-k}$).
\end{proof}

To obtain a lower bound on the minimum Hamming distance of these
codes we need the following theorem known as the Tanner
bound~\cite{Tan01}.

\begin{theorem}
\label{Tannerbound}
The minimum distance, $d_{\textmd{min}}$,
of  a  linear code defined by an $m\times n$ parity-check matrix
$\mathcal{H}$ with constant row weight $\rho$ and constant column weight $\gamma$ satisfy

\begin{enumerate}
  \item $d_{\textmd{min}}\geq \frac{n(2\gamma-\mu_{2})}{\gamma\rho-\mu_{2}};$
  \item $d_{\textmd{min}}\geq \frac{2n(2\gamma+\rho-2-\mu_{2})}{\rho(\gamma\rho-\mu_{2})},$
\end{enumerate}

\noindent where $\mu_{2}$ is the second largest eigenvalue
of $\mathcal{H}^{T}\mathcal{H}$.
\end{theorem}
\vspace{0.3cm}

To obtain a lower bound on $d$ and $d^T$ we need to find
the second largest eigenvalue of $H^TH$ and $HH^T$, respectively.
Note that since the set of eigenvalues of $H^TH$ and $HH^T$ is the same,
it is sufficient to find only the eigenvalues of $H^TH$.

The following lemma is derived from~\cite[p. 563]{HandCD}.

\begin{lemma}\label{lm:spectrum}
Let $\cH$ be an incidence matrix for $\emph{TD}_{\lambda}(k,m)$. The
eigenvalues of $\cH^T\cH$ are $rk$, $r$, and $rk-km\lambda$ with
multiplicities $1, k(m-1)$, and $k-1$, respectively, where $r$ is
a number of blocks that are incident with a given point.
\end{lemma}
\vspace{0.3cm}

By Corollary \ref{cor:2},
$r=q^{(n-k)(k-\delta)}$ in  $TD_{\lambda}(\frac{q^k-1}{q-1},\;q^{n-k})$
with $\lambda=q^{(n-k)(k-\delta-1)}$.
Thus, from Lemma \ref{lm:spectrum} we obtain the spectrum of $H^TH$.

\begin{corollary}
\label{cor:spectrum HTH} The eigenvalues of $H^TH$ are
$q^{(n-k)(k-\delta)}\frac{q^k-1}{q-1}$, $q^{(n-k)(k-\delta)}$, and
$0$ with multiplicities $1$, $\frac{q^k-1}{q-1}(q^{n-k}-1)$, and
$\frac{q^k-1}{q-1}-1$, respectively.
\end{corollary}

\vspace{0.3cm}

Now, by Theorem~\ref{Tannerbound} and Corollary~\ref{cor:spectrum HTH}, we have

\begin{corollary}\label{cor:mindist}
\[d\geq \frac{q^{n-k}(q^k-1)}{q^k-q} ,
\]
\[d^T\geq \left\{\begin{array}{cc}
                2^k & \delta=k-1,~q=2,~k=n-k \\
                4q^{(n-k)(\delta-k+1)} & \emph{otherwise} \\
              \end{array}\right..
\]
\end{corollary}

\begin{proof}
By Corollary~\ref{cor:spectrum HTH},
the second largest eigenvalues of $H^TH$ is $\mu_2=q^{(n-k)(k-\delta)}$.
We apply Theorem \ref{Tannerbound} and obtain lower bounds on $d$:

\begin{equation}
\label{eq:Tanner d 1}
d\geq \begin{footnotesize}\frac{q^{n-k}\frac{q^k-1}{q-1}(2q^{(n-k)(k-\delta)}-q^{(n-k)(k-\delta)})}
      {q^{(n-k)(k-\delta)}\frac{q^k-1}{q-1}-q^{(n-k)(k-\delta)}}
      =\frac{q^{n-k}(q^k-1)}{q^k-q},\end{footnotesize}
\end{equation}

\[d\geq \begin{footnotesize}\frac{2q^{n-k}\frac{q^k-1}{q-1}(2q^{(n-k)(k-\delta)}+
  \frac{q^k-1}{q-1}-2-q^{(n-k)(k-\delta)})}
      {\frac{q^k-1}{q-1}(q^{(n-k)(k-\delta)}\frac{q^k-1}{q-1}-q^{(n-k)(k-\delta)})}
      \end{footnotesize}
\]
\begin{equation}
\label{eq:Tanner d 2}
\begin{footnotesize}
      =\frac{q^{n-k}(q^k-1)}{q^k-q}2\left(\frac{q^{(n-k)(k-\delta)}+
      \frac{q^k-1}{q-1}-2)}{q^{(n-k)(k-\delta)}\frac{q^k-1}{q-1}}\right).\end{footnotesize}
\end{equation}

The expression in~(\ref{eq:Tanner d 1}) is larger than the expression in~(\ref{eq:Tanner d 2}).
Thus, we have that
$d\geq \frac{q^{n-k}(q^k-1)}{q^k-q}$ for all $\delta\leq k-1$.

In a similar way, by using Theorem \ref{Tannerbound} we obtain lower bounds on $d^T$:
\begin{equation}
\label{eq:Tanner d^T 1}
d^T\geq \frac{q^{n-k}(2\frac{q^k-1}{q-1}-q^{(n-k)(k-\delta)})}{\frac{q^k-1}{q-1}-1},
\end{equation}
\begin{equation}
\label{eq:Tanner d^T 2}
d^T\geq 4q^{(n-k)(\delta-k+1)}.
\end{equation}

Note that the expression in~(\ref{eq:Tanner d^T 1}) is negative
for $\delta<k-1$. For $\delta=k-1$ with $k=n-k$ and $q=2$, the
bound in~(\ref{eq:Tanner d^T 1}) is larger than the bound
in~(\ref{eq:Tanner d^T 2}). Thus, we have $d^T\geq 2^k$, if
$\delta=k-1,\;q=2$, and $k=n-k$; and $d^T\geq
4q^{(n-k)(\delta-k+1)}$, otherwise.
\end{proof}

\vspace{0.3cm}
A \emph{stopping set} $S$ in a code $C$ is a subset of the variable
nodes, related to the columns of $H$, in a Tanner graph of
$C$ such that all the neighbors of $S$ are
connected to $S$ at least twice. The size of the smallest stopping set is
called the \emph{stopping distance} of a code $C$. The stopping distance
depends on the specific Tanner graph, and therefore, on the specific
parity-check matrix $H$,  and it is denoted by $s(H)$.
 The stopping distance plays a role in iterative
decoding over the binary erasure channel similar to the role of
the minimum distance in maximum
likelihood decoding~\cite{DPTRU02}. It is easy to see that $s(H)$ is less or equal
to the minimum distance of the code $C$.

It was shown in~\cite[Corollary 3]{ZC05} that the Tanner lower bound on
the minimum distance is also the lower bound on the stopping distance of
a code with  a parity-check matrix $H$, then from Corollary \ref{cor:mindist}
we have the following result.

\begin{corollary}\label{cor:mindistLDPC}
The stopping distance $s(H)$ of $C$ and the stopping distance $s(H^T)$ of $C^T$ satisfy
\[s(H)\geq \frac{q^{n-k}(q^k-1)}{q^k-q},
\]
\[s(H^T)\geq \left\{\begin{array}{cc}
                2^k & \delta=k-1,~q=2,~k=n-k \\
                4q^{(n-k)(\delta-k+1)} & \emph{otherwise} \\
              \end{array}\right..
\]
\end{corollary}

We use the following result proved in~\cite[Theorem 1]{KV03} to
improve the lower bound on $s(H^T)$ and, therefore, on $d^T$.

\begin{lemma}\label{lm:KashVar}
Let $\cH$ be an incidence matrix of blocks (rows) and points
(columns) such that each block contains exactly $\kappa$ points,
and each pair of distinct blocks intersects in at most $\gamma$
points. If $\Sigma$ is a stopping set in the Tanner graph
corresponding to $\cH^T$, then
\[
\left|\Sigma\right|\geq\frac{\kappa}{\gamma}+1.\]

\end{lemma}

\begin{corollary}$ s(H^T)\geq \frac{q^k-1}{q^{k-\delta}-1}+1$.
\end{corollary}

\begin{proof}
By Lemma \ref{lm:KashVar}, with $\kappa=\frac{q^k-1}{q-1}$ and
$\gamma = \frac{q^{k-\delta}-1}{q-1}$, since any two codewords in
a lifted MRD code intersect in at most $(k-\delta)$-dimensional
subspace, we have the following lower bound on the size of every
stopping set of $C^T$ and, particulary, for the smallest stopping set
of $C^T$
\[s(H^T)\geq \frac{(q^k-1)/(q-1)}{(q^{k-\delta}-1)/(q-1)}+1
=\frac{q^k-1}{q^{k-\delta}-1}+1.
\]
Obviously, for all $\delta \leq k-1$, this bound is larger or
equal than the bound of Corollary \ref{cor:mindistLDPC}, and thus
the result follows.
\end{proof}

We summarize all the  results about the minimum distances and the
stopping distances of $C$ and $C^T$ obtained above in the
following theorem.
\begin{theorem}
\label{trm:minDist linear codes}
\[2q^{n-k}\geq d\geq s(H)\geq\frac{q^{n-k}(q^k-1)}{q^k-q},\]
\[2q^{n-k}\geq d^T\geq s(H^T)\geq\frac{q^k-1}{q^{k-\delta}-1}+1.\]
\end{theorem}

\vspace{0.3cm}
Let $\dim(C)$  and $\dim(C^T)$ be the dimensions of $C$ and $C^T$, respectively.
To obtain the lower and upper bounds on $\dim(C)$  and $\dim(C^T)$ we need
the following basic results from linear algebra~\cite{HoJo85}. For a matrix
$A$ over a field $\F$,
let $\rank_{\F}(A)$ denotes the rank  of  $A$ over~$\F$.
\begin{lemma}
\label{lm:algebra}
Let $A$ be a  $\rho \times \eta$ matrix,  and
let $\mathbb{R}$ be the field of real numbers. Then
\begin{itemize}
  \item $\rank_{\mathbb{R}}(A)=\rank_{\mathbb{R}}(A^T)=\rank_{\mathbb{R}}(A^TA)$.
  \item If $\rho=\eta$ and $A$ is a symmetric matrix with the eigenvalue  $0$ of multiplicity $t$,
then $\rank_{\mathbb{R}}(A)=\eta - t$.
\end{itemize}
\end{lemma}

\begin{theorem}\label{trm:lower bound on code dimension}
\[\dim(C)\geq \frac{q^k-1}{q-1}-1,\]
\[\dim(C^T)\geq q^{(n-k)(k-\delta+1)}-\frac{q^k-1}{q-1}(q^{n-k}-1)-1.\]
\end{theorem}

\begin{proof}
First, we observe that
$\dim(C)=\frac{q^k-1}{q-1}q^{n-k}-\rank_{\F_{2}}(H)$, and
$\dim(C^T)=q^{(n-k)(k-\delta+1)}-\rank_{\F_{2}}(H^T)$.
Now, we obtain an upper bound on $\rank_{\F_2}(H)=\rank_{\F_2}(H^T)$.
Clearly, $\rank_{\F_{2}}(H)\leq \rank_{\mathbb{R}}(H)$.
By Corollary~\ref{cor:spectrum HTH}, the multiplicity of an
eigenvalue~$0$ of $H^TH$ is
$\frac{q^k-1}{q-1}-1$. Hence by Lemma~\ref{lm:algebra},
$\rank_{\F_2}(H) \leq \rank_{\mathbb{R}} (H)=\rank_{\mathbb{R}}( H^TH)=
\frac{q^k-1}{q-1}q^{n-k}-(\frac{q^k-1}{q-1}-1)$.
Thus, $\dim(C)\geq \frac{q^k-1}{q-1}q^{n-k}-(\frac{q^k-1}{q-1}q^{n-k}
-(\frac{q^k-1}{q-1}-1))=\frac{q^k-1}{q-1}-1$, and $\dim(C^T)\geq q^{(n-k)(k-\delta+1)}-\frac{q^k-1}{q-1}q^{n-k}+\frac{q^k-1}{q-1}-1$.
\end{proof}

\vspace{0.3cm}
Now, we obtain an upper bound on the dimension of the codes $C$ and $C^T$
for odd $q$.

\begin{theorem}
\label{trm:upper bound odd q}
Let $q$ be a power of an odd prime number.
\begin{itemize}
  \item If $\frac{q^k-1}{q-1}$
is odd, then
$$\dim(C)\leq \frac{q^k-1}{q-1}-1 \textmd{~~and~~}
\dim(C^T)\leq q^{(n-k)(k-\delta+1)}-\frac{q^k-1}{q-1}(q^{n-k}-1)-1.$$

\item If $\frac{q^k-1}{q-1}$ is even, then
$$\dim(C)\leq \frac{q^k-1}{q-1}, \textmd{~~and~~}
\dim(C^T)\leq q^{(n-k)(k-\delta+1)}-\frac{q^k-1}{q-1}(q^{n-k}-1).$$
 \end{itemize}
 \end{theorem}

\begin{proof}
We compute the lower bound on $\rank_{\F_2}(H)$ to obtain the
upper bound on the dimension of the codes $C$ and $C^T$. First, we
observe that $\rank_{\F_2}(H)\geq \rank_{\F_2}(H^TH)$.
By~\cite{BrEi92}, the rank over $\F_2$ of an integral
diagonalizable square matrix $A$ is lower bounded by the sum of
the multiplicities of the eigenvalues of $A$ that do not vanish
modulo $2$. We consider now $\rank_{\F_2}(H^TH)$. By
Corollary~\ref{cor:spectrum HTH}, the second eigenvalue of $H^TH$
is always odd for odd $q$. If $\frac{q^k-1}{q-1}$ is odd, then the
first eigenvalue of $H^TH$ is also odd. Hence, we sum the
multiplicities  of the first two eigenvalues to obtain
$\rank_{\F_2}(H^TH)\geq 1+\frac{q^k-1}{q-1}(q^{n-k}-1)$. If
$\frac{q^k-1}{q-1}$ is even, then the first eigenvalue is even,
and hence we take only the multiplicity of the second eigenvalue
to obtain $\rank_{\F_2}(H^TH)\geq \frac{q^k-1}{q-1}(q^{n-k}-1)$.
The result follows now from the fact that the dimension of a code
is equal to the difference between its length and
$\rank_{\F_2}(H)$.
\end{proof}

\begin{remark}
For even values of $q$ the method used in the proof for
Theorem~\ref{trm:upper bound odd q} leads to a trivial result,
since in this case all the eigenvalues of $H^TH$ are even and thus
by~\cite{BrEi92} we have $\rank_{\F_2}(H^TH)\geq 0$. But clearly,
by Lemma~\ref{lm:partition to permutation blocks} we have
$\rank_{\F_2}(H)\geq q^{n-k}$. Thus, for even $q$, $\dim(C)\leq
\frac{q^k-1}{q-1}q^{n-k}-q^{n-k}=q^{n-k}( \frac{q^k-1}{q-1}-1)$,
and $\dim(C^T)=q^{(n-k)(k-\delta+1)}-q^{n-k}$.
\end{remark}

Note that for odd $q$ and odd $\frac{q^k-1}{q-1}$ the lower and
the upper bounds on the dimension of $C$ and $C^T$ are the same.
Therefore, we have the following corollary.

\begin{corollary} For odd $q$ and odd $\frac{q^k-1}{q-1}$ the dimensions
$\dim(C)$ and $\dim(C^T)$ of the codes $C$ and $C^T$, respectively, satisfy
$\dim(C)=\frac{q^k-1}{q-1}-1$, and
$\dim(C^T)= q^{(n-k)(k-\delta+1)}-\frac{q^k-1}{q-1}q^{n-k}+\frac{q^k-1}{q-1}-1$ .
\end{corollary}

\begin{remark} Some of the results presented in this subsection generalize
the results given in~\cite{JoWe04}. In particular, the lower bounds on the
minimum distance and the bounds on the dimension
of LDPC codes (with girth 6) derived from lifted MRD codes coincide with
the bounds on  LDPC codes from partial geometries considered in~\cite{JoWe04}.
\end{remark}


\subsection{LDPC Codes Derived from $\CMRD$}
\label{subsec:LDPC}

Low-density parity check (LDPC) codes, introduced  by Gallager in 1960's~\cite{Gal62},
are known as Shannon limit approaching codes~\cite{Sha48}.
Kou, Lin, and Fossorier~\cite{KLF01} presented the first systematic construction of LDPC codes
based on finite geometries. Their work started  a new  research  direction
of algebraic constructions of LDPC codes. Many LDPC codes were obtained from
different combinatorial designs, such that balanced incomplete block designs,
Steiner triple systems, orthogonal arrays, and Latin squares
~\cite{AHKXL04,JoWe01,JoWe04,KV03,KLF01,LaMi07,LTLMH08,VKK02,VaMi04}.

LDPC codes are characterized by a sparse parity-check matrix with
constant weight of rows and constant weight of columns; and Tanner
graph without cycles of length 4. Next, we discuss LDPC codes
derived from $\CMRD$. Hence, in this subsection we consider only
$\textmd{TD}_1(\frac{q^k-1}{q-1},q^{n-k})$, obtained from an
$(n,q^{2(n-k)},2(k-1),k)_q$ lifted MRD code.

\begin{remark}
It was pointed out in~\cite{TXLA-G05} that the codes based on
finite geometries can perform well under iterative decoding
despite many cycles of length 4 in their Tanner graphs. Hence,
also the codes mentioned in the previous subsection can be of
interest from this point of view.
\end{remark}

Some parameters of LDPC codes obtained from lifted MRD codes
compared with the LDPC codes based on finite
geometries~\cite{KLF01} (FG in short) can be found in
Table~\ref{tab:CMRD vs FG}.

\begin{table}[h]
\centering \caption{LDPC codes from $\CMRD$ vs. LDPC codes from finite geometries}
\label{tab:CMRD vs FG}
\begin{tabular}{|c|c||c|c|}
\hline
\multicolumn{2}{|c||}{LDPC codes from FG }
&\multicolumn{2}{|c|}{LDPC codes from $\CMRD$ } \\
\hline
  $[N,K,d]$ & $K/N$&$[N,K,d]$ & $K/N$
 \tabularnewline \hline\hline
$[273,191,18]$& 0.699& $[240,160,18]$& 0.667 \\\hline
  $[4095,3367,65]$& 0.822 & $[4096,3499,\geq 64]$& 0.854  \\\hline
  $[4161,3431,66]$& 0.825& $[4032,3304,\geq 66]$& 0.819 \\\hline
\end{tabular}
\end{table}


A code is called \emph{quasi-cyclic} if there is an integer $p$ such that every
cyclic shift of a codeword by $p$ entries is again a codeword.

Let $N(K,d)$ denotes the length of the shortest binary linear code of dimension
$K$ and minimum distance $d$.
Then by Griesmer bound~\cite{MWSl78} ,
\begin{equation}
\label{eq:Griesmer bound}
N(K,d)\geq \sum_{i=0}^{K-1}\left\lceil\frac{d}{2^i}\right\rceil.
\end{equation}

\begin{theorem}\label{trm:QC+Griesmer}
An LDPC code $C$ obtained from an $(n,2^{2(n-2)},2,2)_2$ lifted MRD code
$\mathbb{C^{\emph{MRD}}}$ is a $[2^n-2^{n-2},n,\frac{2^n-2^{n-2}}{2}]$
quasi-cyclic code with  $p=2^{n-2}$, which attains the Griesmer bound.
\end{theorem}

\begin{proof}
First we prove that $C$ is quasi-cyclic.
Let $T$ be the $\textmd{TD}_1(3,2^{n-2})$ obtained from an $(n,2^{2(n-2)},2,2)_2$
code $\mathbb{C^{\textmd{MRD}}}$. $T$ has 3 groups
$\mathbb{V}_i = \{\Span{v}|v=a_iz,z\in \F_q^{n-2}\}$, $1\leq i\leq 3$, where
$a_1=\Span{01}$, $a_2 = \Span{10}$, and $a_3 = \Span{11}$.
The first $2^{n-2}$ columns of the incidence matrix $H$ of $T$ correspond
to the points of $\mathbb{V}_1$, the next $2^{n-2}$ columns correspond to the points of $\mathbb{V}_2$,
and the last $2^{n-2}$ columns correspond to the points of $\mathbb{V}_3$.
The suffices $z\in \F_2^{n-2}$ for the points of $T$ are ordered lexicographically.

Let $X=\{\textbf{0},v,u,w=v+u\}$ be  a codeword of $\mathbb{C^{\textmd{MRD}}}$, where $\textbf{0}$ is
the all-zero vector of length $n$, and $v,u,w\in \F_2^n$.
By Corollary~\ref{cor:exactlyone},
each codeword contains exactly one point
from each group, hence w.l.o.g.  we write $v=01v'$, $u=10u'$, and $w=11w'$,
for $v',u',w'\in \F_2^{n-2}$.
Let $X'$ be a set of points corresponding to the  cyclic shift of the incidence
vector for $X$, by $2^{n-2}$ entries to the left.
Then $X'=\{ 01u', 10w', 11v'\}$. Obviously, $v'+u'=w'$, and since a code
$\mathbb{C^{\textmd{MRD}}}$ contains all the 2-dimensional
subspaces with vectors from $\mathbb{V}$, we have that
$X'\cup\{\textbf{0}\}$ is also a codeword of $\mathbb{C^{\textmd{MRD}}}$.
Therefore, if $c$ is a codeword in $C$, and $c'$ is obtained by the cyclic shift
of $c$ by $2^{n-k}$ entries to the left, then $Hc=0$ implies  that $Hc'=0$.

Now we prove that $C$ is a $[2^n-2^{n-2},n,\frac{2^n-2^{n-2}}{2}]$ code.
Let $K$ be the dimension of~$C$. First assume that $K > n$.
By Theorem~\ref{trm:minDist linear codes},
$d\geq \frac{3\cdot2^{n-2}}{2}$. Hence,
by Griesmer bound~(\ref{eq:Griesmer bound}) we have:

\begin{equation}
\label{eq:Griesmer proof}
2^n-2^{n-2}=3\cdot2^{n-2}\geq \sum_{i=0}^{K-1}\left\lceil\frac{d}{2^i}\right\rceil\geq \sum_{i=0}^{K-1}\left\lceil\frac{3\cdot2^{n-3}}{2^i}\right\rceil $$
$$\geq\sum_{i=3}^{n}3\cdot2^{n-i}+\left\lceil\frac{3}{2}\right\rceil+\left\lceil\frac{3}{4}\right\rceil+(K-n)
=3\cdot2^{n-2}-3+3+(K-n),
\end{equation}
contradiction, thus $K\leq n$. Now assume that $K < n$.
We form an $n\times (3\cdot2^{n-2})$ matrix~$G$, such that its columns are the vectors for the
points of $T$, where the first $2^{n-2}$ columns correspond to the vectors of $\mathbb{V}_1$,
next $2^{n-2}$ columns correspond to the vectors of $\mathbb{V}_2$,
and the last $2^{n-2}$ columns correspond
to the vectors of $\mathbb{V}_3$, in the lexicographic order.
By the construction of $H$, and since the sum of all the vectors
in a subspace equals to the all-zero vector, we have that
$GH^T=0$, therefore, the rows of $G$ are the codewords of $C$.
Moreover, all the rows of $G$ are linearly independent:
for example, if we take the rows in the following order:
$r_2,r_n,r_{n-1},r_{n-2},...,r_3,r_1$, where $r_i$ denotes the $i$th row of $G$,
we obtain a  matrix in row echelon form. Therefore, $K \geq n$,
and thus we proved that $K=n$.
Hence, we have the equality in~(\ref{eq:Griesmer proof}), and therefore,
$d= \frac{3\cdot2^{n-2}}{2}$.
Thus, the code $C$ attains the Griesmer bound.
\end{proof}

\begin{remark} The codes of Theorem \ref{trm:QC+Griesmer} are equivalent to the
punctured Hadamard codes~\cite{MWSl78}.
\end{remark}

It has been observed that for most LDPC codes that decoded with
iterative message-passing algorithms there exists a phenomenon,
called error-floor~\cite{Ric03}. This is a region, where the error probability
does not approaches zero as quickly at high SNRs as it does at low SNRs.
It is known that the error-floor of LDPC codes for AWGN channel
is mostly caused by the combinatorial structure, called trapping set~\cite{Ric03}.
A $(\kappa, \tau)$ \emph{trapping set} of  a code $C$ with  a parity-check matrix
$H$ is defined as  a set $T(\kappa, \tau)$  of size $\kappa$ of the
set of columns of $H$, with $\kappa\geq 1$ and $\tau\geq 0$,
such that in the restriction of $H$ to these $\kappa$ columns,
there are exactly $\tau$ rows of odd weight.

It was shown that the trapping sets with small values $\kappa$
and small ratios $\tau/\kappa$ contribute significantly to
high error-floors~\cite{Ric03}. A $(\kappa,\tau)$ trapping set of an LDPC
codes of length $N$ is called \emph{small} if $\kappa\leq \sqrt N $
and $\tau/\kappa \leq 4$~\cite{LM05}.

Let $C_{\gamma,\rho}$ be a $(\gamma,\rho)$-regular LDPC code, i.e., a code
with the constant column weight $\gamma$ and the constant row weight $\rho$
of the parity-check matrix. Assume that its length is $N$ and the girth is at least six.
In~\cite{HDLA10,HDLA11} were proved two following theorems which show that there are no
(small) trapping sets in $C_{\gamma,\rho}$.

\begin{theorem} There is no trapping set $(\kappa,\tau )$ in $C_{\gamma,\rho}$ such that
$\kappa < \gamma +1$ and $ \tau< \gamma$.
\end{theorem}

\begin{theorem} If $3< \gamma\leq \sqrt{N}$, then there is no small trapping set of size
smaller than $\gamma-3$.
\end{theorem}

Now we apply these results on our LDPC codes derived form lifted MRD codes.
Clearly, these codes have girth six and they are regular codes
with constant column weight $\gamma=q^{n-k}$ for $C$ and constant
column weight $\gamma^T=\frac{q^k-1}{q-1}$
for $C^T$; moreover, $\frac{q^k-1}{q-1}\leq \sqrt{q^{2(n-k)}}$. Hence, we obtain the following results.

\begin{corollary} There is no trapping set $(\kappa,\tau )$ in $C$ such that
$\kappa < q^{n-k} +1$ and $\tau < q^{n-k}$, and there is no trapping set
$(\kappa^T,\tau^T )$ in $C^T$ such that
$\kappa^T < \frac{q^k-1}{q-1} +1$ and $\tau^T < \frac{q^k-1}{q-1}$.
\end{corollary}

\begin{corollary} For all parameters
$q,k,n$ except for $q=k=2$,  there is no small trapping set in $C^T$ of size smaller than
$ \frac{q^k-1}{q-1} -3$.
\end{corollary}

\chapter[New Bounds and Constructions for Codes in Projective Space]
{New Bounds and Constructions for Codes in Projective Space
\let\thefootnote\relax\footnotetext{The material
of Section~\ref{sec:multilevel construction} and Section~\ref{sec:punctured}
is based on~\cite{EtSi09} and also was presented in~\cite{EtSi08}.
The material of
Section~\ref{sec:bounds contructions from designs}
was presented in part in~\cite{SiEt11is}.}}
\label{ch:bounds and constr}

\section[Multilevel Construction via Ferrers Diagrams Rank-Metric Codes]
{Multilevel Construction via Ferrers Diagrams\\ Rank-Metric Codes}
\label{sec:multilevel construction}

Our goal in this section is to
generalize the construction of lifted MRD codes~\cite{SKK08} in the sense
that these codes will be sub-codes of our codes and all our
codes can be partitioned into sub-codes, each one of them is a
lifted rank-metric code (where some of the entries of its codewords are forced to be zeroes).
We use a multilevel approach to design our
codes. First, we select a constant weight code. Each codeword
defines a skeleton of a basis for a subspace in reduced row
echelon form. This skeleton contains a Ferrers diagram on which we
design a rank-metric code. Each such rank-metric code is lifted to
a constant dimension code. The union of these codes is our final
constant dimension code.
The rank-metric codes used for this construction form a new class
of rank-metric codes, called Ferrers diagram rank-metric codes.
The multilevel
construction will be applied to obtain error-correcting constant
dimension codes, but it can be adapted to construct
error-correcting projective space codes without any modification.
We will also consider
the parameters and decoding algorithms for our codes.
The efficiency of the decoding depends on the efficiency of the decoding for
the constant weight codes and the rank-metric codes.

\subsection{Ferrers Diagram Rank-Metric Codes}
\label{subsec:FD rank-metric}
In this subsection
we present rank-metric codes which will
be used for our multilevel construction for codes in the projective space.
Our construction requires
rank-metric codes in which some of the entries are forced to be
zeroes due to constraints given by the Ferrers diagram. We
present an upper bound on the size of such codes. We show how to
construct some rank-metric codes which attain this bound.

Let $v$ be a vector of length $n$ and weight $k$ and let $\mbox{EF}(v)$
be its echelon Ferrers form. Let $\cF$ be the Ferrers diagram of
$\mbox{EF}(v)$. $\cF$ is an $m \times \eta$ Ferrers diagram, $m \leq k$,
$\eta \leq n-k$. A code $\cC$ is an $[\cF,\varrho,\delta]$ {\it
Ferrers diagram rank-metric code} if all codewords are $m \times
\eta$ matrices in which all entries not in $\cF$ are {\it zeroes},
it forms a rank-metric code with dimension $\varrho$, and minimum
rank distance $\delta$. Let $\dim (\cF,\delta)$ be the largest
possible dimension of an $[\cF,\varrho,\delta]$ code.

\begin{theorem}
\label{thm:upper_rank} For a given $i$, $0 \leq i \leq \delta -1$,
if $\nu_i$ is the number of dots in $\cF$, which are not contained
in the first $i$ rows and are not contained in the rightmost
$\delta-1-i$ columns, then $\text{min}_i \{ \nu_i \}$ is an upper
bound of $\dim (\cF,\delta)$.
\end{theorem}
\begin{proof}
For a given $i$, $0 \leq i \leq \delta-1$, let $\cA_i$ be the set
of the $\nu_i$ positions of $\cF$ which are not contained in the
first $i$ rows and are not contained in the rightmost $\delta-1-i$
columns. Assume the contrary that there exists an
$[\cF,\nu_i+1,\delta]$ code $\cC$. Let $\cB = \{ B_1,B_2,
\ldots,B_{\nu_i+1} \}$ be a set of $\nu_i+1$ linearly independent
codewords in $\cC$. Since the number of linearly independent
codewords is greater than the number of entries in $\cA_i$, there
exists a nontrivial linear combination $Y=\sum_{j=1}^{\nu_i+1}
\alpha_j B_j$ for which the $\nu_i$ entries of $\cA_i$ are equal
{\it zeroes}. $Y$~is not the all-zero codeword since the $B_i$'s
are linearly independent. $\cF$ has outside $\cA_i$ exactly $i$
rows and $\delta-i-1$ columns. These $i$ rows can contribute at
most $i$ to the rank of $Y$ and the $\delta-i-1$ columns can
contribute at most $\delta-i-1$ to the rank of $Y$. Therefore $Y$
is a nonzero codeword with rank less than $\delta$, a
contradiction.

Hence, an upper bound on $\dim (\cF,\delta )$ is $\nu_i$ for each
$0 \leq i \leq \delta-1$. Thus, an upper bound on the dimension
$\dim (\cF,\delta )$ is $\text{min}_i \{ \nu_i \}$.
\end{proof}

\begin{conjecture}
The upper bound of Theorem~\ref{thm:upper_rank} is attainable for
any given set of parameters $q$, $\cF$, and $\delta$.
\end{conjecture}

A code which attains the bound of Theorem~\ref{thm:upper_rank} will be
called a \emph{Ferrers diagram MRD code.}
This definition generalizes the definition
of MRD codes.

\vspace{0.3cm}
If we use $i=0$ or $i=\delta-1$ in Theorem~\ref{thm:upper_rank} then we
obtain the following result.

\begin{corollary}
\label{cor:upper_rank} An upper bound on $\dim (\cF,\delta )$ is
the minimum number of dots that can be removed from $\cF$ such
that the diagram remains with at most $\delta-1$ rows of dots or
at most $\delta-1$ columns of dots.
\end{corollary}


\begin{remark}
The $[m \times \eta,\varrho,\delta]$ MRD codes are one class of
Ferrers diagram rank-metric codes which attain the bound of
Corollary~\ref{cor:upper_rank} with equality. In this case the
Ferrers diagram has $m \cdot \eta$ dots.
\end{remark}

\begin{example}
Consider the following Ferrers diagram
\begin{align*}
\cF= \begin{array}{cccc}
\bullet & \bullet & \bullet & \bullet \\
& & \bullet & \bullet\\
& &  & \bullet \\
& & & \bullet
\end{array}
\end{align*}
and $\delta=3$. By Corollary~\ref{cor:upper_rank} we have an upper
bound, $\dim (\cF,3) \leq 2$. But, if we use $i=1$ in
Theorem~\ref{thm:upper_rank} then we have a better upper bound,
$\dim (\cF,3) \leq 1$. This upper bound is attained with the
following basis for an $[ \cF ,1,3]$ rank-metric code.
\[
\left(\begin{array}{cccc}
\bf 1 & \bf 0 & \bf 0 & \bf 0\\
0 &  0 & \bf 1 & \bf 0\\
0 & 0 & 0 & \bf 0\\
0 & 0 & 0 & \bf 1\end{array}\right).\]
\end{example}
\vspace{0.6cm}

When the bound of Theorem~\ref{thm:upper_rank} is attained? We
start with a construction of Ferrers diagram rank-metric codes
which attain the bound of Corollary~\ref{cor:upper_rank}.

\vspace{0.3cm}
Assume we have an $m \times \eta$, $m=\eta+\varepsilon$, $\varepsilon
\geq 0$, Ferrers diagram $\cF$, and that the minimum in the bound
of Corollary~\ref{cor:upper_rank} is obtained by removing all the
dots from the $\eta-\delta+1$ leftmost columns of $\cF$. Hence,
only the dots in the $\delta-1$ rightmost columns will remain. We
further assume that each of the $\delta-1$ rightmost columns of
$\cF$ have $m$ dots. The construction which follows is based on
the construction of MRD $q$-cyclic rank-metric codes given by
Gabidulin~\cite{Gab85}.

A code $\cC$ of length $m$ over $\F_{q^m}$ is called a {\it
$q$-cyclic code} if $(c_{0},c_{1},...,c_{m-1}) \in \cC$ implies
that $(c_{m-1}^{q},c_{0}^{q},...,c_{m-2}^{q})\in \cC$.

For a construction of $[m \times m, \varrho , \delta]$ rank-metric
codes, we use an isomorphism between the field with $q^m$
elements, $\F_{q^m}$, and the set of all $m$-tuples over $\F_q$,
$\F_q^m$. We use the obvious isomorphism by the representation of
an element $\alpha$ in the extension field $\F_{q^m}$ as $\alpha =
( \alpha_1 , \ldots, \alpha_m )$, where $\alpha_i$ is an element
in the ground field $\F_q$.

\vspace{0.3cm}
Recall that a codeword $c$ in an $[m \times m, \varrho , \delta]$ rank-metric
code $\cC$, can be represented by a vector $c=(c_0 , c_1 , \ldots
, c_{m-1})$, where $c_i \in \F_{q^m}$ and the generator matrix $G$
of $\cC$ is an $K \times m$ matrix, $\varrho = mK$. It was proved
by Gabidulin~\cite{Gab85} that if $\cC$ is an MRD $q$-cyclic code
then the generator polynomial of $\cC$ is the linearized
polynomial
$G(x)=\overset{m-K}{\underset{i=0}{\sum}}g_{i}x^{q^{i}}$, where
$g_i \in \F_{q^m}$, $g_{m-K}=1$, $m=K+\delta-1$, and its generator
matrix $G$ has the following form

\begin{align*}
\left(\begin{array}{cccccccc}
g_{0} & g_{1} & \cdots & g_{m-K-1} & 1 & 0 & \cdots & \cdots \\
0 & g_{0}^q & g_{1}^q & \cdots & g_{m-K-1}^q & 1 & \cdots & \cdots \\
0 & 0 & g_{0}^{q^2} & \cdots & \cdots & g_{m-K-1}^{q^2} & 1 & \cdots \\
\cdots & \cdots & \cdots & \cdots & \cdots & \cdots & \cdots  & \cdots \\
0 & \cdots & \cdots & \cdots & \cdots & \cdots &
g_{m-K-1}^{q^{K-1}} & 1
\end{array}\right).
\end{align*}

\vspace{0.2cm}
Hence, a codeword $c \in \cC$, $c \in (\F_{q^m})^m$, derived from
the information word $(a_0,a_1, \ldots,a_{K-1})$, where $a_i \in
\F_{q^m}$, i.e. $c = (a_0,a_1, \ldots,a_{K-1})G$, has the form

\begin{equation*}
c= (a_0 g_0 , a_0 g_1 + a_1 g_0^q , \ldots,a_{K-2}+a_{K-1}
g_{m-K-1}^{q^{K-1}} , a_{K-1})~.
\end{equation*}
\vspace{0.2cm}

We define an $[m \times \eta, m(\eta-\delta+1) , \delta]$
rank-metric code $\cC'$, $m = \eta + \varepsilon$, derived from
$\cC$ as follows:
$$
\cC' = \{ (c_0 , c_1 , \ldots , c_{\eta-1}) ~:~ (0,\ldots,0,c_0 ,
c_1 , \ldots , c_{\eta-1}) \in \cC \} .
$$
\begin{remark}
$\cC'$ is also an MRD code.
\end{remark}
We construct an $[\cF, \ell , \delta]$ Ferrers diagram rank-metric
code $\cC_{\cF} \subseteq \cC'$, where $\cF$ is an $m \times \eta$
Ferrers diagram. Let $\gamma_i$, $1 \leq i \leq \eta$, be the
number of dots in column $i$ of $\cF$, where the columns are
indexed from left to right. A codeword of $\cC_{\cF}$ is derived
from a codeword of $c \in \cC$ by satisfying a set of $m$
equations implied by

\begin{small}
\begin{equation}
\label{eq:equations}
\begin{array}{c}\left(a_{0}g_{0},a_{0}g_{1}+a_{1}g_{0}^{q},
\ldots,a_{K-2}+a_{K-1}g_{m-K-1}^{q^{K-1}},a_{K-1}\right)
=\left(\overset{\varepsilon}{\overbrace{\begin{array}{c}
0\\
\vdots\\
0\end{array}~\ldots~\begin{array}{c}
0\\
\vdots\\
0\end{array}}~}f_{1}~\ldots
f_{K-\varepsilon}\overset{\delta-1}{~\overbrace{\begin{array}{c}
\bullet\\
\vdots\\
\bullet\end{array}~\ldots~\begin{array}{c}
\bullet\\
\vdots\\
\bullet\end{array}}}\right) \end{array}
\end{equation}
\end{small}

\noindent where $f_i = ( \underset{\gamma_i} {\underbrace{\bullet
\cdots \bullet}} ~ \underset{m-\gamma_i}{\underbrace{0 \cdots
0}})^T$ is a column vector of length $m$, $1 \leq i \leq
K-\varepsilon$, and $u^T$ denotes the transpose of the vector $u$.
It is easy to verify that $\cC_{\cF}$ is a linear code.

By (\ref{eq:equations}) we have a system of $m=K+\delta-1$
equations with $K$ variables, $a_0,a_1, \ldots, a_{K-1}$. The
first $\varepsilon$ equations implies that $a_i=0$ for $0 \leq i
\leq \varepsilon -1$. The next $K-\varepsilon=\eta-\delta+1$
equations determine the values of the $a_i$'s, $\varepsilon \leq i
\leq K-1$, as follows. From the next equation $a_\varepsilon
g_0^{q^{\varepsilon}}=( \underset{\gamma_1}{\underbrace{\bullet
\cdots \bullet}}~ \underset{m-\gamma_1}{\underbrace{00...0}})^T$
(this is the next equation after we substitute $a_i=0$ for $0 \leq
i \leq \varepsilon-1$), we have that $a_\varepsilon$ has
$q^{\gamma_1}$ solutions in $\F_{q^m}$, where each element of
$\F_{q^m}$ is represented as an $m$-tuple over $\F_q$. Given a
solution of $a_\varepsilon$, the next equation $a_\varepsilon
g_0^{q^{\varepsilon}}+a_{\varepsilon+1} g_1^{q^{\varepsilon+1}}=
(\underset{\gamma_2} {\underbrace{\bullet \cdots
\bullet}}~\underset{m-\gamma_2}{\underbrace{00...0}})^T$ has
$q^{\gamma_2}$ solutions for $a_{\varepsilon+1}$. Therefore, we
have that $a_0,a_1, \ldots, a_{K-1}$ have
$q^{\sum_{i=1}^{K-\varepsilon} \gamma_i}$ solutions and hence the
dimension of $\cC_{\cF}$ is $\sum_{i=1}^{K-\varepsilon} \gamma_i$
over $\F_q$. Note, that since each of the $\delta-1$ rightmost
columns of $\cF$ have $m$ dots, i.e. $\gamma_i=m$,
$K-\varepsilon+1 \leq i \leq \eta$ (no {\it zeroes} in the related
equations) it follows that any set of values for the $a_i$'s
cannot cause any contradiction in the last $\delta-1$ equations.
Also, since the values of the $K$ variables $a_0,a_1, \ldots,
a_{K-1}$ are determined for the last $\delta-1$ equations, the
values for the related $(\delta-1)m$ dots are determined. Hence,
they do not contribute to the number of solutions for the set of
$m$ equations. Thus, we have

\begin{theorem}
\label{thm:bound_attain}
Let $\cF$ be an $m \times \eta$, $m \geq \eta$, Ferrers diagram.
Assume that each one of the rightmost $\delta-1$ columns of $\cF$
has $m$ dots, and the $i$-th column from the left of $\cF$ has
$\gamma_i$ dots. Then $\cC_{\cF}$ is an $[ \cF , \sum_{i=1}^{\eta
- \delta +1} \gamma_i , \delta ]$ code which attains the bound of
Corollary~\ref{cor:upper_rank}.
\end{theorem}

\begin{remark}
For any solution for variables $a_0,a_1, \ldots, a_{K-1}$ we have that
$(a_0,a_1, \ldots,a_{K-1})G $  $=(0,\ldots,0,c_0 , c_1 , \ldots ,
c_{\eta-1}) \in \cC$ and $(c_0 , c_1 , \ldots , c_{\eta-1}) \in
\cC_{\cF}$.
\end{remark}
\begin{remark}
For any $[m \times \eta, m(\eta-\delta+1) , \delta]$ rank-metric
code $\cC'$, the codewords which have {\it zeroes} in all the
entries which are not contained in $\cF$ form an $[ \cF ,
\sum_{i=1}^{\eta - \delta +1} \gamma_i , \delta ]$ code. Thus, we
can use also any MRD codes, e.g., the codes described
in~\cite{Rot91}, to obtain a proof for
Theorem~\ref{thm:bound_attain}.
\end{remark}
\begin{remark}
Since $\cC_{\cF}$ is a subcode of an MRD code then we can use the
decoding algorithm of the MRD code for the decoding of our code.
Also note, that if $\cF$ is an $m \times \eta$, $m < \eta$,
Ferrers diagram then we apply our construction for the $\eta
\times m$ Ferrers diagram of the conjugate partition.
\end{remark}

When $\delta = 1$ the bounds and the construction are trivial. If
$\delta=2$ then by definition the rightmost column and the top row
of an $m \times \eta$ Ferrers diagram always has $m$ dots and
$\eta$ dots, respectively. It implies that the bound of
Theorem~\ref{thm:upper_rank} is always attained with the
construction if $\delta =2$. This is the most interesting case
since in this case the improvement of our constant dimension codes
compared to the lifted MRD codes in~\cite{KK,SKK08} is the most impressive (see
Subsection~\ref{subsec:parameters}). If $\delta >2$ the improvement
is relatively small, but we will consider this case as it is of
interest also from a theoretical point of view.
We will give two simple examples for constructions of Ferrers diagram
MRD codes with $\delta=3$.

\vspace{0.2cm}
\begin{example}
\label{ex:ferrers1} Consider the following Ferrers diagram
\begin{align*}
\cF= \begin{array}{cccc}
\bullet & \bullet & \bullet & \bullet \\
& \bullet & \bullet & \bullet\\
& & \bullet & \bullet \\
& & & \bullet
\end{array}.
\end{align*}
The upper bound on $\dim (\cF,3)$ is $3$. It is attained with the
following basis with three $4 \times 4$ matrices.

\[
\left(\begin{array}{cccc}
\bf 0 & \bf 1 & \bf 0 & \bf 0\\
0 & \bf 0 & \bf 1 & \bf 0\\
0 & 0 & \bf 0 & \bf 0\\
0 & 0 & 0 & \bf 1\end{array}\right),\left(\begin{array}{cccc}
\bf 0 & \bf 0 & \bf 0 & \bf 1\\
0 & \bf 1 & \bf 0 & \bf 0\\
0 & 0 & \bf 1 & \bf 0\\
0 & 0 & 0 & \bf 0\end{array}\right),\left(\begin{array}{cccc}
\bf 1 & \bf 0 & \bf 0 & \bf 0\\
0 & \bf 1 & \bf 0 & \bf 0\\
0 & 0 & \bf 0 & \bf 1\\
0 & 0 & 0 & \bf 1\end{array}\right).\]
\end{example}
\begin{example}
Consider the following Ferrers diagram
\begin{align*}
\cF= \begin{array}{cccc}
\bullet & \bullet & \bullet & \bullet \\
& \bullet & \bullet & \bullet \\
& \bullet & \bullet & \bullet \\
& & & \bullet
\end{array}.
\end{align*}
The upper bound on $\dim (\cF,3)$ is $4$. It is attained with the
basis consisting of four $4 \times 4$ matrices, from which three
are from Example~\ref{ex:ferrers1} and the last one is
\[
\left(\begin{array}{cccc}
\bf 1 & \bf 0 & \bf 1 & \bf 0\\
0 & \bf 0 & \bf 0 & \bf 1\\
0 & \bf 1 & \bf 0 & \bf 1\\
0 & 0 & 0 & \bf 0\end{array}\right).\]
\end{example}


\subsection{Lifted Ferrers Diagram Rank-Metric Codes}
\label{subsec:lifted FD rank-metric codes}

Usually a lifted MRD code $\C^{\textmd{MRD}}$ is not maximal and it can be
extended. This extension requires to design rank-metric codes,
where the shape of a codeword is a Ferrers diagram rather than an
$k \times (n-k)$ matrix. We would like to use the largest possible
Ferrers diagram rank-metric codes. In the appropriate cases, e.g.,
when $\delta=2$,  for this purpose we will use the Ferrers diagram MRD codes constructed
in the previous subsection.

Assume we are given an echelon Ferrers form $\mbox{EF}(v)$ of a binary
vector $v$, of length $n$ and weight $k$, with a Ferrers diagram
$\cF$ and a Ferrers diagram rank-metric code $\cC_{\cF}$.
$\cC_{\cF}$ is \emph{lifted} to a constant dimension code $\C_v$ by
substituting each codeword $A \in \cC_{\cF}$ in the columns of
$\mbox{EF}(v)$ which correspond to the {\it zeroes} of $v$, to
obtain the generator matrix for a codeword in $\C_v$. Note, that
depending on $\cF$ it might implies conjugating $\cF$ first.
Unless $v$ starts with an {\it one} and ends with a {\it zero}
(the cases in which $\cF$ is a $k \times (n-k)$ Ferrers diagram)
we also need to expand the matrices of the Ferrers diagram
rank-metric code to $k \times (n-k)$ matrices (which will be
lifted), where $\cF$ is in their upper right corner (and the new
entries are {\it zeroes}). As an immediate consequence
from Theorem~\ref{trm:CDC from rank-metric} we have the following.

\begin{lemma}
\label{lem:dist_lift} If $\cC_{\cF}$ is an $[ \cF , \varrho ,
\delta ]$ Ferrers diagram rank-metric code then its lifted code
$\C_v$, related to an $k \times n$ echelon Ferrers form $\emph{EF}(v)$,
is an $(n, q^\varrho , 2\delta , k)_q$ constant dimension code.
\end{lemma}


\begin{example}
For the word $v= 1110000 $, its echelon Ferrers form
\begin{align*}
\emph{EF}(v)=\left[ \begin{array}{ccccccc}
1 & 0 & 0 & \bullet & \bullet & \bullet & \bullet \\
0 & 1 & 0 & \bullet & \bullet & \bullet & \bullet \\
0 & 0 & 1 & \bullet & \bullet & \bullet & \bullet
\end{array}
\right]~,
\end{align*}
the $3 \times 4$ matrix
\[
\left(\begin{array}{cccc}
\bf 1 & \bf 0 & \bf 1 & \bf 0\\
\bf 0 & \bf 0 & \bf 0 & \bf 1\\
\bf 0 & \bf 0 & \bf0 & \bf 0
\end{array}\right)\]
is lifted to the $3$-dimensional subspace with the $3 \times 7$ generator matrix
\begin{align*}
\left( \begin{array}{ccccccc}
1 & 0 & 0 & \bf 1 & \bf 0 & \bf 1 & \bf 0\\
0 & 1 & 0 & \bf 0 & \bf 0 & \bf 0 & \bf 1\\
0 & 0 & 1 & \bf 0 & \bf 0 & \bf0 & \bf 0
\end{array}
\right)~.
\end{align*}
\vspace{0.3cm}
For the word $v= 1001001 $, its echelon Ferrers form
\begin{align*}
\emph{EF}(v)=\left[ \begin{array}{ccccccc}
1 & \bullet & \bullet & 0 & \bullet & \bullet & 0 \\
0 & 0 & 0 & 1 & \bullet & \bullet & 0 \\
0 & 0 & 0 & 0 & 0 & 0 & 1
\end{array}
\right]~,
\end{align*}
the $2 \times 4$ matrix
\[
\left(\begin{array}{cccc}
\bf 1 & \bf 0 & \bf 1 & \bf 0\\
0 &  0 & \bf 0 & \bf 1
\end{array}\right)\]
is lifted to the $3$-dimensional subspace with the $3 \times 7$ generator matrix
\begin{align*}
\left( \begin{array}{ccccccc}
1 & \bf 1 & \bf 0 & 0 & \bf 1 & \bf 0 & 0\\
0 & 0 & 0 & 1 & \bf 0 & \bf 1 & 0\\
0 & 0 & 0 & 0 & 0 & 0 & 1
\end{array}
\right)~.
\end{align*}
\end{example}

A lifted MRD code described in~\cite{SKK08} can be considered as a
lifted Ferrers diagram MRD code, where its identifying vector is $(1
\ldots 1 0 \ldots 0)$. If our lifted codes are the codes constructed
in  Section~\ref{subsec:FD rank-metric} then the same decoding algorithm can
be applied. Therefore, the decoding in~\cite{SKK08} for the
corresponding constant dimension code can be applied directly to
each of our lifted Ferrers diagram MRD codes in this case, e.g. it
can always be applied when $\delta=2$. It would be worthwhile to
permute the coordinates in a way that the identity matrix $I_k$
will appear in the first $k$ columns, from the left, of the reduced row echelon
form, and $\cF$ will appear in the upper right corner of the $k
\times n$ matrix. The reason is that the decoding of~\cite{SKK08} is
described on such matrices.

\subsection{Multilevel Construction}

Assume we want to construct an $(n,M,2\delta,k)_q$ constant
dimension code $\C$.

\begin{itemize}

\item The first step in the construction is to choose a binary constant
weight code ${\bf C}$ of length $n$, weight $k$, and minimum
distance $2 \delta$. This code will be called the {\it skeleton
code}. Any constant weight code can be chosen for this purpose,
but different skeleton codes will result in different constant
dimension codes with usually different sizes. The best choice for
the skeleton code ${\bf C}$ will be discussed in the next
subsection. The next three steps are performed for each codeword
$c \in {\bf C}$.

\item The second step is to construct the echelon Ferrers form $\mbox{EF}(c)$.

\item The third step is to construct an $[\cF,\varrho,\delta]$ Ferrers
diagram rank-metric code $\cC_{\cF}$ for the Ferrers diagram $\cF$
of $\mbox{EF}(c)$. If possible we will construct a code as described in
Subsection~\ref{subsec:FD rank-metric}.

\item The fourth step is to lift $\cC_{\cF}$ to a constant dimension
code $\C_c$, for which the echelon Ferrers form of $X \in \C_c$ is
$\mbox{EF}(c)$.

\end{itemize}

Finally,
$$
\C = \bigcup_{c \in {\bf C}} \C_c ~.
$$
\vspace{0.5cm}

Recall that  by Corollary~\ref{cor:distance}, for any two subspaces
$X,Y\subseteq \F_q^n$, we have that
$d_{S}(X,Y)\geq d_{H}(v(X),v(Y))$. Hence,
as an immediate consequence of Corollary~\ref{cor:distance}
and Lemma~\ref{lem:dist_lift} we have the following theorem.
\begin{theorem}
$\C$ is an $(n,M,2\delta,k)_q$ constant dimension code,
where $M= \sum_{c \in {\bf C}} |\C_c| $.
\end{theorem}

\vspace{0.2cm}
\begin{example}
\label{ex:n-6_k=3} Let $n=6$, $k=3$, and ${\bf C}= \{ 111000,~
100110,~ 010101,~ 001011 \}$ be a constant weight code of length $6$,
weight $3$, and minimum Hamming distance $4$. The echelon Ferrers
forms of these $4$ codewords are

\begin{align*}
\emph{EF}(111000)=\left[ \begin{array}{cccccc}
1 & 0 & 0 & \bullet & \bullet & \bullet \\
0 & 1 & 0 & \bullet & \bullet & \bullet \\
0 & 0 & 1 & \bullet & \bullet & \bullet
\end{array}
\right],\;
%
\emph{EF}(100110)=\left[ \begin{array}{cccccc}
1 & \bullet & \bullet & 0 & 0 & \bullet \\
0 & 0 & 0 & 1 & 0 & \bullet \\
0 & 0 & 0 & 0 & 1 & \bullet
\end{array}
\right],
\end{align*}
\begin{align*}
\emph{EF}(010101)=\left[ \begin{array}{cccccc}
0 & 1 & \bullet & 0 & \bullet & 0 \\
0 & 0 & 0 & 1 & \bullet & 0 \\
0 & 0 & 0 & 0 & 0 & 1
\end{array}
\right],\;
%
\emph{EF}(001011)=\left[ \begin{array}{cccccc}
0 & 0 & 1 & \bullet & 0 & 0 \\
0 & 0 & 0 & 0 & 1 & 0 \\
0 & 0 & 0 & 0 & 0 & 1
\end{array}
\right] ~.
\end{align*}
By Theorem~\ref{thm:bound_attain}, the Ferrers diagrams of these four
echelon Ferrers forms yield Ferrers diagram MRD codes of
sizes $64$, $4$, $2$, and $1$, respectively. Hence, we obtain a
$(6,71,4,3)_2$ constant dimension code $\C$.
\end{example}
\begin{remark} A $(6,77,4,3)_2$ code was obtained by computer
search~\cite{KoKu08}. Similarly, we obtain a $(7,289,4,3)_2$ code.
A $(7,304,4,3)_2$ code was obtained by computer
search~\cite{KoKu08}.
\end{remark}
\begin{example}
\label{ex:c844} Let ${\bf C}$ be the codewords of weight $4$ in the
$[8,\;4,\;4]$ extended Hamming code with the following parity-check
matrix.
\begin{align*}
\left( \begin{array}{cccccccc}
0 & 0 & 0 & 0 & 1 & 1 & 1 & 1 \\
0 & 0 & 1 & 1 & 0 & 0 & 1 & 1 \\
0 & 1 & 0 & 1 & 0 & 1 & 0 & 1 \\
1 & 1 & 1 & 1 & 1 & 1 & 1 & 1
\end{array}
\right)
\end{align*}
${\bf C}$ has $14$ codewords with weight $4$. Each one of these
codewords is considered as an identifying vector for the echelon
Ferrers forms from which we construct the final $(8,4573,4,4)_2$
code $\C$. The fourteen codewords of ${\bf C}$ and their
contribution for the final code $\C$ are given in
Table~\ref{tab:8-4573-4-4}. The codewords are taken in lexicographic order.

\begin{table}[h]
\centering \caption{The $(8, 4573, 4, 4)_2$ code $\C$}
\label{tab:8-4573-4-4}
\begin{tabular}{|c|>{\centering}p{3cm}|>{\centering}p{3cm}|}
\hline
 & $c \in \bf C$ & \textmd{size of} $\C_c$ \tabularnewline
\hline \hline 1 & 11110000 & 4096\tabularnewline \hline 2 &
11001100 & 256\tabularnewline \hline 3 & 11000011 &
16\tabularnewline \hline 4 & 10101010 & 64\tabularnewline \hline 5
& 10100101 & 16\tabularnewline \hline 6 & 10011001 &
16\tabularnewline \hline 7 & 10010110 & 16\tabularnewline \hline 8
& 01101001 & 32\tabularnewline \hline 9 & 01100110&
16\tabularnewline \hline 10 & 01011010 & 16\tabularnewline \hline
11 & 01010101 & 8\tabularnewline \hline 12 & 00111100 &
16\tabularnewline \hline 13 & 00110011& 4\tabularnewline \hline
14 & 00001111 & 1\tabularnewline \hline
\end{tabular}
\end{table}
\end{example}
\vspace{0.6cm}

\subsection{Code Parameters}
\label{subsec:parameters}

Now, we discuss the size of our constant dimension codes obtained by the multilevel construction,
the required choice for the skeleton code ${\bf C}$, and compare the size
of our codes with the size of the lifted MRD codes constructed
in~\cite{SKK08}.

The size of the final constant dimension code $\C$ depends on the
choice of the skeleton code ${\bf C}$. The identifying vector with
the largest size of corresponding rank-metric code is
$\underset{k}{\underbrace{1 \cdots 1}}\underset{n-k}{\underbrace{0
\cdots 0}}$. The corresponding $[k \times (n-k),\ell,\delta]$
MRD code has dimension $\ell=(n-k)(k-\delta+1)$ and hence
it contributes $q^{(n-k)(k-\delta+1)}$ $k$-dimensional subspaces
to our final code $\C$. These subspaces form the lifted MRD codes
of~\cite{SKK08}. The next identifying vector which contributes
the most number of subspaces to $\C$ is
$\underset{k-\delta}{\underbrace{11...1}}\underset{\delta}
{\underbrace{0 \cdots 0}}\underset{\delta}{\underbrace{11...1}}
\underset{n-k-\delta}{\underbrace{000...00}}$. The number of
subspaces it contributes depends on the bounds presented in
Subsection~\ref{subsec:FD rank-metric}. The rest of the code $\C$ usually has
less codewords from those contributed by these two. Therefore, the
improvement in the size of the code compared to the lifted MRD code
is not dramatic. But, for most parameters our codes
are larger than the best known codes. In some cases, e.g. when
$\delta=k$ our codes are as good as the best known codes
(see~\cite{EV08}) and suggest an alternative construction.  When
$k=3$, $\delta=4$, and $n\leq 12$, the cyclic codes
constructed in~\cite{EV08,KoKu08} are larger.

Two possible alternatives for the best choice for the skeleton
code ${\bf C}$ might be of special interest. The first one is for
$k=4$ and $n$ which is a power of two. We conjecture that the best
skeleton code is constructed from the codewords with weight 4 of
the extended Hamming code for which the columns of the
parity-check matrix are given in lexicographic order. We
generalize this choice of codewords from the Hamming code by
choosing a constant weight lexicode~\cite{CoSl86}. Such a code is
constructed as follows. All vectors of length $n$ and weight $k$
are listed in lexicographic order. The code ${\bf C}$ is generated
by adding to the code ${\bf C}$ one codeword at a time. At each
stage, the first codeword of the list that does not violate the
distance constraint with the other codewords of ${\bf C}$, is
joined to ${\bf C}$. Lexicodes are not necessarily the best
constant weight codes. For example, the largest constant weight
code of length 10 and weight 4 is 30, while the lexicode with the
same parameters has size 18. But, the constant dimension code
derived from the lexicode is larger than any constant dimension
code derived from any related code of size 30.

Table~\ref{tab:ML vs MRD} summarized the sizes of some of our codes,
denoted by $\C^{\textmd{ML}}$, obtained by the multilevel construction
compared to the sizes of lifted MRD codes $\C^{\textmd{MRD}}$.
In all these codes we have
started with a constant weight lexicode in the first step of the
construction.

\begin{table}[h]
\centering \caption{$\C^\textmd{{ML}}$ vs. $\CMRD$}
\label{tab:ML vs MRD}
\begin{tabular}{|c|c|c|c|c|c|}
\hline $q$ & $d_S (\C)$ & $n$ & $k$ & $|\C^{\textmd{MRD}}|$ & $|\C^{\textmd{ML}}|$
\tabularnewline \hline \hline 2 & 4 & 9 & 4 & $2^{15}$ &
$2^{15}$+4177\tabularnewline \hline 2 & 4 & 10 & 5 & $2^{20}$ &
$2^{20}$+118751\tabularnewline \hline 2 & 4 & 12 & 4 & $2^{24}$ &
$2^{24}$+2290845\tabularnewline \hline 2 & 6 & 10 & 5 & $2^{15}$ &
$2^{15}$+73\tabularnewline \hline
2 & 6 & 13 & 4 & $2^{18}$ & $2^{18}$+4357\tabularnewline \hline 2
& 8 & 21 & 5 & $2^{32}$ & $2^{32}$+16844809 \tabularnewline \hline
3 & 4 & 7 & 3 & $3^{8}$ & $3^{8}$+124\tabularnewline \hline 3 & 4
& 8 & 4 & $3^{12}$ & $3^{12}$+8137\tabularnewline \hline 4 & 4 & 7
& 3 & $4^{8}$ & $4^{8}$+345\tabularnewline \hline 4 & 4 & 8 & 4 &
$4^{12}$ & $4^{12}$+72529\tabularnewline \hline
\end{tabular}
\end{table}

\subsection{Decoding}

The decoding of our codes is quite straightforward and it mainly
consists of known decoding algorithms. As we used a multilevel
coding we will also need a multilevel decoding. In the first step
we will use a decoding for our skeleton code and in the
second step we will use a decoding for the rank-metric codes.

Assume the received word was a $k$-dimensional subspace $Y$. We
start by generating its reduced row echelon form $\mbox{RE}(Y)$. Given
$\mbox{RE}(Y)$ it is straightforward to find the identifying vector
$v(Y)$. Now, we use the decoding algorithm for the constant weight
code to find the identifying vector $v(X)$ of the submitted
$k$-dimensional subspace $X$. If no more than $\delta -1$ errors
occurred then we will find the correct identifying vector. This
claim is an immediate consequence of
Corollary~\ref{cor:distance}.

In the second step of the decoding we are given the received
subspace $Y$, its identifying vector $v(Y)$, and the identifying
vector $v(X)$ of the submitted subspace $X$. We consider the
echelon Ferrers form $\mbox{EF}(v(X))$, its Ferrers diagram $\cF$, and
the $[\cF, \varrho, \delta ]$ Ferrers diagram rank-metric code
associated with it. We can permute the columns of $\mbox{EF}(v(X))$, and
use the same permutation on $Y$, in a way that the identity matrix
$I_k$ will be in the left side. Now, we can use the decoding of
the specific rank-metric code. If our rank-metric codes are those
constructed in Subsection~\ref{subsec:FD rank-metric} then we can use the
decoding as described in~\cite{SKK08}. It is clear now that the
efficiency of our decoding depends on the efficiency of the
decoding of our skeleton code and the efficiency of the decoding
of our rank-metric codes. If the rank-metric codes are MRD codes
then they can be decoded efficiently~\cite{Gab85,Rot91}. The same is
true if the Ferrers diagram metric codes are those constructed in
Subsection~\ref{subsec:FD rank-metric} as they are subcodes of MRD codes and the
decoding algorithm of the related MRD code applied to them too.

There are some alternative ways for our decoding, some of which
improve on the complexity of the decoding. For example we can make
use of the fact that most of the code is derived from two
identifying vectors or that most of the rank-metric codes are of
relatively small size. One such case can be when all the identity
matrices of the echelon Ferrers forms are in consecutive columns
of the codeword (see~\cite{Ska10}).

Finally, if we allow to receive a word which is an
$\ell$-dimensional subspace $Y$, $k-\delta+1 \leq \ell \leq
k+\delta-1$, then the same procedure will work as long as $d_S
(X,Y) \leq \delta -1$. This is a consequence of the fact that the
decoding algorithm of~\cite{SKK08} does not restrict the dimension
of the received word.

\section[{Bounds and Constructions for Constant Dimension Codes that
Contain $\CMRD$}]{Bounds and Constructions for Constant Dimension\\ Codes that Contain $\CMRD$}
\label{sec:bounds contructions from designs}
Most of the constructions for constant dimension codes known in
the literature produce codes which contain
$\CMRD$~\cite{EtSi09,GaYa10,MGR08,SiEt10,SKK08,Ska10,TrRo10}. The
only constructions which generate codes that do not contain
$\CMRD$ are given in~\cite{EV08,KoKu08}. These constructions are
either of so called orbit codes or specific constructions for
small parameters. Moreover, only $(n,M,d ,3)_2$ orbit codes
(specifically cyclic codes) with $8 \leq n \leq 12$, and
$(6,77,4,3)_2$ and $(7,304,4,3)_2$ codes are the largest codes for
their specific parameters which do not contain
$\CMRD$~\cite{KoKu08}. This motivates the question, what is the
largest constant dimension code which contain $\CMRD$?

In this section we consider upper bounds and constructions for constant dimension codes
which contain the lifted MRD code.
First, we consider two types of upper bounds on the size of constant dimension codes,
presented in~\cite{EV,EV08,KK,WXS-N03}. We estimate the size of lifted MRD codes
and codes generated by the multilevel construction
relatively to these bounds. Next,
we discuss  upper bounds
on the size of constant dimension codes which contain $\C^\textmd{{MRD}}$.
In particular we prove
that if an $(n,M,2(k-1),k)_q$ code $\mathbb{C}$, $k \geq 3$,
contains the $(n,q^{2(n-k)},2(k-1),k)_q$ lifted MRD code then

$$M\leq q^{2(n-k)}+ \cA_q (n-k,2(k-2),k-1)~.$$

We also present a construction for codes which either
attain this bound or almost attain it for $k=3$. These codes are
the largest known $(n,M,4,3)_q$ codes for $n \geq 13$.

We prove that if an $(n,M,2k,2k)_q$ code $\mathbb{C}$ contains the
$(n,q^{(n-2k)(k+1)},2k,2k)_q$ lifted MRD code then

$$M\leq q^{(n-2k)(k+1)}+ \GaussLarge{n-2k}{k}
\frac{q^n-q^{n-2k}}{q^{2k}-q^k} + \cA_q (n-2k, 2k,2k)~.$$

We present a construction for codes which
attain this bound when $k=2$, $n=8$, and for all $q$. These codes are
the largest known for the related parameters.

\subsection {Upper Bounds for Constant Dimension Codes}
\label{subsec:known bounds}

In this subsection we discuss two upper bounds for constant dimension codes
presented in Theorem~\ref{trm:Singleton} (the Singleton bound) and Theorem~\ref{trm:Steiner}, given
by
\begin{equation}
\label{eq:Singleton2}
\mathcal{A}_{q}(n,2\delta,k)\leq \GaussLarge{n-\delta+1}{k-\delta+1}
\end{equation}
and
\begin{equation}
\mathcal{A}_{q}(n,2\delta,k)\leq\frac{\GaussLarge{n}{k-\delta+1}}{\GaussLarge{k}{k-\delta+1}}.
\label{eq:Johnson}
\end{equation}

It was proved in~\cite{KK} that the ratio of the size of a lifted MRD code
to the Singleton bound~(\ref{eq:Singleton2}) satisfies
$$
\frac{|\C^\textmd{{MRD}}|}{\GaussLarge{n-\delta+1}{k-\delta+1}}\geq Q_0,
$$
where $Q_0$, called probabilistic combinatorial constant, is equal
to $\prod _{i=0}^\infty(1-2^{-i})\approx 0.2887881$.

Now we estimate the bound~(\ref{eq:Johnson}).
$$
\frac{\Gauss{n}{k-\delta+1}}{\Gauss{k}{k-\delta+1}}=
\frac{(q^{n}-1)(q^{n-1}-1)\ldots(q^{n-k+\delta}-1)}{(q^{k}-1)(q^{k-1}-1)\ldots(q^{\delta}-1)}
$$
$$
=q^{(n-k)(k-\delta+1)}\frac{(1-q^{-n})(1-q^{-n+1})\ldots(1-q^{-n+k-\delta})}
{(1-q^{-k})(1-q^{-k+1})\ldots(1-q^{-\delta})}
< \frac{q^{(n-k)(k-\delta+1)}}{\prod_{j=\delta}^{\infty}(1-q^{-j})}.
$$
We define $Q_s(q)=\prod_{j=s+1}^{\infty}(1-q^{-j})$. Thus, we have
\begin{lemma}
\label{lm:ratio} The ratio between the size of a lifted MRD code
and the upper bound on $\cA_q(n,2\delta,k)$ given
in~(\ref{eq:Johnson}) satisfies

$$\frac{|\C^\emph{{MRD}}|}{\GaussLarge{n}{k-\delta+1}/\GaussLarge{k}{k-\delta+1}}
> Q_{\delta-1}(q).
$$
\end{lemma}
The function $Q_s(q)$ is increasing in $q$ and also in $s$. We
provide several values of $Q_s(q)$ for different $q$ and $s$ in Table~\ref{tab:Q-s-q}. For
$q=2$ these values were given in~\cite{Ber80}. $Q_1(q)$ was considered also in~\cite{GaYa10a}.

\begin{table}[h!tab]
\centering \caption{ $Q_{s}(q)$}
\label{tab:Q-s-q}
\begin{tabular}{|c|c|c|c|c|c|}
  \hline
  \backslashbox{s}{q} & 2 & 3 & 4 & 5&7  \\ \hline
  $1$ & 0.5776 & 0.8402 & 0.9181& 0.9504 & 0.9763 \\
  \hline
  $2$ & 0.7701 & 0.9452 & 0.9793 & 0.9900 & 0.9966\\
  \hline
  $3$ &  0.8801 & 0.9816 & 0.9948 & 0.9980 & 0.9995\\
   \hline
  $4$ &  0.9388 & 0.9938 & 0.9987 & 0.9996& 0.9999 \\
  \hline
\end{tabular}
\end{table}

One can see that for $q$ large enough or for $\delta$ large enough
the size of a lifted MRD code approaches the upper
bound~(\ref{eq:Johnson}). Thus an improvement on the lower bound
of $\cA_q(n,2\delta,k)$ is important for small minimum distance
and small $q$.

\vspace{0.1cm}
Note, that the lower bound of Lemma~\ref{lm:ratio} is not precise
for small values of $k$. In Tables~\ref{tab:ratio k3}
and~\ref{tab:ratio k4}, we provide a lower bound on this ratio,
denoted by $Q'_{\delta-1}(q)$, for $k=3$ and $k=4$.

\begin{table}[h]
\centering \caption{$Q'_{\delta-1}(q)$ for $k=3$ }
\label{tab:ratio k3}
\begin{tabular}{|c|c|c|c|c|c|}
  \hline
  $q$ & 2 & 3 & 4 & 5 & 7  \\ \hline
  $Q'_1(q)$ & 0.6563 & 0.8560 & 0.9229 & 0.9523 & 0.9767  \\
  \hline
\end{tabular}
\end{table}

\begin{table}[h]
\centering \caption{$Q'_{\delta-1}(q)$ for $k=4$ }
\label{tab:ratio k4}
\begin{tabular}
{|c|c|c|c|c|c|}
  \hline
  $q$ & 2 & 3 & 4 & 5 & 7  \\ \hline
  $Q'_1(q)$ & 0.6152 & 0.8454 & 0.9192 & 0.9508 & 0.9763  \\
  \hline
  $Q'_2(q)$ & 0.8203 & 0.9511 & 0.9805 & 0.9904 & 0.9967  \\
  \hline
\end{tabular}
\end{table}

For $\delta=2$, the lower bound on the ratio between the size of
a constant dimension code $\C^{\textmd{ML}}$, generated by the
multilevel construction of Section~\ref{sec:multilevel construction}, and the upper bound on
$\cA_q(n,2\delta,k)$ given in~(\ref{eq:Johnson}), is presented in
Table~\ref{tab:ratio C-ML}. In the construction of such a code
$\C^{\textmd{ML}}$ we consider only $\CMRD$ code and the codewords
related to the following three identifying vectors
$\underset{k-2}{\underbrace{11...1}}0011\underset{n-k-2}{\underbrace{000...00}}$,
$\underset{k-3}{\underbrace{11...1}}010101\underset{n-k-3}{\underbrace{000...00}}$,
and
$\underset{k-2}{\underbrace{11...1}}000011\underset{n-k-4}{\underbrace{000...00}}$,
which contain most of the code.
\begin{table}[h!tab]
\centering \caption{Lower bound on $\frac{|\C^{\textmd{ML}}|}{\textmd{upper bound}}$}
\label{tab:ratio C-ML}
\begin{tabular}{|c|c|c|c|c|c|}
  \hline
  \backslashbox{k}{q} & $2$ & $3$ & $4$ & $5$ & $7$  \\ \hline
  $3$ & 0.7101 & 0.8678 & 0.9267 & 0.9539 & 0.9771 \\\hline
  $4$ & 0.6657 & 0.8571 & 0.9231 & 0.9524 & 0.9767 \\\hline
  $8$ & 0.6274 & 0.8519 & 0.9219 & 0.9520 & 0.9767 \\\hline
  $30$& 0.6250 & 0.8518 & 0.9219 & 0.9520 & 0.9767 \\
  \hline
\end{tabular}
\end{table}
\vspace{0.6cm}

\subsection{Upper Bounds for Codes which Contain Lifted MRD Codes}
\label{subsec:upper bounds contained MRD}
In this subsection we
will be interested in upper bounds on the size of a constant dimension code which
contains the lifted MRD code $\C^{\text{MRD}}$. To obtain these
bounds we use the structure of a lifted MRD code as a transversal
design, considered in Chapter~\ref{ch:designs}.

Let $\mathbb{T}$ be a subspace transversal design derived from
$\C^{\text{MRD}}$ by Theorem \ref{trm:MRD=STD}. Recall that $\cL$
is the set of $q^{n}-q^{n-k}$ vectors of length $n$ over $\F_q$ in
which not all the first $k$ entries are {\it zeroes}. Let $\cL_0$
be the set of vectors in $\F_q^n$ which start with $k$
\emph{zeroes}. $\cL_0$ is isomorphic to $\F_q^{n-k}$,
$|\cL_0|=q^{n-k}$, and $\F_q^n=\cL_0\cup\cL$. Note, that
$\mathbb{V}_0$ is the set of one-dimensional subspaces of
$\mathcal{G}_q(n,1)$ which contain only vectors from $\cL_0$. A
codeword of a constant dimension code, in $\mathcal{G}_q(n,k)$,
contains one-dimensional subspaces from $\mathcal{G}_q(n,1)=
\mathbb{V}_0\cup \mathbb{V}$. Let $\mathbb{C}$ be a constant
dimension code such that $\C^{\text{MRD}}\subset \mathbb{C}$. Each
codeword of $\C\setminus\C^{\text{MRD}}$ contains either at least
two points from the same group of $\mathbb{T}$ or only points from
$\mathbb{V}_0$ and hence, it contains vectors of $\cL_0$.

\begin{theorem}\label{trm:upper bound from Steiner Structure}
If an $(n,M,2(k-1),k)_q$ code $\mathbb{C}$, $k \geq 3$, contains
the $(n,q^{2(n-k)},2(k-1),k)_q$ lifted MRD code then $M\leq
q^{2(n-k)}+ \cA_q (n-k,2(k-2),k-1)$.
\end{theorem}

\begin{proof}
Let $\mathbb{T}$ be an $\text{STD}_q(2,k, n-k)$  obtained from an
$(n,q^{2(n-k)},2(k-1),k)_q$ code $\C^{\text{MRD}} \subset \C$.
Since the minimum distance of $\mathbb{C}$ is $2(k-1)$, it follows
that any two codewords of $\mathbb{C}$ intersect in at most an
one-dimensional subspace. Hence, each two-dimensional subspace of
$\F_q^n$ is contained in at most one codeword of $\mathbb{C}$.
Each two-dimensional subspace $X$ of $\F_q^n$, such that $X=\Span
{ \{v,u \} }$, $v\in \mathbb{V}_A$, $u\in \mathbb{V}_B$, where
$A\neq B$, $A,B\in \mathcal{G}_q(k,1)$, is contained in a codeword
of $\C^{\text{MRD}}$ by Theorem~\ref{trm:MRD=STD}. Hence, each
codeword $X\in \mathbb{C\setminus}\C^{\text{MRD}}$ either contains
only points from $\mathbb{V}_0$ or contains points from
$\mathbb{V}_0$ and points from $\mathbb{V}_A$, for some $A\in
\mathcal{G}_q(k,1)$. Clearly, $\textmd{dim}(X\cap\cL_0)=k$ in the
first case and $\textmd{dim}(X\cap\cL_0)=k-1$ in the second case.
Since $k \geq 3$ and two codewords of $\mathbb{C}$ intersect in at
most one-dimensional subspace, it follows that each
$(k-1)$-dimensional subspace of $\cL_0$ can be contained only in
one codeword. Moreover, since the minimum distance of the code is
$2(k-1)$, it follows that if $X_1 , X_2 \in
\mathbb{C\setminus}\C^{\text{MRD}}$ and $\textmd{dim}(X_1
\cap\cL_0)= \textmd{dim}(X_2 \cap\cL_0)=k-1$ then $d_S (X_1
\cap\cL_0, X_2 \cap\cL_0) \geq 2(k-2)$. Therefore, $\C' = \{
X\cap\cL_0 : X \in \mathbb{C\setminus}\C^{\text{MRD}},~
\textmd{dim}(X\cap\cL_0)=k-1 \}$ is an $(n-k,M',2(k-2),k-1)_q$
code. Let $\dS$ be the set of codewords in
$\mathbb{C\setminus}\C^{\text{MRD}}$ such that $\textmd{dim}(X
\cap\cL_0)=k$. For each $X \in \dS$ let $\tilde{X}$ be an
arbitrary $(k-1)$-dimensional subspace of $X$, and let $\dS' =
\{ \tilde{X} : X \in \dS \}$ (note that $|\dS'|=|\dS|$). The
code $\C' \cup \dS'$ is an $(n-k,M'',2(k-2),k-1)_q$ code since
each two codewords of $\mathbb{C}$ intersect in at most
one-dimensional subspace, $d_S(\C') \geq 2(k-2)$, and $k \geq 3$.
This implies the result of the theorem.
\end{proof}

\begin{theorem}
\label{trm:bound 2k-k} If an $(n,M,2k,2k)_q$ code $\mathbb{C}$
contains the $(n,q^{(n-2k)(k+1)},2k,2k)_q$ lifted MRD code then
$M\leq q^{(n-2k)(k+1)}+ \Gauss{n-2k}{k} \frac{q^n-q^{n-2k}}{q^{2k}-q^k} +
\cA_q (n-2k, 2k,2k)$.
\end{theorem}

\begin{proof}
Let $\mathbb{T}$ be an $\text{STD}_q(k+1,2k,n-2k)$  obtained from
an $(n,q^{(n-2k)(k+1)},2k,2k)_q$ code $\CMRD \subset
\C$. Since the minimum distance of $\mathbb{C}$ is $2k$, it
follows that any two codewords of $\mathbb{C}$ intersect in at
most a $k$-dimensional subspace. Hence, each $(k+1)$-dimensional
subspace of $\F_q^n$ is contained in at most one codeword of
$\mathbb{C}$. Each $(k+1)$-dimensional subspace $Y$ of $\F_q^n$,
such that $Y=\Span { \{v_1,...,v_k ,v_{k+1}\} }$, $v_i\in
\mathbb{V}_{A_i}$, where $A_i\neq A_j$, for $i \neq j$, and $A_i
\in \mathcal{G}_q(k,1)$, $1 \leq i \leq k+1$, is contained in a
codeword of $\C^{\text{MRD}}$ by Theorem~\ref{trm:MRD=STD}. Hence,
each codeword $X\in \mathbb{C}\setminus\mathbb{C}^{MRD}$ has a
nonempty intersection with exactly $\frac{q^{k-\tau}-1}{q-1}$
groups of $\T$, for some $0\leq \tau \leq k$ and therefore $\dim
(X \cap \cL_0)=k+\tau$. Let $\dS_\tau$ be the set of codewords for
which $X \in \dS_\tau$ if $\dim (X \cap \cL_0) =k+\tau$.

The set $\dS_k$ forms an $(n-2k,M',2k,2k)_q$ code and hence
$|\dS_k| \leq \cA_q (n-2k, 2k,2k)$.

Let $Y$ be a $k$-dimensional subspace of $\cL_0$. If $X_1$ and
$X_2$ are two codewords which contain $Y$ then $Y = X_1 \cap X_2$.
Let $N_{\tau,Y}$ be the number of codewords from $\dS_\tau$ which
contain $Y$. Clearly, for each $\tau$, $0 \leq \tau \leq k$, we
have
\begin{equation}
\label{eq:NtauY} \sum_{Y \in \mathcal{G}_q(n-2k,k)} N_{\tau,Y} =
|\dS_\tau| \Gauss{k+\tau}{k}. \end{equation}

There are $\frac{q^n-q^{n-2k}}{q-1}$ points in $\mathbb{V}$ and
each $X \in \dS_\tau$ contains exactly
$\frac{q^{2k}-q^{k+\tau}}{q-1}$ points from $\mathbb{V}$. Hence,
each $k$-dimensional subspace $Y$ of $\cL_0$ can be a subspace of
at most $\frac{q^n-q^{n-2k} - \sum_{\tau=1}^{k-1} N_{\tau,Y}
(q^{2k}-q^{k+\tau})  }{q^{2k}-q^k}$ codewords of $\dS_0$.
Therefore,

\begin{equation*}| \C | \leq q^{(n-2k)(k+1)}+\sum_{\tau=1}^{k}|\dS_\tau| +
\sum_{Y \in \mathcal{G}_q(n-2k,k)} \frac{q^n-q^{n-2k} -
\sum_{\tau=1}^{k-1} N_{\tau,Y} (q^{2k}-q^{k+\tau})
}{q^{2k}-q^k}\end{equation*}
\begin{equation*}
= q^{(n-2k)(k+1)}+\sum_{\tau=1}^{k}|\dS_\tau|+
(\Gauss{n-2k}{k} \frac{q^n-q^{n-2k}}{q^{2k}-q^k}-
\sum_{\tau=1}^{k-1}|\dS_\tau|\Gauss{k+\tau}{k}\frac{q^{2k}-q^{k+\tau}}{q^{2k}-q^k}),
\end{equation*}
where the equality is derived from~(\ref{eq:NtauY}).
\vspace{0.2cm}

One can easily verify that
$\Gauss{k+\tau}{k}\frac{q^{2k}-q^{k+\tau}}{q^{2k}-q^k}\geq1$ for
$1 \leq \tau \leq k-1$; recall also that $|\dS_k| \leq \cA_q
(n-2k, 2k,2k)$; thus we have

$$|\C|\leq q^{(n-2k)(k+1)}+ \Gauss{n-2k}{k}
\frac{q^n-q^{n-2k}}{q^{2k}-q^k} + \cA_q (n-2k, 2k,2k)~.$$

\end{proof}

\vspace{0.3cm}

\subsection{Construction for $(n, M, 4,3)_q$ Codes }
\label{subsec:construct}

In this subsection we  discuss and present a construction of codes
which contain $\C^{\text{MRD}}$ and attain the bound of
Theorem~\ref{trm:upper bound from Steiner Structure}.
 Such a
construction is  presented only for $k=3$ and $q$ large enough. If
$q$ is not large enough then the codes obtained by a modification of this construction almost attain the bound.

For $k=3$, the upper bound of Theorem~\ref{trm:upper bound from
Steiner Structure} on the size of a code which contains
$\C^{\text{MRD}}$ is $q^{2(n-3)}+\Gauss{n-3}{2}$.
The construction
which follows is inspired by the multilevel construction of
Section~\ref{sec:multilevel construction} and the constriction method described
in~\cite{TrRo10}.
We first choose a binary constant weight code $\bf C$ of length
$n$, weight $k=3$, and minimum Hamming distance $2\delta-2=2$. For
each codeword in $\bf C$ a corresponding lifted Ferrers diagram MRD code
(with the minimum subspace distance $4$)
is constructed. However, since for some pairs of identifying
vectors the Hamming distance is $2$, we need to use
appropriate lifted Ferrers diagram MRD codes to make sure that the
final subspace distance of the code will be $4$. For this
purpose we use a method based on pending dots in Ferrers
diagram~\cite{TrRo10}.

The \emph{pending dots} of a Ferrers diagram $\cF$ are the leftmost dots in the
first row of $\cF$ whose removal has no impact on the size of the
corresponding Ferrers diagram rank-metric code. The following lemma is proved in~\cite{TrRo10}.

\vspace{0.3cm}
\begin{lemma}\cite{TrRo10}\label{lm:pending dots}
Let $X$ and $Y$  be two subspaces in $\Gr$
with $d_H(v(X),v(Y))=2\delta-2$, such that the leftmost \emph{one} of $v(X)$
is in the same position as the leftmost  \emph{one} of $v(Y)$.
If $X$ and $Y$ have the same
set $P$ of the pending dots and the entries in $P$ (of their
Ferrers tableaux forms) are assigned with different values in at least one position,
then $d_S(X,Y)\geq 2\delta.$
\end{lemma}

\begin{example}
Let $X$ and $Y$  be subspaces in $\mathcal{G}_q(8,3)$ which are given by
the following generator matrices in RREF:

\[\emph{RE}(X)=\left(\begin{array}{cccccccc}
1 &\textcircled{\raisebox{-0.9pt}{0}} &\textcircled{\raisebox{-0.9pt}{0}} & 0  & v_1 & v_2 & 0 & v_3\\
0 & \:0 &\: 0 & 1  & v_4 & v_5 & 0 & v_6 \\
0 & \:0 & \:0 & 0 & 0 & 0 &  1 & v_7
\end{array}\right)
\]
\[\;\emph{RE}(Y)=\left(\begin{array}{cccccccc}
1 &\textcircled{\raisebox{-0.9pt}{0}} &\textcircled{\raisebox{-0.9pt}{1}} & v_1 & 0 & v_2 & 0 & v_3\\
0 & \:0 & \:0 & 0 & 1  & v_4 & 0 & v_5 \\
0 & \:0 & \:0 & 0 & 0 & 0 &  1 & v_6
\end{array}\right),
\]
where $v_i\in \F_q$, and the pending dots are emphasized by
circles. Their identifying vectors are $v(X)=10010010$ and
$v(Y)=10001010$, respectively. Clearly, $d_H(v(X), v(Y))=2$, while {$d_S(X,Y)=4$}.
\end{example}

The following result is the direct consequence from Theorem~\ref{thm:upper_rank}.

\begin{lemma}\label{lm:FD MRD bound}
Let $n\geq 8$, $k=3$, $\delta=2$, and let $v$ be a
vector in which the leftmost \emph{one} appears in one of the first three
entries.
Let $\cF$ be the corresponding Ferrers diagram of $\emph{EF}(v)$ and
$[\cF,\varrho,\delta]$ be a Ferrers diagram rank-metric code. Then
$\varrho$ is at most the number of dots in $\cF$, which are not
contained in its first row.
\end{lemma}

A code which attains the bound of Lemma~\ref{lm:FD MRD bound} is
a Ferrers diagram MRD code. A construction for such
codes is presented in Subsection~\ref{subsec:lifted FD rank-metric codes}.

The following results from the graph theory will be useful for our construction.

A \emph{matching} in a graph $G$ is a set of pairwise-disjoint
edges in $G$.  An \emph{one-factor}  is a matching  such that
every vertex of $G$ occurs in exactly one edge of the matching. A
partition of the edge set of $G$ into one-factors is called an
\emph{one-factorization}. Let $K_n$ be a complete graph with $n$
vertices. The following result is given in~\cite[p. 476]{vLWi92}

\begin{lemma}\label{lm:one-factor even}
$K_{2n}$ has an one-factorization for all $n$.
\end{lemma}

A near-one-factor in $K_{2n-1}$ is a matching with $n-1$ edges
which contain all but one vertex. A set of near-one-factors which
contains each edge in $K_{2n-1}$ precisely once is called a
near-one-factorization. The following corollary is the direct
consequence from Lemma~\ref{lm:one-factor even}.

\begin{corollary}
$K_{2n-1}$ has a near-one-factorization for all $n$.
\end{corollary}

\begin{corollary}\cite{vLWi92}\label{lm:1-factorization}
Let $D$ be a set of all binary vectors  of length $m$ and weight
$2$.
\begin{itemize}
  \item If $m$ is even, $D$ can be partitioned into $m-1$ classes,
  each one has $\frac{m}{2}$ vectors with pairwise disjoint positions of \emph{ones};
  \item If $m$ is odd, $D$ can be partitioned into $m$ classes,
  each one has $\frac{m-1}{2}$ vectors with pairwise disjoint positions of \emph{ones}.
\end{itemize}
\end{corollary}

\subsubsection{The Construction}

\textbf{Construction I: } Let $n\geq 8$ and $q^2+q+1\geq n-4$ for
odd  $n$, (or $q^2+q+1\geq n-3$ for even~$n$). First we describe
our choice of identifying vectors for the code. The identifying
vector $v_0=11100\ldots 0$ corresponds to the lifted MRD code
$\C^{\text{MRD}}$. The other identifying vectors are of the form
$xy$, where $x$ is of length 3 and weight 1, and $y$ is of length
$n-3$ and weight~$2$. We use all the $\binom{n-3}{2}$ vectors of
weight 2 in the last $n-3$ coordinates of the identifying vectors.
By Corollary~\ref{lm:1-factorization}, there is a partition of the set
of vectors of length $n-3$  and weight~$2$ into $s=n-4$ classes if
$n-3$ is even (or into $s=n-3$ classes if $n-3$ is odd),
$F_1,F_2,\ldots,F_s$. We define $\cA_1=\{(001y):y\in F_1\}$,
$\cA_2=\{(010y):y\in F_i, 2\leq i\leq \min\{q+1,s\}\}$, and if
$s>q+1$ then $\cA_3=\{(100y):y\in F_i,~ q+2\leq i\leq s\}$ (If
$s\leq q+1$ then $\cA_3=\varnothing$).


All the Ferrers diagrams  which correspond to the identifying
vectors from $\cA_2$ have one common pending dot in the first
entry of the first row. We assign the same value of $\F_q$ in this
entry of the Ferrers tableaux form for each vector in the same
class. Two subspaces with identifying vectors from different
classes of $\cA_2$ have different values in the entry of the
pending dot. On the remaining dots of Ferrers diagrams we
construct Ferrers diagram MRD codes and lift them.

Similarly, all the Ferrers diagrams  which correspond to the
identifying vectors from $\cA_3$, have two common pending dots in
the first two entries of the first row. We assign the same value
of $\F_q$ in these two entries in the Ferrers tableaux form for
each vector in the same class. Two subspaces with identifying
vectors from different classes of $\cA_3$ have different values in
at least one of these two entries. On the remaining dots of
Ferrers diagrams we construct Ferrers diagram MRD codes and lift
them.

Our code $\C$ is a union of $\C^{\text{MRD}}$ and the lifted codes
corresponding to the identifying vectors in $\cA_1$, $\cA_2$, and
$\cA_3$.



\begin{theorem}\label{trm:existence of codes for ContrA} For $q$ satisfying
$q^2+q+1\geq s$, where
\[s= \left\{\begin{array}{cc}
                n-4, & n \emph{ is odd }\\
                n-3, & n \emph{ is even }\\
              \end{array},\right.
\]
the code $\C$ obtained by Construction I
attains the bound of Theorem \ref{trm:upper bound from Steiner Structure}.

\end{theorem}
\begin{proof}
First, we prove that the minimum subspace distance of $\C$ is 4.

Note, that for $X,Y \in \C$, such that $v(X)\in \cA_i$, $v(Y)\in
\cA_j$, for $i\neq j$, we have $d_H(v(X),v(Y))=4$. Hence, by Corollary~\ref{cor:distance},
$d_S(X,Y)\geq 4$.

If $X,Y \in \C$ with identifying vectors $v(X) = zw$, $v(Y)=zw'$,
where $z$ is of length 3, and $w,w' \in F_i$, $1 \leq i \leq s$
then $d_H(v(X),v(Y))=4$ which implies, by Corollary~\ref{cor:distance}, that $d_S(X,Y)\geq 4$.

Let $X,Y \in \C$ with identifying vectors $v(X) = zw$, $v(Y)=zw'$,
where $z$ is of length 3, $w \in F_i$, $w' \in F_j$, $i \neq j$.
If $d_H(v(X),v(Y))=4$ then $d_S(X,Y)\geq 4$. If $d_H(v(X),v(Y))=2$
then by Lemma~\ref{lm:pending dots} we have $d_S(X,Y)\geq 4$.

Next, we calculate the size of $\C$. Note, that since $q^2+q+1\geq
s$ it follows that all the $\binom{n-3}{2} $ vectors of weight $2$
and length $n-3$ are taken as the suffices of the identifying
vectors. There are also $\binom{n-3}{2}$ different Ferrers
diagrams for subspaces in $\mathcal{G}_q(n-3,2)$. Therefore, by
Lemma~\ref{lm:FD MRD bound} the size
of $\C$ is $q^{2(n-3)}+\Gauss{n-3}{2}$.
\end{proof}

\begin{remark}
The code for $n=6$ whose size attains the upper bound of
Theorem~\ref{trm:upper bound from Steiner Structure} is
constructed in~\cite{EtSi09} and the code for $n=7$ whose size
attains this bound is constructed in~\cite{TrRo10}.
\end{remark}

Now we describe a construction of codes for the other values of
$q$, which does not satisfy the condition of
Theorem~\ref{trm:existence of codes for ContrA}.

\textbf{Construction II: }
 Let $n\geq 8$ and $q^2+q+1 < n-4$ for
odd $n$ (or $q^2+q+1 < n-3$ for even $n$). We modify  Construction~I as follows.

The identifying vector $v_0=11100\ldots 0$ corresponds to the
lifted MRD code $\C^{\text{MRD}}$. Let
$\alpha=\floorenv{\frac{n-3}{q^2+q+2}}$ and
$r=n-3-\alpha(q^2+q+2)$. We partition the last $n-3$ coordinates,
of the other identifying vectors, into $\alpha$~sets, where each set consists of $q^2+q+2$
consecutive coordinates and the last set consists of $r<q^2+q+2$
consecutive coordinates. Since $q^2+q+2$ is always an even
integer, it follows from Corollary~\ref{lm:1-factorization} that
there is a partition of vectors of length $q^2+q+2$ and weight 2,
corresponding to the $i$th set,  $1\leq i\leq \alpha$, into
$q^2+q+1$ classes $F_1^i,F_2^i,\ldots,F_{q^2+q+1}^i$. We define
$Y_1^i=\{(0^{(i-1)(q^2+q+2)}y0^{n-3-i(q^2+q+2)}):y\in F_1^i\}$,
$Y_2^i=\{(0^{(i-1)(q^2+q+2)}y0^{n-3-i(q^2+q+2)}):y\in F_j^i, 2\leq
j\leq q+1\}$, and
$Y_3^i=\{(0^{(i-1)(q^2+q+2)}y0^{n-3-i(q^2+q+2)}):y\in F_j^i,
q+2\leq j\leq q^2+q+1\}$. Let
$$\cA_1^i=\{(001y):y\in Y_1^i\},\;1\leq i\leq \alpha,$$
$$\cA_2^i=\{(010y):y\in Y_2^i\},\;1\leq i\leq \alpha,$$
$$\cA_3^i=\{(100y):y\in Y_3^i\},\;1\leq i\leq \alpha.$$

The identifying vectors (excluding $v_0$), of the code that we
construct, are partitioned into the following three sets:
$$\cA_1=\cup_{i=1}^{\alpha}\cA_1^i,~
\cA_2=\cup_{i=1}^{\alpha}\cA_2^i,~
\cA_3=\cup_{i=1}^{\alpha}\cA_3^i.$$

As in Construction~I, we construct lifted
Ferrers diagram MRD code for each identifying vector, by using pending dots.
Our code $\C$ is a union of $\C^{\text{MRD}}$ and the
lifted codes corresponding to the identifying vectors
in $\cA_1$, $\cA_2$, and $\cA_3$.

\begin{remark} The identifying vectors
with two \emph{ones} in the last $r$ entries can be also used in
Construction~II, but their contribution to the final code is
minor.
\end{remark}

In the similar way to the proof of Theorem~\ref{trm:existence of codes for ContrA} one can prove
the following theorem, based on the fact that the size of the lifted Ferrers diagram MRD code obtained
from identifying vectors in $\cA_1^i\cup \cA_2^i\cup \cA_3^i$, $1\leq i\leq \alpha$, is
$\Gauss{q^2+q+2}{2}q^{2(n-3-(q^2+q+2)i)}$.

\begin{theorem}\label{trm:codes for ContrA'} For $q$ satisfying
$q^2+q+1<s$, where

\[s = \left\{\begin{array}{cc}
                n-4, & n \emph{ is odd }\\
                n-3, & n \emph{ is even } \\
              \end{array}\right.,
\]
Construction~II generates an $(n,M,4,3)_q$ constant dimension code
with $M=q^{2(n-3)}+\sum_{i=1}^{\alpha}\Gauss{q^2+q+2}{2}q^{2(n-3-(q^2+q+2)i)}$,
which  contains $\C^\emph{{MRD}}$.
\end{theorem}

For all admissible values of $n$, the ratio
$(|\C| - |\CMRD|) /$$\begin{tiny}\left[\begin{array}{c} n-3\\
2\end{array}\right]_{q}\end{tiny}$,
for the code $\C$ generated by
Construction~II, is greater than 0.988 for $q=2$ and 0.999 for
$q>2$.

In Table~\ref{tab:size new codes} we compare the size of codes obtained by
 Constructions I and II (denoted by $\C_{new}$) with the size of the
largest previously known codes (denoted by $\C_{old}$) and with the
upper bound~(\ref{eq:Johnson}) (for $k=3$).

\begin{table}[h]
\centering \caption{The size of new codes vs. the previously
known codes and the upper bound~(\ref{eq:Johnson})}
\label{tab:size new codes}
\begin{tabular}{|c|c|c|c|c|}
\hline $q$ & $n$ &  $|\C_{old}|$ & $|\C_{new}|$ &upper bound~(\ref{eq:Johnson})
\tabularnewline \hline \hline 2 & 13 & $1192587$~\cite{EtSi09} &
$1221296$&1597245
 \tabularnewline \hline
2 & 14 &  $4770411$~\cite{EtSi09} & $4885184$& 6390150
\tabularnewline \hline 5 & 9 &  $244644376$~\cite{EtSi09}&
$244649056$& 256363276 \tabularnewline \hline
\end{tabular}
\end{table}

The new ratio between the new best lower bound and the upper
bound~(\ref{eq:Johnson}) for constant dimension codes with $k=3$
and $\delta=2$, is presented in Table~\ref{tab:full new ratio}.
One should compare it with Tables~\ref{tab:ratio k3}
and~\ref{tab:ratio C-ML}.

\begin{table}[h]
\centering \caption {Lower bounds on ratio between $|\C_{new}|$ and
the bound in~(\ref{eq:Johnson}) }
\label{tab:full new ratio}
\begin{tabular}{|c|c|c|c|c|c|}
  \hline
  $q$ & 2 & 3 & 4 & 5 & 7  \\ \hline
  $|\C_{new}|/\textmd{bound}$ & 0.7657 & 0.8738 & 0.9280 & 0.9543 & 0.9772  \\
  \hline
\end{tabular}
\end{table}


\subsection {Construction for  $(8,M,4,4)_q$ Codes}
\label{subsec:constructionB}

In this section we introduce a construction of $(8,M,4,4)_q$ codes
which attain the upper bound of Theorem~\ref{trm:bound 2k-k}.
They are based on 2-parallelism of subspaces in
$\mathcal{G}_q(4,2)$.

A $k$\emph{-spread} in $\Gr$ is a set of $k$-dimensional subspaces
which partition $\F_q^n$ (excluding the all-zero vector).
We say that two subspaces are disjoint if they have only trivial intersection.
A $k$-spread in $\Gr$  exists if and only if $k$ divides $n$.
Clearly, a $k$-spread is a constant dimension code in $\Gr$
with maximal possible minimum distance $d_S=2k$.
A partition of all $k$-dimensional
subspaces of $\Gr$ into disjoint
$k$-spreads is called a $k$\emph{-parallelism}.

\textbf{Construction III:}
 Let $\C^{\text{MRD}}$ be an
$(8,2^{12},4,4)_2$ lifted MRD code, and let $\mathbb{T}$
be the corresponding $\text{STD}_2(3,4,4)$.
We generate the following new codewords (blocks) of
$\mathbb{C\setminus}\C^{\text{MRD}}$.
Let $\cB_1, \cB_2,\ldots, \cB_7$ be a partition of all the
subspaces of $\mathcal{G}_2(4,2)$ into seven $2$-spreads, each one
of size 5, i.e., a well known 2-parallelism in
$\mathcal{G}_2(4,2)$~\cite{Beu74}. For each $i$, $1\leq i\leq 7$,
and each pair $B,B'\in \cB_i$, we can write $B=\{v_0=\textbf{0},
v_1, v_2, v_3\}$ and $B'=\{ v'_0=\textbf{0}, v'_1, v'_2, v'_3\}$,
where $ v_t, v'_t \in \F_2^4$, $0\leq t\leq 3$, and
$\textbf{0}=(0000)$. The
$2$-dimensional subspace $B$ has four cosets  $B_0=B, B_1, B_2,
B_3$ in $\F_2^4$.

We define  the following four codewords in
$\mathbb{C\setminus}\C^{\text{MRD}}$, as blocks with fifteen
points:
$$\{\Span{\textbf{0}u}: u \in B\setminus \{\textbf{0}\}\}\cup
\{\Span{v'_1y}:y\in B\}\cup
\{\Span{v'_2y}:y\in B\}\cup
\{\Span{v'_3y}:y \in B\},
$$
$$\{\Span{\textbf{0}u}: u \in B\setminus \{\textbf{0}\}\}\cup
\{\Span{v'_1y}:y\in B_1\}\cup
\{\Span{v'_2y}:y\in B_2\}\cup
\{\Span{v'_3y}:y\in B_3\},
$$
$$\{\Span{\textbf{0}u}: u \in B\setminus \{\textbf{0}\}\}\cup
\{\Span{v'_1y}:y\in B_2\}\cup
\{\Span{v'_2y}:y\in B_3\}\cup
\{\Span{v'_3y}:y\in B_1\},
$$
$$\{\Span{\textbf{0}u}: u \in B\setminus \{\textbf{0}\}\}\cup
\{\Span{v'_1y}:y\in B_3\}\cup
\{\Span{v'_2y}:y\in B_1\}\cup
\{\Span{v'_3y}:y\in B_2\}.
$$

In addition to these codewords we add a codeword which contains
all the points of $\mathbb{V}_0$.


\begin{theorem}
\label{trm:construction b} Construction~III generates an
$(8,2^{12}+701,4,4)_2$ constant dimension code~$\C$ which attains
the bound of Theorem~\ref{trm:bound 2k-k} and contains an
$(8,2^{12},4,4)_2$ lifted MRD code.
\end{theorem}

\begin{proof}
First, we observe that the four types of codewords given in the
construction are indeed $4$-dimensional subspaces of $\F_2^8$.
Each one of the codewords contains 15 different one-dimensional
subspaces, and hence each codeword contains 15 different nonzero
vectors of $\F_2^8$. It is easy to verify that all these vectors
are closed under addition in $\F_2$, thus each constructed
codeword is a $4$-dimensional subspace of $\F_2^8$.

To prove that for each two codewords $X,Y \in \C$, we have
$d_S(X,Y)\geq 4$, we distinguish between three cases:
\begin{itemize}
  \item Case 1: $X,Y\in \C^{\text{MRD}}$. Since the minimum distance
  of $\C^{\text{MRD}}$ is 4, we have that $d_S(X,Y)\geq 4$.
  \item Case 2: $X\in \C^{\text{MRD}}$ and $Y\in \mathbb{C\setminus}\C^{\text{MRD}}$.
  The codewords of $\C^{\text{MRD}}$ forms the
blocks of a  subspace transversal design $\T$, and hence meet each group in
exactly one point. Each codeword of $\mathbb{C\setminus}\C^{\text{MRD}}$
meets exactly three groups of $\T$. Hence, $\dim(X\cap Y)\leq 2$ for
each $X\in \C^{\text{MRD}}$ and $Y\in \mathbb{C\setminus}\C^{\text{MRD}}$,
therefore,  $d_S(X,Y)\geq 4$.
  \item Case 3: $X,Y\in \mathbb{C\setminus}\C^{\text{MRD}}$.
  If $X$ and $Y$ have three common points
in $\mathbb{V}_0$ (which correspond to a 2-dimensional subspace
contained in $\cL$), then they are disjoint in all the groups of
$\mathbb{T}$, since there points in $\mathbb{V}$ correspond to
the different cosets, or different blocks in the same spread.
If $X$ and $Y$ have only one common point in $\mathbb{V}_0$, then they have at most
two common points in at most one group of $\T$. Thus, $d_S(X,Y)\geq 4$.
\end{itemize}

$\C^{\text{MRD}}$ contains $2^{12}$ codewords. There are
$\GaussBin{4}{2}$ $2$-dimensional subspaces contained in
$\mathcal{G}_2(4,2)$, and hence there are 35 different choices for
$B$. Since the size of a spread is~5, it follows that there are 5
different choices for $B'$, and for each such pair $B,B'$ there
are 4~codewords based on the 4 different cosets of $B$ as defined
in Construction~III. With the additional codeword which contains all
the points of $\mathbb{V}_0$ we obtain $35\cdot5\cdot4+1=701$
codewords. Thus in the constructed code $\mathbb{C}$ there are
$2^{12}+701=4797$ codewords.

Thus, the code attains the bound of Theorem~\ref{trm:bound 2k-k}.
\end{proof}

\begin{remark} Theorem~\ref{trm:construction b} implies that $A_2(8,4,4)\geq 4797$
(the previously known  largest code of size $4573$ was obtained by the multilevel construction.)
\end{remark}

\begin{remark}\label{rm:gen q} Construction~III  can be easily generalized for all $q\geq2$,
since there is a $2$-parallelism  in $\mathcal{G}_q(n,2)$ for all $q$, where  $n$ is
power of $2$~\cite{Beu74}. Thus from this construction we can obtain
a $(8,M,4,4)_q$ code with $M=q^{12}+\Gauss{4}{2}(q^2+1)q^2+1$, since
the size of a $2$-spread in $\mathcal{G}_q(4,2)$ is $q^2+1$ and there are
$q^2$ different cosets of a $2$-dimensional subspace in $\F_q^4$.
\end{remark}

In Table~\ref{tab:new codes k=4} we compare the size of codes obtained by
the Constructions III (denoted by $\C_{new}$) with the size of $\C^\textmd{{MRD}}$,
the size of the
largest previously known codes (denoted by $\C_{old}$) and with the
upper bound~(\ref{eq:Johnson}) (for $n=8$ and $k=4$).

\begin{table}[h]
\centering \caption{The size of new codes vs. previously known codes and bound~(\ref{eq:Johnson})}
\label{tab:new codes k=4}
\begin{tabular}{|c|c|c|c|c|}
\hline $q$ & $|\CMRD|$&  $|\C_{old}|$ & $|\C_{new}|$&upper bound~(\ref{eq:Johnson})
\tabularnewline \hline \hline
2 &$2^{12}$ & $2^{12}+477$~\cite{EtSi09} & $2^{12}+701$&$2^{12}+2381$
\tabularnewline \hline
3 & $3^{12}$ &$3^{12}+8137$~\cite{EtSi09} & $3^{12}+11701$&$3^{12}+95941$
\tabularnewline \hline
4 & $4^{12}$ &$4^{12}+72529$~\cite{EtSi09} & $4^{12}+97105$&$4^{12}+1467985$
 \tabularnewline \hline
\end{tabular}
 \end{table}

\begin{remark}
In general, the existence of $k$-parallelism  in $\Gr$ is an open
problem. It is known that $2$-parallelism exists for $q=2$, and
all $n$~\cite{ZZS71}. Recently it has been proved that there is a
$3$-parallelism for $q=2$ and $n=6$~\cite{ToZh10}. Thus we believe
that  Construction~III can be generalized to a larger family of
parameters assuming that there exists a corresponding parallelism.

\end{remark}

\section{Error-Correcting Projective Space Codes}
\label{sec:punctured}
In this section our goal will be to construct large codes in $\Ps$
which are not constant dimension codes. We first note that the
multilevel coding described in Section~\ref{sec:multilevel construction} can be
used to obtain a code in $\Ps$. The only difference is that we
should start in the first step with a general binary code of
length $n$ in the Hamming space as a skeleton code. The first question which will
arise in this context is whether the method is as good as for
constructing codes in $\Gr$. The answer can be inferred from the
following example.

\vspace{0.2cm}
\begin{example}
\label{ex:c394} Let $n=7$ and $d=3$, and consider the $[7,\; 4,\; 3]$
Hamming code with the parity-check matrix
\begin{align*}
\left( \begin{array}{ccccccc}
0 & 0 & 0 & 1 & 1 & 1 & 1 \\
0 & 1 & 1 & 0 & 1 & 1 & 0 \\
1 & 0 & 1 & 1 & 0 & 1 & 0
\end{array}
\right) .
\end{align*}
By using the multilevel construction with this Hamming code we obtain a
code with minimum distance $3$ and size $394$ in
$\smash{{\sP\kern-2.0pt}_2\kern-0.5pt(7)}$.
\end{example}
\vspace{0.6cm}

As we shall see in the sequel this code is much smaller than a
code that will be obtained by puncturing. We have also generated
codes in the projective space based on the multilevel
construction, where the skeleton code is a lexicode. The
constructed codes appear to be much smaller than the codes
obtained by puncturing. Puncturing of a constant dimension code $\C$
(or union of
codes with different dimensions and the required minimum distance)
results in a projective
space code $\C'$. If the minimum distance of $\C$ is $2 \delta$
then the minimum distance of $\C'$ is $2 \delta -1$.
If $\C$ is a code
obtained by the multilevel
construction, then  $\C'$ has a
similar structure in the sense that the identifying
vectors of the codewords can form a skeleton code. But the
artificial "skeleton code" can be partitioned into pair of
codewords with Hamming distance one, while the distance between
two codewords from different pairs is at least $2 \delta -1$. This
property yields larger codes by puncturing, sometimes with double
size, compared to codes obtained by the multilevel construction.

\subsection{Punctured Codes}
\label{sec:punct}

Puncturing and punctured codes are well known in the Hamming
space.
Let ${\bf
C}$ be an $(n,M,d)$ code in the Hamming space. Its punctured code
${\bf C}'$ is obtained by deleting one coordinate of ${\bf C}$.
Hence, there are $n$ punctured codes and each one is an
$(n-1,M,d-1)$ code. In the projective space there is a very large
number of punctured codes for a given code $\C$ and in contrary to
the Hamming space the sizes of these codes are usually different.

Let $X$ be an $\ell$-dimensional subspace of $\F_q^n$ such that the unity
vector with an {\it one} in the $i$th coordinate is not an element
in $X$. The {\it $i$-coordinate puncturing} of $X$, $\Delta_i(X)$,
is defined as the $\ell$-dimensional subspace of $\F_q^{n-1}$
obtained from $X$ by deleting coordinate $i$ from each vector in
$X$. This puncturing of a subspace is akin to puncturing a code
${\bf C}$ in the Hamming space by the $i$th coordinate.

Let $\C$ be a code in $\Ps$ and let $Q$ be an $(n-1)$-dimensional
subspace of $\F_q^n$. Let $\mbox{RE}(Q)$ be the $(n-1) \times n$ generator
matrix of $Q$ (in RREF) and let $\tau$ be the
position of the unique {\it zero} in its identifying vector $v(Q)$. Let $v \in \F_q^n$ be a
vector such that $v \notin Q$. We define the {\it punctured} code

$$\C'_{Q,v} = \C_Q \cup \C_{Q,v}~,$$
where
$$\C_Q=\left\{ \Delta_\tau (X) ~:~ X \in \C, ~ X \subseteq Q\right\})$$
and
$$\C_{Q,v}= \left\{ \Delta_\tau (X \cap Q)~:\: X \in \C,~v\in X \right\}~.$$

\begin{remark} If $\C$ was constructed by the multilevel construction of
Section~\ref{sec:multilevel construction} then the codewords of $\C_Q$ and
$\C_{Q,v}$ can be partitioned into related lifted codes of Ferrers
diagram rank-metric codes. Some of these codes are cosets of the
linear Ferrers diagram rank-metric codes.
\end{remark}

The following theorem can be easily verified.

\begin{theorem}
The punctured code $\C'_{Q,v}$ of an $(n,M,d)_q$ code $\C$ is an
$(n-1,M',d-1)_q$ code.
\end{theorem}

\begin{remark} The code $\tilde{\C} = \left\{ X ~:~ X \in \C, ~ X \subseteq
Q\right\}) \cup \left\{ X \cap Q~:\: X \in \C,~v\in X \right\}$ is
an $(n,M',d-1)_q$ code whose codewords are contained in $Q$. Since
$Q$ is an $(n-1)$-dimensional subspace it follows that there is an
isomorphism $\varphi$ such that $\varphi(Q) = \F_q^{n-1}$. The
code $\varphi (\tilde{\C}) = \{ \varphi (X) ~:~ X \in \tilde{\C}
\}$ is an $(n-1,M',d-1)_q$ code. The code $\C'_{Q,v}$ was obtained
from $\tilde{\C}$ by such isomorphism which uses the
$\tau$-coordinate puncturing on all the vectors of $Q$.
\end{remark}

\begin{example}
\label{ex:c573} Let $\C$ be the $(8,4573,4,4)_2$ constant dimension code given in
Example~\ref{ex:c844}. Let $Q$ be the $7$-dimensional subspace whose
$7 \times 8$ generator matrix is

\[
\left(\begin{array}{cccccc}
1 & 0 & \ldots & 0 & 0\\
0 & 1 & \ldots & 0 & 0\\
\vdots & \vdots & \ddots & \vdots & \vdots\\
0 & 0 & \ldots & 1 & 0\end{array}\right).\]

By using puncturing with $Q$ and $v=10000001$ we obtained a code
$\C'_{Q,v}$ with minimum distance $3$ and size $573$.  By adding to
$\C'_{Q,v}$ two codewords, the null space $\{ 0 \}$ and $\F_2^7$,
we obtained a $(7,575,3)_2$ code in $\mathcal{P}_{2}(7)$. In Table~\ref{tab:punctured}
we show the number of codewords which were obtained
from each of the identifying vectors with weight $4$ of
Example~\ref{ex:c844}.
\begin{table}[h]
\centering \caption{ The punctured $(7,573,3)_q$ code $\C'_{Q,v}$} \label{tab:punctured}
\begin{tabular}{|c|c|}
\hline
\multicolumn{2}{|c|}{$\mathbb{C}_{Q}$} \\
\hline \hline identifying
vector & addition to $\mathbb{C}_{Q}$ \tabularnewline \hline
11110000 & 256\tabularnewline \hline 11001100 & 16\tabularnewline
\hline10101010 & 8\tabularnewline \hline 10010110 &
2\tabularnewline \hline 01100110 & 4\tabularnewline \hline
01011010 & 2\tabularnewline \hline 00111100 & 1\tabularnewline
\hline
\end{tabular}\;\;
\begin{tabular}{|c|c|}
\hline \multicolumn{2}{|c|}{$\mathbb{C}_{Q,v}$, $\; v=10000001$}\\
\hline \hline
identifying vector & \multicolumn{1}{p{1.1in}|}{addition to
$\mathbb{C}_{Q,v}$}\tabularnewline \hline 11110000 &
256\tabularnewline \hline 11001100 & 16\tabularnewline \hline
11000011 & 1\tabularnewline \hline 10101010 & 4\tabularnewline
\hline 10100101 & 2\tabularnewline \hline 10011001 &
4\tabularnewline \hline 10010110 & 1\tabularnewline \hline
\end{tabular}
\end{table}
\vspace{0.3cm}

\end{example}

\begin{example}
\label{ex:punct582}
Let $\C$ be a $(8,M,4,4)_q$ code with
$M=q^{12}+\Gauss{4}{2}(q^2+1)q^2+1$, obtained from Construction~III of
Subsection~\ref{subsec:constructionB}. Let $Q$ be the same $7$-dimensional
subspace as in Example~\ref{ex:c573}, and let $v\in \F_q^8$ be
a vector in which not all the first~$4$~entries are {\it zeroes}, i.e., $v\in \cL$.
From the construction of $\C$, it follows that
$|\C_{Q,v}|= |\C_Q| = q^8+\Gauss{4}{2}$, hence we obtained the punctured code
$\C'_{Q,v}$ with minimum distance $3$ and size $2\left(q^8+\Gauss{4}{2}\right)$.

%
%
%
%
\vspace{0.2cm}

For $q=2$, the size of this punctured code  is equal to $582$. By adding to
$\C'_{Q,v}$ two codewords, the null space $\{ 0 \}$ and $\F_2^7$,
we obtained a $(7,584,3)_2$ code in $\mathcal{P}_{2}(7)$.
\end{example}
\vspace{0.2cm}
The large difference between the sizes of the codes of
Example~\ref{ex:c394} and Examples~\ref{ex:c573} and~\ref{ex:punct582} shows the strength of
puncturing when applied on codes in $\Ps$.

\subsection{Code Parameters}

First we ask, what is the number of punctured codes which can be derived in
this way from~$\C$? $Q$ is an $(n-1)$-dimensional subspace of
$\F_q^n$ and hence it can be chosen in $\frac{q^n-1}{q-1}$
different ways. There are $\frac{q^n-q^{n-1}}{q-1}=q^{n-1}$
distinct way to choose $v \notin Q$ after $Q$ was chosen. Thus, we
have that usually puncturing of a code $\C$ in $\Ps$ will result
in $\frac{q^{2n-1}-q^{n-1}}{q-1}$ different punctured codes.

\begin{theorem}
If $\C$ is an $(n,M,d,k)_q$ code then there exists an
$(n-1,M',d-1)_q$ code $\C'_{Q,v}$ such that $M' \geq
\frac{M(q^{n-k}+q^k-2)}{q^n-1}$.
\end{theorem}
\begin{proof}
As before, $Q$ can be chosen in $\frac{q^n-1}{q-1}$ different
ways. By using basic enumeration, it is easy to verify that each
$k$-dimensional subspace of $\Ps$ is contained in
$\frac{q^{n-k}-1}{q-1}$~~$(n-1)$-dimensional subspaces of $\Ps$.
Thus, by a simple averaging argument we have that there exists an
$(n-1)$-dimensional subspace $Q$ such that $|\C_Q|\geq
M\frac{q^{n-k}-1}{q^n-1}$.

There are $M- |\C_Q|$ codewords in $\C$ which are not contained in
$Q$. For each such codeword $X \in \C$ we have $\dim (X \cap Q)=k-1$.
Therefore, $X$ contains $q^k-q^{k-1}$ vectors which do not belong
to $Q$. In $\F_q^n$ there are $q^n-q^{n-1}$ vectors which do not
belong to $Q$. Thus, again by using simple averaging argument we
have that there exist an $(n-1)$-dimensional subspace $Q \subset
\F_q^n$ and $v\notin Q$ such that $|\C_{Q,v}|\geq
\frac{(M-|\C_Q|)(q^k-q^{k-1})}{q^n-q^{n-1}}=\frac{M-|\C_Q|}{q^{n-k}}$.

Therefore, there exists an $(n-1,M',d-1)_q$ code $\C'_{Q,v}$ such
that $M'=|\C_Q|+|\C_{Q,v}|\geq
\frac{|\C_Q|q^{n-k}+M-|\C_Q|}{q^{n-k}}=
\frac{(q^{n-k}-1)|\C_Q|+M}{q^{n-k}} \geq
\frac{(q^{n-k}-1)M(q^{n-k}-1)+M(q^n-1)}{(q^n-1)q^{n-k}}=
\frac{M(q^{n-k}+q^k-2)}{q^n-1}$.
\end{proof}

\vspace{0.3cm}

Clearly, choosing the $(n-1)$-dimensional subspace $Q$ and the
element $v$ in a way that $\C'_{Q,v}$ will be maximized is
important in this context. Example~\ref{ex:c573} can be
generalized in a very simple way. We start with a
$(4k,q^{2k(k+1)},2k,2k)_q$ code obtained from the skeleton codeword
$\underset{2k}{\underbrace{1 \cdots 1}}\underset{2k}{\underbrace{0
\cdots 0}}$ in the multilevel approach. We apply puncturing with
the $(4k-1)$-dimensional subspace $Q$ whose $(4k-1) \times (4k)$
generator matrix is

\[
\left(\begin{array}{cccccc}
1 & 0 & \ldots & 0 & 0\\
0 & 1 & \ldots & 0 & 0\\
\vdots & \vdots & \ddots & \vdots & \vdots\\
0 & 0 & \ldots & 1 & 0\end{array}\right).\]

It is not difficult to show that in the $[(2k) \times (2k) ,
2k(k+1), k]$ rank-metric code $\cC$ there are $q^{2k^2}$ codewords
with {\it zeroes} in the last column and $q^{2k^2}$ codewords with
{\it zeroes} in the first row. There is also a codeword whose
first row ends with a {\it one}. If $u$ is this first row which
ends with a {\it one} there are $q^{2k^2}$ codewords whose first
row is $u$. We choose $v$ to be $v=1\underset{2k-1}{\underbrace{0
\cdots 0}}u$. By using puncturing with $Q$ and $v$ we have
$|\C_Q|=q^{2k^2}$ and $|\C_{Q,v}|=q^{2k^2}$. Hence, $\C'_{Q,v}$ is
a $(4k-1,2q^{2k^2},2k-1)_q$ code in $\mathcal{P}_{q}(4k-1)$. By
using more codewords from the constant weight code in the
multilevel approach and adding the null space and $\F_q^{4k-1}$ to
the code we construct a slightly larger code with the same
parameters.

\begin{remark}If $n$ is odd  then the size of a punctured code (in $\mathcal{P}_q(n-1)$) can be
 smaller than the size of a constant dimension code in
$\mathcal{G}_q(n-1, \left\lfloor\frac{n-1}{2}\right\rfloor)$ obtained by
the multilevel construction, which  gives the lower
bound on $\mathcal{A}_q(n-1, d)$~\cite{GaYa10a}.
\end{remark}

\subsection{Decoding}

We assume that $\C$ is an $(n,M,d)_q$ code and that all the
dimensions of the subspaces in $\C$ have the same parity which
implies that $d=2 \delta$. This assumption makes sense as these
are the interesting codes on which puncturing is applied,
similarly to puncturing in the Hamming space. We further assume
for simplicity that w.l.o.g. if $\mbox{RE}(Q)$ is the $(n-1) \times n$
generator matrix of $Q$ then the first $n-1$ columns are linearly
independent, i.e., $\mbox{RE}(Q)= [ I ~ u ]$, where $I$ is an $(n-1)
\times (n-1)$ identity matrix and $u$ is a column vector of length
$n-1$.

Assume that the received word from a codeword $X'$ of $\C'_{Q,v}$
is an $\ell$-dimensional subspace $Y'$ of $\F_q^{n-1}$. The first
step will be to find a subspace $Z$ of $\F_q^n$ on which we can apply
the decoding algorithm of $\C$. The result of this decoding will be reduced
to the $(n-1)$-dimensional subspace $Q$ and punctured
to obtain the codeword of $\C'_{Q,v}$. We start
by generating from $Y'$ an $\ell$-dimensional subspace $Y \subset
Q$ of $\F_q^n$. This is done by appending a symbol to the end of
each vector in $Y'$ by using the generator matrix $\mbox{RE}(Q)$ of $Q$.
If a generator matrix $\mbox{RE}(Y')$ is given we can do this process only
to the rows of $\mbox{RE}(Y')$ to obtain the generator matrix $\mbox{RE}(Y)$ of
$Y$. This generator
matrix of $Y$ is formed in its reduced row echelon form.

\begin{remark} If the {\it zero} of $v(Q)$ is in coordinate $\tau$ then
instead of appending a symbol to the end of the codeword we insert
a symbol at position $\tau$.
\end{remark}

Assume that $p$ is the parity
of the dimension of any subspace in $\C$, where $p=0$ or $p=1$.
Once we have $Y$ we distinguish between two cases to form a new
subspace $Z$ of $\F_q^n$.

\noindent {\bf Case 1:} $\delta$ is even.
\begin{itemize}
\item If $\ell \equiv p~(mod~2)$ then $Z= Y \cup (v + Y)$.

\item If $\ell \not\equiv p~(mod~2)$ then $Z=Y$.
\end{itemize}

\noindent {\bf Case 2:} $\delta$ is odd.
\begin{itemize}
\item If $\ell \equiv p~(mod~2)$ then $Z= Y$.

\item If $\ell \not\equiv p~(mod~2)$ then $Z= Y \cup (v + Y)$.
\end{itemize}

Now we use the decoding algorithm of the code $\C$ with the word
$Z$. The algorithm will produce as an output a codeword $\cX$. Let
$\tilde{X} = \cX \cap Q$ and $\tilde{X}'$ be the subspace of
$\F_q^{n-1}$ obtained from $\tilde{X}$ by deleting the last entry
of $\tilde{X}$. We output $\tilde{X}'$ as the submitted codeword
$X'$ of $\C'_{Q,v}$. The correctness of the decoding algorithm is
an immediate consequence from the following theorem.

\begin{theorem}
If $d_S (X',Y') \leq \delta-1$ then $\tilde{X}'=X'$.
\end{theorem}
\begin{proof}
Assume that $d_S (X',Y') \leq \delta-1$. Let $X \subseteq Q$ be
the word obtained from $X'$ by appending a symbol to the end of
each vector in $X'$ (this can be done by using the generator
matrix $\mbox{RE}(Q)$ of $Q$). If $u \in X' \cap Y'$ then we append the same
symbol to $u$ to obtain the element of $X$ and to obtain the element
of $Y$. Hence, $d_S (X,Y)=d_S (X',Y') \leq \delta
-1$. If $d_S (X,Y) \leq \delta -2$ then $d_S (X,Z) \leq d_S (X,Y)
+1 \leq \delta -1$. Now, note that if $\delta-1$ is odd then $Z$
does not have the same parity as the dimensions of the subspaces
in $\C$ and if $\delta-1$ is even then $Z$ has the same parity as
the dimensions of the subspaces in $\C$. Therefore, if $d_S
(X,Y)=\delta-1$ then by the definition of $Z$ we have $Z=Y$ and
hence $d_S (X,Z)=\delta-1$. Therefore, the decoding algorithm of
$\C$ will produce as an output the unique codeword $\cX$ such that
$d_S ( \cX , Z) \leq \delta -1$, i.e., $X = \cX$. $X'$ is obtained
by deleting the last entry is each vector of $X \cap Q$;
$\tilde{X}'$  is obtained by deleting the last entry is each
vector of $\cX \cap Q$. Therefore, $\tilde{X}'=X'$.
\end{proof}

\begin{remark}
The constant dimension codes constructed in
Section~\ref{sec:multilevel construction} have the same dimension for all
codewords. Hence, if $\C$ was constructed by our multilevel
construction, then its decoding algorithm can be applied on the
punctured code $\C'_{Q,v}$.
\end{remark}

\chapter[Enumerative Coding and Lexicodes in Grassmannian]
{Enumerative Coding and Lexicodes in Grassmannian
\let\thefootnote\relax\footnotetext{The material of
Section~\ref{sec:lexi order} and Section~\ref{sec:enum coding}
 was published in~\cite{SiEt09} and also was presented in~\cite{SiEt09c}; the material
of Section~\ref{sec:lexicodes} was published in~\cite{SiEt10}.}}
\label{ch:Enum and Lexi}

In this chapter we consider enumerative coding and lexicodes in the Grassmannian.
Two different lexicographic orders for the
Grassmannian induced by different representations of
$k$-dimensional subspaces of $\F_q^n$ are given in Section~\ref{sec:lexi order}.
The main goal of Section~\ref{sec:enum coding} is to present efficient enumerative
encoding and decoding techniques for the Grassmannian which are
based on these two orders for $\Gr$. One enumerative coding
method is based on a Ferrers diagram representation and on an
order for $\Gr$ based on this representation. The complexity of
this enumerative coding is $O(k^{5/2} (n-k)^{5/2})$ digit
operations. Another order of the Grassmannian is based on a
combination of an identifying vector and a reduced row echelon
form representation of subspaces. The complexity of the
enumerative coding, based on this order, is $O(nk(n-k)\log
n\log\log n)$ digits operations. A combination of the two methods
reduces the complexity on average by a constant factor.

Constant dimension lexicodes are considered in Section~\ref{sec:lexicodes}.
A computer search for large constant dimension codes is usually
inefficient since the search space domain is extremely large. Even
so, we found that some constant dimension lexicodes are larger
than other known codes. We show how to make the computer search
more efficient.
\section{Lexicographic Order for Grassmannian}
\label{sec:lexi order}

In this section we present two different lexicographic orders for the Grassmannian.
First is based on the the extended representation of a subspace in $\Gr$, and
the second one is based on Ferrers Tableaux Form representation of a subspace in $\Gr$.
We will see in  the sequel that the first order will result in more efficient
enumerative coding in $\Gr$,
while the second order will lead to large constant dimension lexicodes.

\subsection{Order for $\Gr$ Based on Extended Representation}
\label{subsec:ext order}
Let $\{ x \}$ denotes the value of
$x=(x_1,x_2,...,x_r) \in \Z_q^r$ (or $x=(x_1,x_2,...,x_r)^T \in
\Z_q^r$), where the vector $x$ is viewed as a number in base-$q$
notation. Let $\{i\}_{q}$ be the base-$q$ representation of the
nonnegative integer $i$. The resulting vector is either a row
vector or a column vector depending on the context.

Let $X$ and $Y\in \Gr$ be two $k$-dimensional subspaces and
$\mbox{EXT}(X)$ and $\mbox{EXT}(Y)$ be the extended representations
of $X$ and $Y$, respectively. Let $i$ be the least index (from the right) such that
$\mbox{EXT}(X)$ and $\mbox{EXT}(Y)$ have different columns. We say
that $X<Y$ if $\begin{footnotesize}\left\{
\begin{array}{c} v(X)_i \\X_i
\end{array}\right\}<\left\{ \begin{array}{c} v(Y)_i \\Y_i
\end{array}\right\}.\end{footnotesize}$ Clearly,
this definition induces an order for $\Gr$.

\vspace{0.1cm}

\begin{example}
For $X,Y,Z\in\mathcal G_2(6,3)$ whose extended representations are given by

$$
\emph{EXT}(X)=\left(\begin{array}{cccccc}
1 & 1 & 1 & 0 & \emph{\textbf{0}} & 0 \\
1 & 0 & 0 & 0 & \emph{\textbf{1}} & 0 \\
0 & 1 & 0 & 0 & \emph{\textbf{0}} & 0 \\
0 & 0 & 1 & 1 & \emph{\textbf{0}} & 0 \\
\end{array}
\right),\;
\emph{EXT}(Y)=\left( \begin{array}{cccccc}
1 & 1 & 0 & 1 & \emph{\textbf{0}} & 0 \\
1 & 0 & 0 & 0 & \emph{\textbf{0}} & 0 \\
0 & 1 & 1 & 0 & \emph{\textbf{0}}& 0 \\
0 & 0 & 0 & 1 & \emph{\textbf{0}} & 0
\end{array}
\right),\;
$$
$$ \textmd{and }\;
\emph{EXT}(Z)=\left( \begin{array}{cccccc}
1 & 1 & 0 & 0 & \emph{\textbf{1}} & 0 \\
1 & 0 & 0 & 0 & \emph{\textbf{0}} & 0 \\
0 & 1 & 0 & 0 & \emph{\textbf{0}} & 0 \\
0 & 0 & 0 & 0 & \emph{\textbf{1}} & 0
\end{array}
\right) ,
$$
we have $Y<X<Z$.
\end{example}

\subsection{Order for $\Gr$ Based on Ferrers Tableaux Form}
\label{subsec:FD order}

Let $\cF_X$ be a Ferrers diagram  of a subspace $X\in \Gr$.  $\cF_X$ can be
embedded in  a
$k\times (n-k)$ box. We represent $\cF_X$ by an integer vector of
length $n-k$, $(\cF_{n-k},...,\cF_2,\cF_1)$, where $\cF_i$ is
equal to the number of dots in the $i$th column of $\cF_X$, $1\leq
i\leq n-k$, where we number the columns from right to left. Note
that $\cF_{i+1} \leq \cF_i$, $1 \leq i \leq n-k-1$.

To define an order of all the subspaces in the Grassmannian we
need first to define an order of all the Ferrers diagrams embedded
in the $k\times (n-k)$ box.

For two Ferrers diagrams $\cF$ and
$\widetilde{\cF}$, we say that $\cF < \widetilde{\cF}$ if one of
the following two conditions holds.
\begin{itemize}
\item $|\cF| > |\widetilde{\cF}|$;

\item $|\cF| = |\widetilde{\cF}|$,  and  $\cF_i > \widetilde{\cF}_i$
for the least index $i$ where the two diagrams $\cF$  and
$\widetilde{\cF}$ have a different number of dots.
\end{itemize}

\begin{example} If three Ferrers diagrams are given by
\begin{align*}
\cF_1=
\begin{array}{ccc}
\bullet  & \bullet  & \bullet  \\
& \bullet  & \bullet   \\
& \bullet  & \bullet
\end{array},~~~
\cF_2=
\begin{array}{ccc}
\bullet & \bullet &\bullet \\
\bullet & \bullet  & \bullet   \\
& & \bullet
\end{array},~~~
\cF_3=
\begin{array}{ccc}
\bullet  & \bullet  & \bullet  \\
& \bullet  & \bullet   \\
& & \bullet
\end{array},
\end{align*}
then $\cF_1< \cF_2 < \cF_3$.
\end{example}

\begin{remark}
Note, that this order for Ferrers diagrams is similar to the lexicographic order
defined in the literature for unrestricted partitions,
e.g.~\cite{NMS71},\cite[pp. 93-98]{Rus}.

\end{remark}

Now, we define the following order of subspaces in the
Grassmannian based on the Ferrers tableaux form representation.
Let $X$, $Y\in\Gr$ be two $k$-dimensional subspaces,  and  $\cF_X,~\cF_Y$ their Ferrers
diagrams, respectively. Let
$x_1,x_2,...,x_{|\cF_X|}$ and $y_1,y_2,...,y_{|\cF_Y|}$ be the
entries of Ferrers tableaux forms $\cF(X)$ and $\cF(Y)$,
respectively. The entries of a Ferrers tableaux form are numbered
from right to left, and from top to bottom.

We say that $X < Y$ if one of the following two conditions holds.
\begin{itemize}
\item $\cF_X < \cF_Y; $

\item $\cF_X = \cF_Y $,  and $(x_1,x_2,...,x_{|\cF_X|}) <
(y_1,y_2,...,y_{|\cF_Y|}).$
\end{itemize}
\vspace{0.2cm}
\begin{example}
Let $X,Y,Z,W\in\mathcal G_2(6,3)$ be given by
\begin{align*}
\cF(X)=
\begin{array}{ccc}
1 & 1 & 1 \\
1 & 1 & 1  \\
&  & 1
\end{array},~~~
\cF(Y)=
\begin{array}{ccc}
1 & 0 & 1 \\
& 0 & 0  \\
& 1 & 1
\end{array},~~~
\cF(Z)=
\begin{array}{ccc}
1 & 1 & 1 \\
& 1 & 1  \\
& & 0
\end{array},~~~
\cF(W)=
\begin{array}{ccc}
1 & 1 & 1 \\
& 1 & 1  \\
& & 1
\end{array}.
\end{align*}
By the definition, we have that $\cF_Y < \cF_X < \cF_Z =\cF_W $.
Since $(z_1,z_2,...,z_{|\cF_Z|})=(1,1,0,1,1,1)<
(w_1,...,w_{|\cF_W|})=(1,1,1,1,1,1)$, it follows that $Y<X<Z<W$.
\end{example}

\section{Enumerative Coding for Grassmannian}
\label{sec:enum coding}

In this section we consider the problem of encoding/decoding of subspaces in the
Grassmannian in an efficient way. By encoding we mean a
transformation of an information word into a $k$-dimensional
subspace. Decoding is the inverse transformation of the
$k$-dimensional subspace into the information word.

To solve this coding problem, we will use the general enumerative
coding method which was presented by Cover~\cite{Cover}. Let
$\{0,1\}^n$ denote the set of all binary vectors of length $n$.
Let $S$ be a subset of $\{0,1\}^n$. Denote by
$n_S(x_1,x_2,\ldots,x_k)$ the number of elements of $S$ for which
the first $k$ coordinates are given by $(x_1,x_2,\ldots,x_k)$,
where $x_1$ is the most significant bit. A lexicographic order of
$S$ is defined as follows. We say that for $x,y\in \{0,1\}^n$,
$x<y$, if $x_k<y_k$ for the least index $k$ such that $x_k\neq
y_k$. For example, $00101<00110$.

\begin{theorem}\cite{Cover}
\label{thm:cover} The lexicographic index (decoding) of $x\in S$
is given by
$$\emph{ind}_S(x)=\sum_{j=1}^{n}x_j \cdot n_S(x_1,x_2,\ldots,x_{j-1},0).$$
\end{theorem}

Let $S$ be a given subset and let $i$ be a given index. The
following algorithm finds the unique element $x$ of the subset $S$
such that $\text{ind}_S(x)=i$ (encoding).

\textit{Inverse algorithm}~\cite{Cover}:
For $k=1,\ldots,n$, if $i\geq n_S(x_1,x_2,\ldots,x_{k-1},0)$ then
set $x_k=1$ and $i=i-n_S(x_1,x_2,\ldots,x_{k-1},0)$; otherwise set
$x_k=0$.

\begin{remark}
The coding algorithms of Cover are efficient if
$n_S(x_1,x_2,\ldots,x_{j-1},0)$ can be calculated efficiently.
\end{remark}

Cover~\cite{Cover} also presented the extension of these results
to arbitrary finite alphabets. For our purpose this extension is
more relevant as we will see in the sequel. The formula for
calculating the lexicographic index of $x\in
S\subseteq\{1,2,3,\ldots,M\}^n$ is given as follows.
\begin{equation}\text{ind}_S(x)=\sum_{j=1}^{n}\sum_{m<x_j}n_S(x_1,x_2,\ldots,x_{j-1},m).\label{cover}
\end{equation}



Enumerative coding has various applications and it was considered
in many papers, e.g.~\cite{BrIm00,Imm99,Kur02}. Our goal in this
section is to apply this scheme to the set of all subspaces in a
Grassmannian, using different lexicographic orders, presented in the
previous section.

\subsection[Enumerative Coding for $\Gr$ Based on Extended Representation]
{Enumerative Coding for $\Gr$ Based on \\Extended Representation}
\label{subsec:Ext enum coding}
We present an
enumerative coding technique for the Grassmannian using the extended representation
and discuss its complexity.
Let $\begin{footnotesize} N\left( \begin{array}{ccc}
 v_j & \ldots &  v_1 \\
 X_j & \ldots &  X_1
\end{array}
\right)
\end{footnotesize}$
be the number of elements in $\Gr$  for which the  first $j$
columns in the extended representation are given by
$\begin{footnotesize} \left( \begin{array}{ccc}
 v_j & \ldots &  v_1 \\
 X_j & \ldots &  X_1
\end{array}
\right).
\end{footnotesize}$

\begin{remark} We view all the $q$-ary vectors of length ${k+1}$
as our finite alphabet. Let $S$ be the set of all $q$-ary
${(k+1)\times n}$ matrices which form extended representations of
some $k$-dimensional subspaces. Now, we can use Cover's method to
encode/decode the Grassmannian. In this setting note that
$\begin{footnotesize} N\left( \begin{array}{ccc}
 v_j & \ldots &  v_1 \\
 X_j & \ldots &  X_1
\end{array}
\right)
\end{footnotesize}$ is equivalent to
$n_S(x_1,x_2,\ldots,x_j)$, where $\left( \begin{array}{c}
 v_i \\
 X_i
\end{array}
\right)$ has the role of $x_i$.
\end{remark}

Let $w_j$ denotes the weight of the first $j$ entries of $v(X)$,
i.e., $w_j=\sum_{\ell=1}^j v_\ell$.
\begin{lemma}
\label{lem:N} For $1 \leq j \leq n$ we have
\[
N\left( \begin{array}{ccc}
 v_j & \ldots &  v_1 \\
 X_j & \ldots &  X_1
\end{array}\right)=\begin{small}\left[\begin{array}{c}
n-j\\k- w_j \end{array}\right]_{q}\end{small}.
\]

\end{lemma}
\begin{proof}
Let $X$ be a $k$-dimensional subspace in $\Gr$  for which the
first $j$ columns in the extended representation are given by
$\begin{footnotesize} \left( \begin{array}{ccc}
 v_j & \ldots &  v_1 \\
 X_j & \ldots &  X_1
\end{array}
\right).
\end{footnotesize} $
Then in the last $n-j$ entries of $v(X)$ there are $k-w_j$
\textit{ones}, and the $w_j$ last rows of $n-j$ last columns of
$\mbox{EXT}(X)$ have only \textit{zeroes}. Therefore, restriction
of $\mbox{EXT}(X)$ to the first $(k+1)-w_j$ rows of the last $n-j$
columns defines a subspace in $\mathcal G_q(n-j,k- w_j ).$ Hence,
we have
\[
N\left( \begin{array}{ccc}
 v_j & \ldots &  v_1 \\
 X_j & \ldots &  X_1
\end{array}\right)=\begin{small}\left[\begin{array}{c}
n-j\\k-w_j\end{array}\right]_{q}\end{small}~.
\]
\end{proof}
\begin{theorem}
\label{thm:Ind} Let $X\in\mathcal{G}_{q}(n,k)$ be a subspace,
where
\[
\emph{EXT}(X)=\left( \begin{array}{cccc}
 v_n & \ldots & v_2 & v_1 \\
 X_n & \ldots & X_2 & X_1
\end{array}
\right).
\]
Then the lexicographic index (decoding) of $X$,
$\emph{I}_{\begin{tiny}\emph{EXT}\end{tiny}}(X)$, is given by
\begin{equation}
\label{eq:Ind2}
\emph{I}_{\begin{tiny}\emph{EXT}\end{tiny}}(X)=\sum_{j=1}^{n}(v_{j}q^{k-w_{j-1}}+(1-v_{j})
\frac{\left\{ X_{j}\right\} }{q^{w_{j-1}}})
\begin{footnotesize}\left[\begin{array}{c}n-j\\
k-w_{j-1}\end{array}\right]_{q}\end{footnotesize}.
\end{equation}
\end{theorem}
\begin{proof}
By (\ref{cover}) we have that
$\mbox{I}_{\begin{tiny}\mbox{EXT}\end{tiny}}(X)$ is equal to
\begin{equation} \sum_{j=1}^{n}\sum_{\tiny\tiny
\left( \begin{array}{c} u \\W
\end{array}\right)<\left( \begin{array}{c} v_j \\X_j
\end{array}\right)}N\left( \begin{array}{cccc}
u & v_{j-1} & \ldots &  v_1 \\
W & X_{j-1} & \ldots &  X_1
\end{array}\right).\label{ind_2_proof}
\end{equation}

To compute the $j$th summand of (\ref{ind_2_proof}), we
distinguish between two cases.\\
Case 1: $v_j=1$. It implies that $X_j$ has weight one, and its
bottom $w_{j-1} +1$ entries (as a column vector) are an {\it one}
followed by $w_{j-1}$ {\it zeroes}, i.e., $X_j =
\{q^{w_{j-1}}\}_q$. Hence, $\mbox{EXT}(X)$ has the form
\[ \left( \begin{array}{ccccccc}
v_n &\ldots & v_{j+1} &1 & v_{j-1} & \ldots &  v_1 \\
X_n &\ldots & X_{j+1} & \{q^{w_{j-1}}\}_q & X_{j-1} & \ldots &  X_1\\
\end{array}\right)~.
\]\\
Therefore, a subspace $Y \in \Gr$ is lexicographically preceding
$X$, where $\mbox{EXT}(Y)$ has the same first $j-1$ columns as
$\mbox{EXT}(X)$, if and only if $\mbox{EXT}(Y)$ has the form
$$\left( \begin{array}{ccccccc}
v'_n &\ldots & v'_{j+1} & 0 & v_{j-1} & \ldots &  v_1 \\
Y_n & \ldots & Y_{j+1}& Y_j & X_{j-1} & \ldots &  X_1
\end{array}\right)~.
$$
Note, that $Y_j$ has {\it zeroes} in the last $w_{j-1}$ entries
(since the leading coefficients of the last $w_{j-1}$ rows are
contained in $( X_{j-1} ~ \cdots ~ X_1 )$). The first $k-w_{j-1}$
entries of $Y_j$ can have any values.
Therefore, in this case the $j$th summand of (\ref{ind_2_proof})
is equal to
\[\sum_{s=0}^{q^{k-w_{j-1}}-1}N\left( \begin{array}{cccc}
 0 & v_{j-1} & \ldots &  v_1 \\
 \{s\cdot q^{w_{j-1}}\}_q & X_{j-1} & \ldots &  X_1
\end{array}\right)
\]\\
which is equal by Lemma~\ref{lem:N} to
\begin{equation}q^{k-w_{j-1}}\begin{small}\left[\begin{array}{c}
n-j\\
k-w_{j-1}\end{array}\right]_{q}\end{small}.\label{firstSummand}
\end{equation}\\
Case 2: $v_j=0$. Since $w_{j-1}=\sum_{\ell=1}^{j-1}v_\ell$, it
follows that the last $w_{j-1}$ entries of $X_j$ are {\it zeroes},
i.e., $\{X_j\}$ is a multiple of $q^{w_{j-1}}$. Hence,
$\mbox{EXT}(X)$ has the form
\[ \left( \begin{array}{ccccccc}
v_n &\ldots & v_{j+1} &0 & v_{j-1} & \ldots &  v_1 \\
X_n &\ldots & X_{j+1} & X_j & X_{j-1} & \ldots &  X_1\\
\end{array}\right).
\]\\
Therefore, a subspace $Y \in \Gr$ is lexicographically preceding
$X$, where $\mbox{EXT}(Y)$ has the same first $j-1$ columns as
$\mbox{EXT}(X)$, if and only if $\mbox{EXT}(Y)$ has the form
\[ \left( \begin{array}{ccccccc}
v'_n &\ldots & v'_{j+1} &0 & v_{j-1} & \ldots &  v_1 \\
Y_n &\ldots & Y_{j+1} &\{s\cdot q^{w_{j-1}}\}_q & X_{j-1} & \ldots
&  X_1
\end{array}\right),
\]\\
where $0\leq s\leq \frac {\{X_j\}}{q^{w_j-1}}-1$.
Thus, in this case the $j$th summand of (\ref{ind_2_proof}) is
equal to
\[\sum_{s=0}^{\frac {\{X_j\}}{q^{w_{j-1}}}-1}N\left(
\begin{array}{cccc}
 0 & v_{j-1} & \ldots &  v_1 \\
 \{s\cdot q^{w_{j-1}}\}_q & X_{j-1} & \ldots &  X_1
\end{array}\right),
\]\\
which is equal by Lemma~\ref{lem:N} to
\begin{equation}\frac{\left\{ X_{j}\right\} }{q^{w_{j-1}}}\left[
\begin{array}{c}
n-j\\
k-w_{j-1}\end{array}\right]_{q}.\label{secondSummand}
\end{equation}
Finally, combining equations (\ref{firstSummand}) and
(\ref{secondSummand}) in Case 1 and Case~2 implies equation~(\ref{eq:Ind2}).
\end{proof}
\begin{example}
\label{exm:X_0} Let $X\in \mathcal G_2(6,3)$ be a subspace
represented by
\begin{align*}
\emph{EXT}(X)=\left( \begin{array}{cccccc}
0 & 1 & 0 & 1 & 1 & 0 \\
0 & 1 & 1 & 0 & 0 & 1 \\
0 & 0 & 0 & 1 & 0 & 0 \\
0 & 0 & 0 & 0 & 1 & 1
\end{array}
\right) .
\end{align*}
By Theorem~\ref{thm:Ind} we have that
$$\emph{I}_{\begin{tiny}\emph{EXT}\end{tiny}}(X)=5\cdot\begin{footnotesize}\left[\begin{array}{c}5\\
3\end{array}\right]_{2}\end{footnotesize}
+2^3\cdot\begin{footnotesize}\left[\begin{array}{c}4\\3\end{array}\right]_{2}\end{footnotesize}
+2^2\cdot\begin{footnotesize}\left[\begin{array}{c}3\\2\end{array}\right]_{2}\end{footnotesize}
+1\cdot\begin{footnotesize}\left[\begin{array}{c}2\\1\end{array}\right]_{2}\end{footnotesize}
+2\cdot\begin{footnotesize}\left[\begin{array}{c}1\\1\end{array}\right]_{2}\end{footnotesize}
+0\cdot\begin{footnotesize}\left[\begin{array}{c}0\\0\end{array}\right]_{2}\end{footnotesize}=928.
$$
\end{example}

\vspace{0.2cm}

Now, suppose that an index $0 \leq i <
\begin{footnotesize}\left[\begin{array}{c}n\\k\end{array}\right]_q\end{footnotesize}$
is given. Encoding Algorithm A finds $X \in \Gr$ such that
${\mbox{I}_{\begin{tiny}\mbox{EXT}\end{tiny}}(X)=i}$.

\begin{figure}[hbt]
\centering
\begin{algorithm}
\noindent \textit{Encoding Algorithm A:}

\noindent Set $i_{0}=i$, $w_0=0$.

\noindent For $j=1,2,...,n$ do

\begin{itemize}
\item if $w_{j-1}= k$ then set $v_j=v(X)_j=0$, $w_j=w_{j-1}$,
$X_j{=\{0\}_q}$, and $i_j=i_{j-1}$;

\item otherwise

\begin{itemize}
\item if $i_{j-1}\geq q^{k-w_{j-1}}
\begin{footnotesize}\left[\begin{array}{c}
n-j\\k-w_{j-1}\end{array}\right]_{q}\end{footnotesize}$ then set
$v_j{=v(X)_j=1}$, $w_j=w_{j-1}+1$, $X_{j}=\{q^{w_{j-1}}\}_{q}$,
and $i_{j}=i_{j-1}-q^{k-w_{j-1}}\begin{footnotesize}
\left[\begin{array}{c}
n-j\\k-w_{j-1}\end{array}\right]_{q}\end{footnotesize}$;

\item otherwise let $val=\left\lfloor
i_{j-1}/\begin{footnotesize}\left[\begin{array}{c}
n-j\\k-w_{j-1}\end{array}\right]_{q}\end{footnotesize}\right
\rfloor $ and set $v_j{=v(X)_j=0}$, $w_j=w_{j-1}$, $X_{j}=\left\{
val*q^{w_{j-1}} \right\} _{q}$, and
$i_{j}=i_{j-1}-val*\begin{footnotesize}\left[
\begin{array}{c}
n-j\\k-w_{j-1}\end{array}\right]_{q}\end{footnotesize}.$
\end{itemize}
\end{itemize}

\noindent Form the output \[ \mbox{EXT}(X)=\left(
\begin{array}{cccc}
v_n & \ldots & v_2 & v_1 \\
X_n & \ldots & X_2 & X_1
\end{array}
\right).
\]
\end{algorithm}
\end{figure}

\begin{theorem}
Encoding Algorithm A finds the $k$-dimensional subspace $X \in \Gr$, such that
${\emph{I}_{\begin{tiny}\emph{EXT}\end{tiny}}(X)=i}$.
\end{theorem}
\begin{proof}
First we will show that the output of the algorithm is  a
$k$-dimensional subspace. In other words, we will prove that the
weight $w_n$ of identifying vector of the resulting subspace $X$
is equal to $k$. We observe that the first "if" of the algorithm
implies that $w_n\leq k$. Note also that $i_j\geq 0$ for all
$1\leq j\leq n$. Suppose that $w_n=k-t$ for some $t>0.$ Let
$n-k+t\leq j'\leq n$  be the last index  where $v(X)_{j'}=0.$ Then
$w_{j'}=k-t-n+j'=w_{j'-1}.$ According to the algorithm,
${i_{j'-1}< q^{k-w_{j'-1}}{\scriptstyle
\begin{footnotesize}\left[\begin{array}{c}
n-j'\\k-w_{j'-1}\end{array}\right]_{q}\end{footnotesize}}=q^{t+n-j'}
{\scriptstyle \begin{footnotesize}\left[\begin{array}{c}
n-j'\\t+n-j'\end{array}\right]_{q}\end{footnotesize}}=0}$ (since
$t>0$), which contradicts the observation that $i_j\geq 0$ for
each $1\leq j\leq n$.

Let $S_j$ be the $j$th summand of
$\mbox{I}_{\begin{tiny}\mbox{EXT}\end{tiny}}(X),$ given in
(\ref{eq:Ind2}), i.e.,
$\mbox{I}_{\begin{tiny}\mbox{EXT}\end{tiny}}(X)=\sum_{t=1}^nS_t.$
To prove the theorem it is sufficient to show that
$i_j=i-\sum_{t=1}^jS_t$ for all $1\leq j\leq n$ and $i_n=0$. The
proof will be inductive.

By the algorithm, for each coordinate $1\leq j\leq n-k$,
\[ i_{j} = \left\{
\begin{array}{cc}i_{j-1}- q^{k-w_{j-1}}
\begin{small}\left[\begin{array}{c} n-j\\k-w_{j-1}\end{array}\right]_{q}\end{small},
& \textrm{if }v(X)_{j}=1\\
i_{j-1}-\frac{\{X_j\}}{q^{w_{j-1}}}
\begin{small}\left[\begin{array}{c}
n-j\\k-w_{j-1}\end{array}\right]_{q}\end{small},
 & \textrm{if }v(X)_{j}= 0\end{array}\right.\]
Therefore,
\begin{equation}
\label{index_induction}
i_{j}=i_{j-1}- v(X)_{j} q^{k-w_{j-1}}
\begin{small}\left[\begin{array}{c} n-j\\k-w_{j-1}\end{array}\right]_{q}\end{small}
 -(1-v(X)_{j})\frac{\{X_j\}}{q^{w_{j-1}}}
\begin{small}\left[\begin{array}{c} n-j\\k-w_{j-1}\end{array}\right]_{q}\end{small}
=i_{j-1}-S_{j}
\end{equation}
for all $1\leq j\leq n-k$. Hence, for $j=1$ we have
$i_{1}=i-S_{1}$. We assume that $i_{j}=i-\sum_{t=1}^{j}S_{t}$, for
$j\geq1$.
By (\ref{index_induction}), $i_{j+1}=i_{j}-S_{j+1},$ therefore,
$i_{j+1}=i-\sum_{t=1}^{j}S_{t}-S_{j+1}=i-\sum_{t=1}^{j+1}S_{t}.$

Now, we will show that for all $0\leq j\leq n$, $i_j$ is the
lexicographic index of a subspace in $\mathcal{G}_q(n-j,k-w_j)$
with given $j$ first columns of its representation matrix. It will
complete the proof since $i_n$ is the index of subspace in
$\mathcal{G}_q(0,0)$ and thus it is equal to 0.

It is sufficient to prove that $i_j <
\begin{footnotesize}\left[\begin{array}{c}
n-j\\k-w_{j}\end{array}\right]_{q}\end{footnotesize}$ for all
$0\leq j\leq n$. The proof will be inductive. For $j=0$ we observe
that $i_0=i<\begin{footnotesize}\left[\begin{array}{c}
n\\k\end{array}\right]_{q}\end{footnotesize}$ is given. Assume
that $i_{j-1}<\begin{footnotesize}\left[\begin{array}{c}
n-j+1\\k-w_{j-1}\end{array}\right]_{q}\end{footnotesize}$. We will
show that $i_{j}<\begin{footnotesize}\left[\begin{array}{c}
n-j\\k-w_{j}\end{array}\right]_{q}\end{footnotesize}$. We
distinguish between two cases.\\
Case 1: $i_{j-1}\geq
q^{k-w_{j-1}}\begin{footnotesize}\left[\begin{array}{c}
n-j\\k-w_{j-1}\end{array}\right]_{q}\end{footnotesize}$. Then, by
the algorithm, $v_{j}=1$, $w_{j}=w_{j-1}+1$, and
$i_{j}=i_{j-1}-q^{k-w_{j-1}}\begin{footnotesize}\left[\begin{array}{c}
n-j\\k-w_{j-1}\end{array}\right]_{q}\end{footnotesize}$. By the
assumption, $i_{j} < \begin{footnotesize}\left[\begin{array}{c}
n-j+1\\k-w_{j-1}\end{array}\right]_{q}\end{footnotesize}-q^{k-w_{j-1}}
\begin{footnotesize}\left[\begin{array}{c}
n-j\\k-w_{j-1}\end{array}\right]_{q}\end{footnotesize}$ and thus
by  Lemma~\ref{lem:pascal},  $i_{j}\leq
\begin{footnotesize}\left[\begin{array}{c}
n-j\\k-w_{j-1}-1\end{array}\right]_{q}\end{footnotesize}=
\begin{footnotesize}\left[\begin{array}{c}
n-j\\k-w_{j}\end{array}\right]_{q}\end{footnotesize} $.\\
Case 2: $i_{j-1} <
q^{k-w_{j-1}}\begin{footnotesize}\left[\begin{array}{c}
n-j\\k-w_{j-1}\end{array}\right]_{q}\end{footnotesize}$. Then, by
the algorithm, $v_{j}=0$, $w_{j}=w_{j-1}$, and
$$i_{j}=i_{j-1}-\left\lfloor i_{j-1}/
\begin{footnotesize}\left[\begin{array}{c}
n-j\\k-w_{j-1}\end{array}\right]_{q}\end{footnotesize}\right\rfloor
\begin{footnotesize}\left[\begin{array}{c}
n-j\\k-w_{j-1}\end{array}\right]_{q}\end{footnotesize}$$

$$< (\left\lfloor i_{j-1}/\begin{footnotesize}\left[\begin{array}{c}
n-j\\k-w_{j-1}\end{array}\right]_{q}\end{footnotesize}\right\rfloor+1)
\begin{footnotesize}\left[\begin{array}{c}
n-j\\k-w_{j-1}\end{array}\right]_{q}\end{footnotesize}$$

$$-\left\lfloor i_{j-1}/\begin{footnotesize}\left[\begin{array}{c}
n-j\\k-w_{j-1}\end{array}\right]_{q}\end{footnotesize}\right\rfloor
\begin{footnotesize}\left[\begin{array}{c}
n-j\\k-w_{j-1}\end{array}\right]_{q}\end{footnotesize}=
\begin{footnotesize}\left[\begin{array}{c}
n-j\\k-w_{j-1}\end{array}\right]_{q}\end{footnotesize},
$$
since we can write $\lfloor \frac{a}{b}\rfloor\leq a < (\lfloor
\frac{a}{b}\rfloor+1)b$ for all positive integers $a$  and $b$.
\end{proof}
\vspace{0.1cm}

\begin{example}
Let $q=2$, $n=6$, $k=3$, and $i=928$. By using  Encoding
Algorithm~A we will find the subspace $X\in \mathcal G_2(6,3)$
such that $\emph{I}_{\begin{tiny}\emph{EXT}\end{tiny}}(X)=i$. We
apply the following steps of the algorithm.

\noindent $j=1$: $i_0=928<
2^3\begin{footnotesize}\left[\begin{array}{c}5\\3\end{array}
\right]_{2}\end{footnotesize}=1240$ and hence $v_1=v(X)_1=0,$
$val=\lfloor 928/155 \rfloor=5,$ $\begin{footnotesize} X_1=\left(
\begin{array}{c}1 \\ 0 \\ 1 \\ \end{array} \right)
\end{footnotesize}$, and $i_1=928-5\cdot 155=153$.

\noindent $j=2$: $i_1=153\geq
2^3\begin{footnotesize}\left[\begin{array}{c}4\\3\end{array}
\right]_{2}\end{footnotesize}=120$ and hence ${v_2=v(X)_2=1}$,
$\begin{footnotesize} X_2=\left(
\begin{array}{c}0 \\ 0 \\ 1 \\ \end{array} \right)
\end{footnotesize}$, and $i_2=153-120=33$.

\noindent $j=3$: $i_2=33 \geq
2^2\begin{footnotesize}\left[\begin{array}{c}3\\2\end{array}
\right]_{2}\end{footnotesize}=28$ and hence $v_3=v(X)_3=1$,
$\begin{footnotesize} X_3=\left(
\begin{array}{c}0 \\ 1 \\ 0 \\ \end{array} \right)
\end{footnotesize}$, and $i_3=33-28=5$.

\noindent $j=4$: $i_3=5<
2^1\begin{footnotesize}\left[\begin{array}{c}2\\1\end{array}
\right]_{2}\end{footnotesize}=6$ and hence $v_4=v(X)_4=0,$
$val=\lfloor 5/3\rfloor=1,$ $\begin{footnotesize} X_4=\left(
\begin{array}{c}1 \\ 0 \\ 0 \\ \end{array} \right)
\end{footnotesize}$, and $i_4=5-3=2.$

\noindent $j=5$: $i_4=2\geq
2^1\begin{footnotesize}\left[\begin{array}{c}1\\1\end{array}
\right]_{2}\end{footnotesize}=2$ and hence $v_5=v(X)_5=1$,
$\begin{footnotesize} X_5=\left(
\begin{array}{c}1 \\ 0 \\ 0 \\ \end{array} \right)
\end{footnotesize}$, and $i_5=2-2=0$.

\noindent
$j=6$: $w_5=3=k$ and hence $v_6=v(X)_6=0$,
$\begin{footnotesize} X_6=\left(
\begin{array}{c}0 \\ 0 \\ 0 \\ \end{array} \right)
\end{footnotesize}$, and $i_6=i_5=0$.

\vspace{0.2cm}
Therefore, we obtain a subspace $X\in \mathcal G_2(6,3)$ whose
extended representation is given~by
\begin{align*}
\emph{EXT}(X)=\left( \begin{array}{cccccc}
0 & 1 & 0 & 1 & 1 & 0 \\
0 & 1 & 1 & 0 & 0 & 1 \\
0 & 0 & 0 & 1 & 0 & 0 \\
0 & 0 & 0 & 0 & 1 & 1
\end{array}
\right) .
\end{align*}
\end{example}
\subsubsection{Complexity}
We consider the complexity of computation of lexicographic index
$\mbox{I}_{\begin{tiny}\mbox{EXT}\end{tiny}}( \cdot )$ in
(\ref{eq:Ind2}). Note that all the integers that we use in the
calculations  are $q$-ary integers. Let $M[a,b]$ denotes the
number of operations for the multiplication of two $q$-ary
integers of length $a$ and $b$. It is known~\cite[p. 634]{Knu99},
that for $a>b,$ $M[a,b]=a\log b\log\log b$.

First, we calculate the length of the $q$-ary integer which
represents the largest Gaussian coefficient in (\ref{eq:Ind2}).
This Gaussian coefficient is
$$\begin{small}\left[\begin{array}{c}n-1\\k\end{array}\right]_{q}\end{small}=
\frac{(q^{n-1}-1)\cdots(q^{n-k}-1)} {(q^k-1)\cdots(q-1)},$$ and
hence this length is less than $k(n-k)$.

\noindent If $w_j=w_{j-1}$ then
\begin{equation}
\label{eq:same} \begin{small}\left[\begin{array}{c}n-j\\k-w_{j-1}
\end{array}\right]_{q}\end{small} = \begin{small}
\left[\begin{array}{c}n-(j+1)\\k-w_j\end{array}\right]_{q}\end{small}
\cdot\frac {q^{n-j}-1}{q^{n-k-j+w_j}-1}~.
\end{equation}
If $w_j=w_{j-1}+1$ then
\begin{equation}
\label{eq:samep1}
\begin{small}\left[\begin{array}{c}n-j\\k-w_{j-1}
\end{array}\right]_{q}\end{small} = \begin{small}
\left[\begin{array}{c}n-(j+1)\\k-w_j\end{array}\right]_{q}\end{small}
\cdot\frac{q^{n-j}-1} {q^{k-w_j+1}-1}~.
\end{equation}
The Gaussian coefficients in (\ref{eq:Ind2}) can be derived from
the identifying vector. Their computation is done by
(\ref{eq:same}) and (\ref{eq:samep1}). Hence, the complexity for
computation of all the Gaussian coefficients that we need in
(\ref{eq:Ind2}) is $O(nM[k(n-k),n])$.

Since multiplication or division by $q^{i}$ is done by a shift of
$i$ digits, there are $n-k$ indices where $v_j=0$, and the length
of $\{X_j\}$ is $k$, it follows that the complexity of these
operations is $O((n-k)M[k(n-k),k])$. Finally, in (\ref{eq:Ind2})
there are at most $n$ additions of integers whose length is at
most $k(n-k+1)$, and therefore the complexity of these operations
can be omitted.

Hence, the complexity of computation of
$\mbox{I}_{\begin{tiny}\mbox{EXT}\end{tiny}} (\cdot)$ in
(\ref{eq:Ind2}) is $O(nM[k(n-k),n])$, i.e., $O(nk(n-k)\log
n\log\log n)$.

Therefore, we have proved the following theorem:
\begin{theorem}
\label{thm:complexity2} The computation complexity of the
lexicographic index (decoding) in (\ref{eq:Ind2}) is
$O(nk(n-k)\log n\log\log n)$ digits operations.
\end{theorem}

\vspace{0.3cm}

If $k < \log n\log\log n$ then the Gaussian coefficients in
(\ref{eq:Ind2}) can be computed more efficiently. For their
computation we can use Lemma~\ref{lem:pascal}. To compute
$\begin{footnotesize}\left[\begin{array}{c}
n\\
k\end{array}\right]_{q}\end{footnotesize}$ we need to compute
$\begin{footnotesize}\left[\begin{array}{c}
\eta\\
\kappa\end{array}\right]_{q}\end{footnotesize}$ for all $\eta$ and
$\kappa$ such that $0 \leq \kappa \leq k$ and $0 \leq \eta -
\kappa \leq n-k$. It requires at most $k(n-k)$ additions of
integers whose length is at most $k(n-k)$, and a total of at most
$k(n-k)$ shifts. All other computations do not change and can be
omitted from the total complexity. Thus, we have

\begin{theorem}
\label{thm:complexity3} If $\min \{ k,n-k \} < \log n\log\log n$,
then the computation complexity of the lexicographic index in
(\ref{eq:Ind2}) is $O(n^2 \min \{ k,n-k \}^2 )$ digits operations.
\end{theorem}

Finally, in a similar way we can show that the computation
complexity of Encoding Algorithm A is the same as the computation
complexity given for the decoding in Theorem~\ref{thm:complexity2}
and in Theorem~\ref{thm:complexity3}.

\subsection{Enumerative Coding for $\Gr$ Based on Ferrers Tableaux Form}
\label{subsec:FTF enum coding}

In this section we present an enumerative
coding for the Grassmannian based on the Ferrers tableaux form
representation of $k$-dimensional subspaces. Note, that even so
this enumerative coding is less efficient, it is more intuitive
and might have its own applications. For example, lexicodes based on the
related order, discussed in the next section, were found to be larger than the known
codes.


\subsubsection{Enumerative Coding for Ferrers Diagrams of the Same Size}

Let $\cF=(\cF_{n-k},...,\cF_2,\cF_1)$ be a Ferrers diagram  of size $m$ embedded in a $k\times
(n-k)$ box, where $\cF_i$ is equal to
the number of dots in the $i$-th column of $\cF$, $1\leq i\leq
n-k$.
Let $N_m(\cF_j,...,\cF_2,\cF_1)$ be the number of Ferrers diagrams
of size $m$ embedded in a $k\times (n-k)$ box, for which the first
$j$ columns are given by $(\cF_j,...,\cF_2,\cF_1)$. Recall, that the function
$p(\alpha, \beta, m)$, defined in
Subsection~\ref{subsec:Ferrers Tableaux Form Representation}, is the
number of partitions of $m$ whose Ferrers diagrams can be embedded into an $\alpha\times \beta$ box.


\begin{lemma}
\label{lem:calc_Nm} If $1 \leq j \leq n-k$ and $0 < m \leq k(n-k)$
then
\begin{equation*}
N_m(\cF_j,...,\cF_2,\cF_1)=p(\cF_j,n-k-j,m-\sum_{i=1}^j\cF_i).\label{eq:N_m}
\end{equation*}
\end{lemma}
\begin{proof}
The lemma is an immediate consequence from the fact that $\cF =
(\cF_{n-k},...,\cF_2,\cF_1)$ is a Ferrers diagram with $m$ dots
embedded in a $k \times (n-k)$ box if and only if
$(\cF_{n-k},...,\cF_{j+1})$ is a Ferrers diagram with
$m-\sum_{i=1}^j\cF_i$ dots embedded in an $\cF_j \times (n-k-j)$
box.
\end{proof}

\vspace{0.1cm}

\begin{remark}
We view the set $\Z_{k+1} = \{ 0,1,\ldots , k \}$ as our finite
alphabet since $0 \leq \cF_i \leq k$. Let $S$ be the set of all
$(n-k)$-tuples over $\Z_{k+1}$ which represent Ferrers diagrams
embedded in a $k \times (n-k)$ box. In other words,
${(\cF_{n-k},...,\cF_2,\cF_1) \in S}$ if and only if $0 \leq \cF_i
\leq \cF_{i-1} \leq k$ for each $2 \leq i \leq n-k$. Now, we can
use Cover's method to encode/decode the set of Ferrers diagrams
with $m$ dots embedded in a $k \times (n-k)$ box. In this setting
note that $N_m(\cF_j,...,\cF_2,\cF_1)$ is equivalent to
$n_S(x_1,x_2,\ldots,x_j)$, where $\cF_i$ has the role of $x_i$.
\end{remark}

\vspace{0.1cm}
\begin{theorem}
\label{thm:calc_Nm} Let $\cF = (\cF_{n-k},...,\cF_2,\cF_1)$ be a
Ferrers diagram of size $m$ embedded in a $k \times (n-k)$ box.
Then the lexicographic index (decoding), $\emph{ind}_m$, of $\cF$
among all the Ferrers diagrams with the same size $m$ is given by
\begin{equation}
\label{eq:ind_m} \emph{ind}_m(\cF)=\sum_{j=1}^{n-k}
\sum_{a=\cF_j+1}^{\cF_{j-1}}p(a,n-k-j,m-\sum_{i=1}^{j-1}\cF_i -a),
\end{equation}
where we define $\cF_{0}=k.$
\end{theorem}

\begin{proof}
By (\ref{cover}) we have that
\begin{equation*}
\mbox{ind}_m(\cF)=\sum_{j=1}^{n-k}
\sum_{a=\cF_j+1}^{\cF_{j-1}}N_m(a,\cF_{j-1},...,\cF_2,\cF_1).
\end{equation*}
The theorem follows now from Lemma~\ref{lem:calc_Nm}.
\end{proof}

\vspace{0.2cm}
\begin{remark}
The summation in Theorem~\ref{thm:calc_Nm} is over larger values,
while the summation in (\ref{cover}) is over smaller values, due
to the defined order ($\cF < \widetilde{\cF}$ if $\cF_i
>\widetilde{\cF}_i$ for the least index~$i$).
\end{remark}

\vspace{0.1cm}

Theorem~\ref{thm:calc_Nm} implies that if we can calculate
$p(\alpha,\beta,m)$ efficiently then we can calculate
$\mbox{ind}_m(\cF)$ efficiently for a Ferrers diagram of size $m$
embedded in a $k\times (n-k)$ box.

\vspace{0.8cm}
Now suppose that an index $0 \leq i < p(k,n-k,m)$ is given.
Encoding Algorithm B finds the Ferrers diagram $\cF$  of size $m$
embedded in a $k\times(n-k)$ box, such that $\mbox{ind}_m(\cF)=i$.

\begin{figure}[hbt]
\centering
\begin{algorithm}
\noindent \textit{Encoding Algorithm B:}

\textit{Step 1}: Set $\cF_0=k ,~ \ell_1=0,~h=i, ~ i_0=i$;

\begin{itemize}
\item while $h\geq N_m(\cF_0-\ell_1)$ set $h=h-N_m(\cF_0-\ell_1)$,
$\ell_1=\ell_1+1$;

\item set $\cF_1=\cF_0-\ell_1$, and $ i_1=h$;
\end{itemize}

\textit{Step 2}: For $j=2,...,n-k$ do
\begin{itemize}
\item if $\sum_{i=1}^{j-1}\cF_i=m$ then set $\cF_j=0$;

\item otherwise do

\noindent begin

\begin{itemize}
\item set $\ell_j=0,h=i_{j-1}$;

\item while $h\geq N_m(\cF_{j-1}-\ell_j,\cF_{j-1},...,\cF_1)$ set
$h=h-N_m(\cF_{j-1}-\ell_j,\cF_{j-1},...,\cF_1)$,
$\ell_j=\ell_{j}+1$;

\item set $\cF_j=\cF_{j-1}-\ell_j$, and $ i_j=h$;
\end{itemize}
\end{itemize}
\indent ~~~~~ end \{begin\}

\textit{Step 3}: Form the output $\cF =
(\cF_{n-k},...,\cF_2,\cF_1)$.

\end{algorithm}
\end{figure}
\begin{remark}
We did not join Step 1 and Step 2, since
$N_m(\cF_{j-1}-\ell_j,\cF_{j-1},...,\cF_1)$ is not defined for
$j=1$.
\end{remark}

\begin{theorem}
Encoding Algorithm B finds the Ferrers diagram $\cF$ of size $m$
embedded in a $k\times(n-k)$ box, such that $\emph{ind}_m(\cF)=i$.
\end{theorem}

\begin{proof}
First we define for each $1 \leq j \leq n-k$,
$$S_j=\sum_{a=\cF_j+1}^{\cF_{j-1}}p(a,n-k-j,m-\sum_{i=1}^{j-1}\cF_i -a)$$
and observe that by (\ref{eq:ind_m}) we have
$\textmd{ind}_m(\cF)=\sum_{j=1}^{n-k}S_j$. By the algorithm, for all $1
\leq j\leq n-k$, we have that
$i_j=i_{j-1}-\sum_{\ell=0}^{\ell_j-1}N_m(\cF_{j-1}-\ell,\cF_{j-1},...,\cF_2,\cF_1
)$ and hence by Lemma~\ref{lem:calc_Nm} it follows that $i_j=
i_{j-1}-S_j$. Hence, by using induction we obtain that for all
$1\leq j\leq n-k$, $i_j=i-\sum_{t=1}^j S_t$. Thus,
$i_{n-k}=i-\textmd{ind}_m(\cF)$.

Now observe that by the algorithm, for all $0 \leq j \leq n-k$,
when we set $i_j=h$, we have $h <
N_{m}(\cF_{j},\cF_{j-1},...,\cF_{1})$ and hence $0\leq i_{j}<
N_{m}(\cF_{j},\cF_{j-1},...,\cF_{1})$. Thus, by
Lemma~\ref{lem:calc_Nm},
\begin{equation}
\label{eq:part_loop} 0\leq i_{j}<
p(\cF_{j},n-k-j,m-\sum_{\ell=1}^{j}\cF_\ell)~.
\end{equation}
Note that $\sum_{\ell=1}^{j}\cF_{\ell}\leq m$, for all $1\leq j\leq n-k$, otherwise (\ref
{eq:initial_cond}) and (\ref{eq:part_loop}) imply that $0\leq
i_{j}< 0$, a contradiction. Note also that
$\sum_{\ell=1}^{n-k}\cF_{\ell}= m$, otherwise (\ref
{eq:initial_cond}) implies that $0\leq
i_{n-k}<p(\cF_{n-k},0,\sum_{\ell=1}^{n-k}\cF_{\ell})=0$, a
contradiction. Also, by the algorithm we have $\cF_{j}\leq
\cF_{j-1}$, and therefore the algorithm generates a Ferrers diagram.
It implies that $0\leq i_{n-k}<p(\cF_{n-k},0,0)=1,$ i.e.,
$i_{n-k}=0$ and thus, $i=\textmd{ind}_m(\cF).$
\end{proof}

\vspace{0.6cm}
\subsubsection{Enumerative Coding Based on Ferrers Tableaux Form}

In this subsection, we use the order of Ferrers tableaux
forms given in Sebsection~\ref{subsec:FD order} and the connection
between integer partitions and Gaussian coefficients given in Theorem~\ref{thm:vanLint},
for enumerative coding in $\Gr$.
\begin{theorem}
\label{thm:Ferrers_index} Let $X \in \Gr$, $\cF_X$ be the Ferrers
diagram of $X$, $\cF(X)$ be the Ferrers tableaux form of $X$,
and let $x= (x_1,x_2,...,x_{|\cF_X|})$ be the
entries vector of $\cF(X)$. Then the lexicographic index
(decoding) of $X$, $\emph{Ind}_{\cF} (X)$, defined by the order of $\Gr$
based on Ferrers tableaux form, is given by
\begin{equation}\emph{Ind}_{\cF}(X)=\sum_{i=|\cF_X|+1}^{k(n-k)}\alpha_{i}q^{i}+
\emph{ind}_{|\cF_X|}(\cF_X)q^{|\cF_X|}+\{x\},\label{Ind1}
\end{equation}
where $\alpha _i$, for $|\cF_X|+1 \leq i \leq k(n-k)$, is defined in
Theorem~\ref{thm:vanLint}, and $\textmd{ind}_{|\cF_X|}$ is given in Theorem~\ref{thm:calc_Nm}.
\end{theorem}

\begin{proof} To find $\mbox{Ind}_{\cF}(X)$ we have to calculate the number of
$k$-dimensional subspaces which are preceding $X$ according to the
order defined above.
\begin{enumerate}
\item All the $k$-dimensional subspaces with Ferrers diagrams
which have more dots than $\cF_X$ are preceding $X$. Their number
is $\sum_{i=|\cF_X|+1}^{k(n-k)}\alpha_{i}q^{i}$.
\item There are $\mbox{ind}_{|\cF_X|}(\cF_X)$ Ferrers diagrams
with $|\cF_X|$ dots which are preceding $X$. Hence, there are
$\mbox{ind}_{|\cF_X|}(\cF_X)q^{|\cF_X|}$ $k$-dimensional subspaces
whose Ferrers diagrams have $|\cF_X|$ dots and preceding $X$.
\item Finally, the number of $k$-dimensional subspaces whose
Ferrers diagram is $\cF_X$ which are preceding $X$ is $\{x\}$.
\end{enumerate}
\end{proof}

\begin{example}
Let $X\in\mathcal{G}_2(6,3)$ be the subspace of
Example~\ref{exm:X_0}, whose Ferrers tableaux form and Ferrers
diagram are
$$\begin{footnotesize}
\cF(X)=
\begin{array}{cc}
1 & 1  \\
 & 0  \\
 & 1
\end{array}
\end{footnotesize}\;\mbox{and}\;
\begin{footnotesize}
\cF_X=
\begin{array}{cc}
\bullet & \bullet  \\
 & \bullet   \\
 & \bullet
\end{array}.
\end{footnotesize} $$
By Theorem~\ref{thm:Ferrers_index} we have that
$$\emph{Ind}_\cF(X)=\sum_{i=5}^{9}\alpha_{i}2^{i}+
\emph{ind}_{4}(\cF_X)2^4+\{(1011)\}.$$ Since $\alpha_5=3$,
$\alpha_6=3$, $\alpha_7=2$, $\alpha_8=1$, $\alpha_9=1$
(see~\cite[pp. 326-328]{vLWi92}), $\emph{ind}_4(\cF_X)=0$, and
$\{(1011)\}=11$, it follows that $\emph{Ind}_\cF(X)=1323$.
\end{example}


Now suppose that an index $0 \leq i <
\begin{footnotesize}\left[\begin{array}{c}n\\k\end{array}\right]_{q}\end{footnotesize}$ is given.
Encoding Algorithm C finds a subspace $X \in \Gr$ such that
$\mbox{Ind}_{\cF}(X)=i$.

\begin{figure}[hbt]
\centering
\begin{algorithm}
\noindent \textit{Encoding Algorithm C:}

\noindent Set $i_0=i$.

\noindent For $j=0,\ldots, k(n-k)$ do

\begin{itemize}
\item if $i_j< \alpha_{k(n-k)-j}q^{k(n-k)-j}$ then set
$|\cF_X|{=k(n-k)-j}$,
$\cF_X=\mbox{ind}_{|\cF_X|}^{-1}(\lfloor\frac{i_j}{q^{k(n-k)-j}}\rfloor)$;
$\{i_{j}{-\lfloor\frac{i_j}{q^{k(n-k)-j}}\rfloor
q^{k(n-k)-j}\}_q}$ is assigned to $x$ (the entries vector of
$\cF(X)$) and stop;

\item otherwise set $i_{j+1}=i_j-\alpha_{k(n-k)-j}q^{k(n-k)-j}$.
\end{itemize}

\end{algorithm}
\end{figure}
\vspace{-0.6cm}
\begin{theorem}
Encoding Algorithm C finds a subspace $X$ such that
$\emph{Ind}_{\cF}(X)=i.$
\end{theorem}

\begin{proof} Let $\cF (X)$ be the Ferrers tableaux form generated by
the algorithm,  $x$ the entries vector of $\cF (X)$, and $\cF_X$ the Ferrers
diagram of the corresponding subspace $X$.


Let $j'$ be the value of $j$ in the algorithm for which we have
$i_{j'}<\alpha_{k(n-k)-j'}q^{k(n-k)-j'}$. By the algorithm, for
all $1\leq j\leq j'$, we have
$i_j=i_{j-1}-\alpha_{k(n-k)-(j-1)}q^{k(n-k)-(j-1)}$. Hence,

\begin{equation}
\label{eq:ind1}
i_{j'}=i-\sum_{t=k(n-k)-(j'-1)}^{k(n-k)}\alpha_{t}q^t ~.
\end{equation}

By the algorithm we have ${|\cF_X|=k(n-k)-j'}$,
$\cF_X=\textmd{ind}_{|\cF_X|}^{-1}(\lfloor\frac{i_{j'}}{q^{k(n-k)-j'}}\rfloor),$
and $x=\{i_{j'}-\lfloor\frac{i_{j'}}{q^{k(n-k)-j'}}\rfloor
q^{k(n-k)-j'}\}_q$. Therefore,

$$\textmd{Ind}_{\cF}(X)=\sum_{t=k(n-k)-(j'-1)}^{k(n-k)}\alpha_{t}q^t
+\textmd{ind}_{k(n-k)-j'}(\textmd{ind}_{k(n-k)-j'}^{-1}(\lfloor\frac{i_{j'}}
{q^{k(n-k)-j'}}\rfloor))q^{k(n-k)-j'}$$
\begin{equation}
\label{eq:ind1a} +i_{j'}-\lfloor\frac{i_{j'}}{q^{k(n-k)-j'}}\rfloor
q^{k(n-k)-j'}=\sum_{t=k(n-k)-(j'-1)}^{k(n-k)}\alpha_{t}q^t
+i_{j'},
\end{equation}
where the last equality follows from the observation that
$\textmd{ind}_m(\textmd{ind}_m^{-1}(\cF))=\cF$ for all Ferrers diagrams of size $m$,
$0\leq m\leq k(n-k)$. Therefore, by (\ref{eq:ind1}) and
(\ref{eq:ind1a}) we have
$$\textmd{Ind}_{\cF}(X)=\sum_{t=k(n-k)-(j'-1)}^{k(n-k)}\alpha_{t}q^t +i_{j'}=
\sum_{t=k(n-k)-(j'-1)}^{k(n-k)}\alpha_{t}q^t +i-\sum_{t=k(n-k)-(j'-1)}^{k(n-k)}\alpha_{t}q^t=i.$$
\end{proof}

\subsubsection{Complexity}
We consider the complexity of the computation of the lexicographic
index $\mbox{Ind}_{\cF} (X)$, for $X \in \Gr$, whose Ferrers
diagram is $\cF_X = (\cF_{n-k},...,\cF_2,\cF_1)$.

\begin{theorem}
\label{thm:complexity1} The computation complexity of the
lexicographic index (decoding) in (\ref{eq:Ind_1_p}) is $O(k^{5/2}
(n-k)^{5/2})$ digit operations.
\end{theorem}
\begin{proof}
First, we combine the expressions in (\ref{eq:ind_m}) and
(\ref{Ind1}) to obtain:
\begin{equation*}
\label{eq:Ind_1_p}
\mbox{Ind}_{\cF}(X)=\sum_{i=|\cF_X|+1}^{k(n-k)}p(k,n-k,i)q^{i}+\{x\}
\end{equation*}
\begin{equation}
\label{eq:Ind_1_p} +q^{|\cF_X|}\sum_{j=1}^{n-k}
\sum_{a=\cF_j+1}^{\cF_{j-1}}p(a,n-k-j,|\cF_X|-
\sum_{i=1}^{j-1}\cF_i-a).
\end{equation}
By the recurrence relation of Lemma~\ref{lem: recursion}, we can
compute the table of $p(j,\ell,i)$ for $j\leq k$, $\ell \leq
n-k$, and $i\leq m$ with no more than $mk(n-k)$ additions. By
Lemma~\ref{lem:bound_p} each integer in such addition has $O(
\sqrt{k(n-k)})$ digits. Therefore, the computation of all the
values which are needed from the table takes
${O(k^{5/2}(n-k)^{5/2})}$ digit operations.

The number of additions in~(\ref{eq:Ind_1_p}) is $O(k(n-k))$. Each
integer in this addition has $O(k(n-k))$ digits (as a consequence
of Lemma~\ref{lem:bound_p} and the powers of $q$
in~(\ref{eq:Ind_1_p})). The multiplication by $q^i$ is a shift by
$i$ symbols. Hence, these additions and shifts do not increase the
complexity.
\end{proof}

Similarly, we can prove the following theorem.
\begin{theorem}
\label{thm:complexityD1} The computation complexity of Encoding
Algorithm C is ${O(k^{5/2} (n-k)^{5/2})}$ digit operations.
\end{theorem}

\begin{remark}
\label{rem:large} If $k(n-k)-|\cF_X|$ is a small integer then the
complexity of the computation becomes much smaller than the
complexity given in Theorems~\ref{thm:complexity1}
and~\ref{thm:complexityD1}. For example, if $|\cF_X|=k(n-k)$ then
the complexity of the enumerative decoding is $O(k(n-k))$ since
$\mbox{Ind}_{\cF}(X)=\{x\}$ in~(\ref{eq:Ind_1_p}).
\end{remark}

It is worth to mention in this context that the number of
operations in the algorithms can be made smaller if we will
consider the following two observations~\cite[p. 47]{And84}:

\begin{itemize}
\item If $m_1 < m_2 \leq \frac{\alpha \beta}{2}$ then $p(\alpha, \beta ,m_1)
\leq p(\alpha, \beta ,m_2)$.

\item $p(\alpha,\beta,m)=p(\alpha,\beta,\alpha\beta-m)$ and hence we can assume that
$m\leq \frac{\alpha \beta }{2}$.
\end{itemize}

\subsection{Combination of the Coding Techniques}
\label{subsec:combination}

By Theorems~\ref{thm:complexity2},~\ref{thm:complexity3},
and~\ref{thm:complexity1}, it is clear that the enumerative coding
based on the extended representation is more efficient than the
one based on Ferrers tableaux form. But, for some of
$k$-dimensional subspaces of $\F_q^n$ the enumerative coding based
on Ferrers tableaux form is more efficient than the one based on
the extended representation (see Remark~\ref{rem:large}). This is
the motivation for combining the two methods.

The only disadvantage of the Ferrers tableaux form coding is the
computation of the $\alpha_i$'s and $\mbox{ind}_{|\cF_X|}(\cF_X)$
in Theorem~\ref{thm:Ferrers_index}. This is the reason for its
relatively higher complexity. The advantage of this coding is that
once the values of the $\alpha_i$'s and the value of
$\mbox{ind}_{|\cF_X|}(\cF_X)$ are known, the computation of
$\mbox{Ind}_{\cF} (X)$, for $X \in \Gr$, is immediate. Our
solutions for the computation of the $\alpha_i$'s and
$\mbox{ind}_{|\cF_X|}(\cF_X)$ are relatively not efficient and
this is the main reason why we suggested to use the enumerative
coding based of the RREF and the identifying vector of a subspace.
The only disadvantage of this enumerative coding is the
computation of the Gaussian coefficients in (\ref{eq:Ind2}). It
appears that a combination of the two methods is more efficient
than the efficiency of each one separately. The complexity will
remain $O(nk(n-k)\log n\log\log n)$, but the constant will be
considerably reduced on the average. This can be done if there
won't be any need for the computation of the $\alpha_i$'s and the
computation of $\mbox{ind}_{|\cF_X|}(\cF_X)$ will be efficient.

It was proved in~\cite{KK} that $q^{k(n-k)} <
\begin{footnotesize}\left[\begin{array}{c}n\\k\end{array}\right]_q\end{footnotesize} < 4 q^{k(n-k)}$
for $0 < k < n$. Thus, more than $\frac{1}{4}$ of the
$k$-dimensional subspaces in $\Gr$ have the unique Ferrers diagram
with $k(n-k)$ dots, where the identifying vector consists of $k$
{\it ones} followed by $n-k$ {\it zeroes}. All the codewords of
the Reed-Solomon-like code in~\cite{KK} (or, equivalently, lifted MRD
codes~\cite{SKK08}) have this Ferrers diagram.
Note, that most of the $k$-dimensional subspaces have Ferrers
diagrams with a large number of dots. We will encode/decode these
subspaces by the Ferrers tableaux form coding and the other
subspaces by the extended representation coding. We will choose a
set $S_{\cF}$ with a small number of Ferrers diagrams. $S_{\cF}$
will contain the largest Ferrers diagrams. The Ferrers tableaux
form coding will be applied on these diagrams.


We say that a subspace $X \in \Gr$ is of Type $S_{\cF}$ if $\cF_X
\in S_{\cF}$. In the new order these subspaces are ordered first,
and their internal order is defined as the order of the Ferrers
tableaux forms of Subsection~\ref{subsec:FD order}. The order of the
other subspaces is defined by the order of the extended
representation of Subsection~\ref{subsec:ext order}. We define a new index
function
$\mbox{I}_{\begin{scriptsize}\mbox{comb}\end{scriptsize}}$ as
follows:
\begin{equation}
\label{eq:combination}
\mbox{I}_{\begin{scriptsize}\mbox{comb}\end{scriptsize}}(X)=
\left\{
\begin{array}{cc}
\mbox{Ind}_{\cF}(X) & \cF_X\in S_{\cF}\\
\mbox{I}_{\begin{tiny}\mbox{EXT}\end{tiny}}(X)+\Delta_X (S_{\cF})
& \textrm{otherwise}\end{array},\right.
\end{equation}
where $\Delta_X(S_{\cF})$ is the number of subspaces of Type
$S_{\cF}$, which are lexicographically succeeding $X$ by the
extended representation ordering. These $\Delta_X(S_{\cF})$
subspaces are preceding $X$ in the ordering induced by combining
the two coding methods.

We demonstrate the method for the simple case where $S_{\cF}$
consists of the unique Ferrers diagram with $k(n-k)$ dots.

\begin{lemma}
\label{lem:Delta} Let $S_{\cF}$ be a set of Ferrers diagrams,
embedded in a $k \times (n-k)$ box, which contains only one
Ferrers diagram, the unique one with $k(n-k)$ dots. Let $X \in
\Gr$, $X \not\in S_{\cF}$, $\emph{RE}(X)=(X_n,\ldots,X_1)$, and
let $\ell$, $0\leq \ell \leq n-k-1$, be the number of consecutive
\textit{zeroes} before the first \textit{one} (from the right) in
the identifying vector $v(X)$. Then $\Delta_X(S_{\cF})=
\sum_{i=1}^{\ell}(q^k-1-\{X_i\})q^{k(n-k-i)}$.
\end{lemma}

\begin{proof} If $\ell=0$ then $v(X)_1=1$ and hence
there are no subspaces of Type $S_{\cF}$ which are
lexicographically succeeding $X$ and hence $\Delta_X(S_{\cF})=0$.
For $1\leq \ell \leq n-k-1,$ let $X_1,...,X_\ell$ be the first
$\ell$ columns of $\mbox{RE}(X)$. All the subspaces of Type
$S_{\cF}$ in which the value of the first column is greater than
$\{ X_1 \}$, are lexicographically succeeding $X$. There are
$(q^k-1-\{X_1\})q^{k(n-k-1)}$ such subspaces. All the subspaces of
Type $S_{\cF}$ in which the first $i-1$ columns, $2\leq i\leq
n-k-1$, are equal to the first $i-1$ columns of $\mbox{RE}(X)$,
and the value of the $i$th column is greater than $\{X_i\}$, are
lexicographically succeeding $X$. There are
$(q^k-1-\{X_{i}\})q^{k(n-k-i)}$ such subspaces. Therefore, there
are $\sum_{i=1}^{\ell}(q^k-1-\{X_i\})q^{k(n-k-i)}$ subspaces of
Type $S_{\cF}$ which are lexicographically succeeding $X$ by the
extended representation ordering.
\end{proof}

\begin{example}
Let $X$ be the subspace of Example~\ref{exm:X_0}. By
Example~\ref{exm:X_0} we have
$\emph{I}_{\begin{tiny}\emph{EXT}\end{tiny}}(X)=928$, and by
Lemma~\ref{lem:Delta} we have
$\Delta_X(S_{\cF})=(2^3-1-5)2^{3\cdot 2}=2^7$. Hence,
$\emph{I}_{\begin{scriptsize}\emph{comb}\end{scriptsize}}
(X)=\emph{I}_{\begin{tiny}\emph{EXT}\end{tiny}}(X)+\Delta_X(S_{\cF})=928+128=1056$.
\end{example}
\vspace{0.3cm}

Now, suppose that an index $0 \leq i <
\begin{footnotesize}\left[\begin{array}{c}n\\k\end{array}\right]_q\end{footnotesize}$ is given.
Encoding Algorithm D finds a
subspace $X\in \Gr$ such that
$\mbox{I}_{\begin{scriptsize}\mbox{comb}\end{scriptsize}} (X)=i$,
where $S_{\cF}$ consists of the unique Ferrers diagram with
$k(n-k)$ dots.

\vspace{0.3cm}

\begin{figure}[hbt]
\centering
\begin{algorithm}
\noindent \textit{Encoding Algorithm D:}
\begin{itemize}
\item if $i<q^{k(n-k)}$ then apply Encoding Algorithm C on $i$ and stop;

\item otherwise set $i_{0}=i$.

\end{itemize}
\vspace{0.2cm}

\noindent For $j=1,2,...,n$ do

\noindent begin

$w_{j-1}=\sum_{i=1}^{j-1}v(X)_i$.
\begin{itemize}
\item
If $w_{j-1}= k$ then set $v_j=v(X)_{j}=0$, $w_j=w_{j-1}$,
$X_{j}=\{0\}_q$, $i_j=i_{j-1}$;
\item otherwise: set\\
\noindent \[ A_j = \left\{
\begin{array}{cc} q^{k-w_{j-1}}
\begin{footnotesize}\left[\begin{array}{c}
n-j\\k-w_{j-1}\end{array}\right]_{q}\end{footnotesize}
& \textrm{if }w_{j-1}\neq 0\\
q^{k-w_{j-1}}
\begin{footnotesize}\left[\begin{array}{c}
n-j\\k-w_{j-1}\end{array}\right]_{q}\end{footnotesize}-
q^{k(n-k-j+1)} & \textrm{if }w_{j-1}= 0\end{array}\right.\]

\[ B_j =\left\{ \begin{array}{cc}
\begin{footnotesize}\left[\begin{array}{c}
n-j\\k-w_{j-1}\end{array}\right]_{q}\end{footnotesize}
& \textrm{if }w_{j-1}\neq 0\\
\begin{footnotesize}\left[\begin{array}{c}
n-j\\k-w_{j-1}\end{array}\right]_{q}\end{footnotesize}-
q^{k(n-k-j)} & \textrm{if }w_{j-1} = 0\end{array}\right.\]

\begin{itemize}
\item if $i_{j-1}\geq A_j$ then set $v_j=v(X)_{j}=1$, $w_j=w_{j-1}$+1,
$X_{j}=\{q^{w_{j-1}}\}_{q}$, $i_{j}=i_{j-1}-A_j$;

\item otherwise set $val=\left\lfloor i_{j-1}/B_j\right\rfloor $,
$v_j=v(X)_{j}=0$, $w_j=w_{j-1}$, $X_{j}=\left\{ val*q^{w_{j-1}}\right\} _{q}$,
$i_{j}=i_{j-1}-val*B_j .$
\end{itemize}
\end{itemize}
\noindent end \{begin\}

\end{algorithm}
\end{figure}

The correctness of the Encoding Algorithm D
follows from Lemma~\ref{lem:Delta} and the correctness of Encoding Algorithm A
and Encoding Algorithm C.

\begin{remark} If the size of $S_{\cF}$ is greater than $1$ then
the calculations of $\Delta_X(S_{\cF})$ should be changed. It
becomes more and more mathematically complicated to find the
formula of $\Delta_X(S_{\cF})$ as the size of $S_{\cF}$ is larger.
\end{remark}

\section{Constant Dimension Lexicodes}
\label{sec:lexicodes}

\textit{Lexicographic codes}, or \textit{lexicodes}, are greedily
generated error-correcting codes which were first developed by
Levenshtein~\cite{Lev60}, and rediscovered by Conway and
Sloane~\cite{CoSl86}. The construction  of a lexicode with a
minimum distance $d$ starts with the set $ \mathcal {S}= \{S_0\}$,
where $S_0$ is the first element in a lexicographic order, and
greedily adds the lexicographically first element whose distance
from all the elements of $\mathcal {S}$ is at least $d$. In the
Hamming space, the lexicodes include the optimal codes, such as
the Hamming codes and the Golay codes.

In this section we consider lexicodes in the Grassmannian.
It turns out that
the lexicodes which were formed
based on the Ferrers tableaux form representation and related order
of $\Gr$ are always larger than the ones formed based on the extended representation
and hence we consider only these codes.
We describe a search method for constant dimension lexicodes
based on their multilevel structure. Some of the lexicodes obtained by
this search are the largest known constant dimension codes with
their parameters. We also describe several ideas to make this search
more efficient.

\vspace{0.3cm}
\subsection{Analysis of Constant Dimension Codes}
\label{subsec:analysis}

In this subsection we introduce some properties of constant dimension
codes which will help us to simplify the search for lexicodes.
First, we consider the multilevel structure of a code in the
Grassmannian.

In Chapter~\ref{ch:representation and distance} we mentioned that
all the  binary vectors of the length $n$ and weight $k$
can be considered as  the identifying vectors of all the subspaces
in  $\Gr$. These  $\binom{n}{k}$ vectors partition  $\Gr$
into the $\binom{n}{k}$ different classes, where each class
consists of all subspaces in  $\Gr$ with the same identifying
vector.
According to this  partition all the constant dimension
codes have a multilevel structure: we can partition  all
the codewords  of a code into different classes (sub-codes),
where all the codewords in each such a class
have the same identifying vector. Therefore, the first level of this
structure is the set of different identifying vectors, and the second
level is the subspaces corresponding to these vectors. The multilevel construction
presented in Section~\ref{sec:multilevel construction} is based on this approach.

\vspace{0.3cm}
Let $\C\subseteq \Gr$ be a constant dimension code, and let
$\{v_1,v_2,\ldots,v_{t}\}$ be all the different identifying
vectors of the codewords in $\C$. Let $\{\C_1,\C_2,\ldots,\C_t\}$
be the partition of $\C$ into $t$ sub-codes induced by these $t$
identifying vectors, i.e., $v(X)=v_i$, for each $X\in \C_i$,
$1\leq i\leq t$.

\begin{remark} We can choose any constant weight code $ \bf{C}$  with  minimum
Hamming distance~$d$ to be the set of identifying vectors. If for
each identifying vector $v\in \bf{C}$  we have a sub-code
$\C_v$ for which $v(X)=v$  for each $X\in \C_v$, and
$d_S(\C_v)=d$, then by Corollary \ref{cor:distance} we obtain a
constant dimension code with the same minimum distance $d$. If for
all such identifying vectors we construct the maximum size
constant dimension sub-codes (lifted Ferrers diagram rank-metric codes)
then we obtain the multilevel
construction  which was described in
Chapter~\ref{ch:bounds and constr}.
\end{remark}

For $X\in \Gr$, we define the $k\times (n-k)$ matrix $R(X)$ as the
sub-matrix of $\mbox{RE}(X)$ with the columns which are indexed
by\textit{ zeroes} of $v(X)$.
\begin{example}Let $X$ be a subspace in $\mathcal{G}_2(7,3)$ given by
\begin{align*}
\emph{RE}(X) =\left( \begin{array}{ccccccc}
1 & 0 & 0 & 0 & 1 & 1 & 0 \\
0 & 0 & 1 & 0 & 1 & 0 & 1 \\
0 & 0 & 0 & 1 & 0 & 1 & 1
\end{array}
\right), \textmd{ then  }R(X) =\left( \begin{array}{cccc}
0 &  1 & 1 & 0 \\
0 &  1 & 0 & 1 \\
0 &  0 & 1 & 1
\end{array}
\right).
\end{align*}
\end{example}

\vspace{0.5cm}
 By Corollary~\ref{cor:distanceSameID}, for any two codewords $X,Y\in
\C_i$, where $\C_i \subseteq \C$, $1\leq i\leq t$, the subspace distance
between $X$ and $Y$ can be calculated in terms of rank distance,
i.e.,
$$d_S(X,Y)=2\rank(\mbox{RE}(X)-\mbox{RE}(Y))=2d_R(R(X),R(Y)).$$


For each sub-code $\C_i \subseteq \C$, $1\leq i\leq t$, we define
a Ferrers diagram rank-metric code
$$ R(\C_i)\,\ \deff\ \,\{R(X):X\in \C_i\}.$$
Note, that such a code is obtained by the inverse operation to the
\textit{lifting} operation, defined in Chapter~\ref{ch:Introduction}. Thus,
$R(\C_i)$ will be called the \textit{unlifted code} of the
sub-code $\C_i$.

We define the subspace distance between two sub-codes $\C_i$, $\C_j$
of $\C$, $1\leq i\neq j \leq t$ as follows:
$$d_S(\C_i,\C_j) =\min\{d_S(X,Y):X\in \C_i, Y\in \C_j\}.$$
By Corollary~\ref{cor:distance},
$$d_S(\C_i,\C_j)\geq d_H(v_i,v_j).$$

The following lemma shows a case in which the last inequality becomes an
equality.
\begin{lemma}
Let $\C_i$  and $\C_j$ be two different sub-codes of $\C\subseteq
\Gr$, each one contains the subspace whose RREF is the
corresponding column permutation of the matrix $(I_k 0_{k\times
(n-k)})$,  where $I_k$ denotes the $k\times k$ identity matrix and
$0_{a\times b}$  denotes an $a\times b$ all-zero matrix. Then
$$ d_S(\C_i,\C_j)=d_H(v_i,v_j).$$
\end{lemma}

\begin{proof} Let $X\in \C_i$ and $Y\in \C_j$ be subspaces whose
RREF equal to some column permutations of the matrix
$(I_k 0_{k\times (n-k)})$. It is easy to verify that
\begin{align}
\rank
\left(\begin{array}{c}
\textmd{RE}(X)\\\textmd{RE}(Y)\end{array}\right)= \rank \left(\begin{array}{c}
\textmd{RE}(X)\\Y_{\mu^C}\\Y_\mu\end{array}\right)=\rank \left(\begin{array}{c}
\textmd{RE}(X)\\Y_{\mu^C}\end{array}\right),
\end{align}
where $\mu$ and $Y_\mu$ are defined in Section~\ref{sec:distance}.

Clearly,  $\rank(Y_{\mu^C})=\frac{d_H(v_i,v_j)}{2}$, and hence,
$\text{rank}(\textmd{RE}(X)*\textmd{RE}(Y))=k+\frac{d_H(v_i,v_j)}{2}$.
By~(\ref{distance-rank}), $d_S(X,Y)=2\text{rank}(\textmd{RE}(X)*\textmd{RE}(Y))-2k=2k+d_H(v_i,v_j)-2k=
d_H(v_i,v_j)$, i.e., $d_S(\C_i,\C_j)\leq d_H(v_i,v_j)$.
By Corollary~\ref{cor:distance}, $ d_S(\C_i,\C_j)\geq d_H(v_i,v_j)$,
and hence, $d_S(\C_i,\C_j)=d_H(v_i,v_j)$.
\end{proof}
\begin{corollary}
\label{cor:linearity} Let $v_i$ and $v_j$ be two identifying
vectors of codewords in an $(n,M,d,k)_q$ code~$\C$. If $d_H(v_i,
v_j) < d$ then at least one of the corresponding sub-codes, $\C_i$
and $\C_j$,  does not contain the subspace with RREF which is a
column permutation of the matrix $(I_k 0_{k\times(n-k)})$. In
other words, the corresponding unlifted code is not linear since
it does not contain the all-zero  codeword.
\end{corollary}

Assume that we can add codewords to a code $\C$, $d_S(\C)=d$,
constructed by the multilevel construction of Chapter~\ref{ch:bounds and constr}
with a maximal constant weight
code (for the identifying vectors) $\bf{C}$,
$d_H(\bf{C})$$=d$. Corollary~\ref{cor:linearity} implies that
any corresponding unlifted Ferrers diagram rank-metric code of any
new identifying vector will be nonlinear.

\vspace{0.1cm}
The next two lemmas reduce the search  domain for constant dimension lexicodes.
\begin{lemma}
\label{lm:first_id} Let $\C$ be an $(n,M,d=2\delta,k)_q$ constant
dimension code. Let $\C_1\subseteq\C$, $v(X)=v_1=11\ldots
100\ldots 0$  for each $X\in\C_1$, be a sub-code for which
$R(\C_1)$ attains the  upper  bound of
Theorem~\ref{thm:upper_rank}, i.e., $|\C_1| = | R(\C_1)|=
q^{(k-\delta+1)(n-k)}$. Then there is no codeword $Y$ in $\C$ such
that $d_H(v(Y),v_1)<d$.
\end{lemma}

\begin{proof}
Let $\C$ be a given $(n,M,d=2\delta,k)_q$ constant dimension code.
Since the minimum distance of the code is $d$, the intersection of any
two subspaces in $\C$ is at most of dimension $k-\frac{d}{2}=k-\delta$.
Therefore, a subspace of dimension $k-\delta+1$
can be contained in at most  one codeword of $\C$.

We define the following set of subspaces:
\[A=\{X\in \cG_q(n,k-\delta+1): \;supp(v(X))\subseteq supp(v_1)\},
\]
where $supp(v)$ is as the set of nonzero entries in $v$. Each
codeword of the sub-code $\C_1$ contains $\begin
{footnotesize}\sbinomq {k}{k-\delta+1}\end{footnotesize}$
subspaces of dimension $k-\delta+1$, and all subspaces of
dimension $k-\delta+1$ which are contained in codewords of $\C_1$
are in $A$. Since $|\C_1| = q^{(k-\delta+1)(n-k)}$, it follows
that $\C_1$ contains $q^{(k-\delta+1)(n-k)}\cdot \begin
{footnotesize}\sbinomq {k}{k-\delta+1}\end{footnotesize}$
subspaces of~$A$.

Now we calculate the size of $A$. First we observe that
\[ A= \{X\in \cG_q(n,k-\delta+1): v(X)=ab,\; |a|=k,\; |b|=n-k,\; w(a)=k-\delta+1,\; w(b)=0\},
\]
where $|v|$ and $w(v)$ are the length and the weight of a vector
$v$, respectively. Thus $\textmd{EF}(v(X))$ of each $v(X)=ab$, such that
$X\in A$, has the form
\begin{equation}\textmd{EF}(v(X))=\left[\textmd{EF}(a)
\begin{footnotesize}
\begin{array}{cccc}
\bullet & \bullet &\ldots & \bullet  \\
\bullet & \bullet & \ldots &\bullet  \\
\bullet & \bullet &\ldots  &\bullet  \\
\end{array}
\end{footnotesize}\\\right].\label{EF_of_A}
\end{equation}
The number of dots in (\ref{EF_of_A}) is $(k-\delta+1)(n-k)$, and
the size of the following set $$\{\textmd{EF}(a):|a|=k,\;
w(a)=k-\delta+1\}$$ is $\begin{footnotesize}\sbinomq
{k}{k-\delta+1}\end{footnotesize}$. Therefore, $|A|= \begin
{footnotesize}\sbinomq {k}{k-\delta+1}\end{footnotesize}\cdot
q^{(k-\delta+1)(n-k)}$. Hence, each subspace of~$A$ is contained
in some codeword from $\C_1$. A subspace $Y\in \Gr$ with
$d_H(v(Y),v_1)=2\delta-2i$, $1\leq i\leq \delta-1$, contains some
subspaces of  $A$, and therefore, $Y\notin\C$.
\end{proof}

\begin{lemma}
\label{lm:second_id} Let $~\C$  be an $(n,M,d=2\delta,k)_q$
constant dimension code, where $\delta-1\leq k-\delta$. Let $\C_2$
be a sub-code of $\C$ which corresponds to the identifying vector
$v_2 = abfg$, where $a=\underset{k-\delta}{\underbrace{11\ldots
1}}$, $b=\underset{\delta}{\underbrace{00\ldots 0}}$,
$f=\underset{\delta}{\underbrace{11\ldots 1}}$, and
$g=\underset{n-k-\delta}{\underbrace{00\ldots 0}}$. Assume further
that $R(\C_2)$ attains the  upper bound of
Theorem~\ref{thm:upper_rank}, i.e.,
$|\C_2|=|R(\C_2)|=q^{(k-\delta+1)(n-k)-\delta^2}$. Then there is
no codeword $Y\in \C$  with $v(Y)=a'b'fg'$, $|a'b'|=k$,
$|g'|=n-k-\delta$, such that $d_H(v(Y),v_2)<d$.
\end{lemma}

\begin{proof}
Similarly to the proof of Lemma~\ref{lm:first_id},
we define the following set of subspaces:
\[B=\{X\in \cG_q(n,k-\delta+1):
v(X)=a''bfg \; \textrm{with}\; |a''|=k-\delta, \; w(a'')=k-2\delta+1 \}.
\]
As in the previous proof, we can see that $\C_2$ contains
$q^{(k-\delta+1)(n-k)-\delta^2}\cdot \begin {footnotesize}\sbinomq
{k-\delta}{k-2\delta+1}\end{footnotesize}$ subspaces of $B$. In
addition, $|B|= \begin {footnotesize}\sbinomq
{k-\delta}{k-2\delta+1}\end{footnotesize}\cdot q^{(k-2
\delta+1)\delta+(k-\delta+1)(n-k-\delta)}=\begin
{footnotesize}\sbinomq {k-\delta}{k-2\delta+1}\end{footnotesize}
\cdot q^{(k-\delta+1)(n-k)-\delta^2}$. Thus each subspace in $B$
is contained in some codeword from $\C_2$. A~subspace $Y\in \Gr$,
such that $v(Y)=a'b'fg'$ ($|a'b'|=k$, $|g'|=n-k-\delta$), with
$d_H(v(Y),v_2)=2\delta-2i$, $1\leq i\leq \delta-1$, contains some
subspaces of $B$, and therefore, $Y\notin \C$.
\end{proof}

\subsection{Search for Constant Dimension Lexicodes}
\label{subsec:search}

In this section we describe our search method for constant
dimension lexicodes, and present some  resulting codes which are
the largest currently known constant dimension codes for their
parameters.

To search for large constant dimension code we use the multilevel
structure of such codes, described in the previous subsection. First,
we order the set of all binary words of length $n$ and weight $k$
by an appropriate order. The words in this order are the
candidates to be the identifying vectors of the final code. In
each step of the construction we have the current code $\C$ and
the set of subspaces not examined yet. For each candidate for an
identifying vector $v$ taken by the given order, we search for a
sub-code in the following way: for each subspace $X$ (according to
the lexicographic order of subspaces associated with $v$) with the
given Ferrers diagram we calculate the distance between $X$ and
$\C$, and add $X$ to $\C$ if this distance is at least $d$. By
Theorem~\ref{thm:distance} and Corollary~\ref{cor:distance} it
follows that in this process, for some subspaces it is enough only
to calculate the Hamming distance between the identifying vectors
in order to determine a lower bound on the subspace distance. In
other words, when we examine a new subspace to be inserted into
the lexicode, we first calculate the Hamming distance between its
identifying vector and the identifying vector of a codeword, and
only if this distance is smaller than $d$, we calculate the rank
of the corresponding matrix, (see (\ref{subspace_distance})).
Moreover, by the multilevel structure of a code, we need only to
examine the Hamming distance between the identifying vectors of
representatives of sub-codes, say the first codeword in  each
sub-code. This approach will speed up the process of the code
generation.

This construction of constant dimension lexicodes is based on the
Ferrers tableaux form ordering of the Grassmannian.
Note that in this construction we order the
identifying vectors by the sizes of corresponding Ferrers
diagrams. The motivation is that usually a larger diagram
contributes more codewords than a smaller one.

\begin{example}
\label{exm:8-4-4}
Table \ref{tab:lexicode8-4-4} shows  the identifying vectors and the sizes
of corresponding sub-codes in the $(8,4605,4,4)_2$ lexicode, denoted by $\C^{lex}$,
and the $(8,4573,4,4)_2$  code, denoted by $\C^{ML}$, obtained by the multilevel
construction considered in Chapter~\ref{ch:bounds and constr}.

\begin{table}[h]
\centering \caption{ $\C^{lex}$  vs. $\C^{ML}$ in $\mathcal G_2(8,4)$ with
$d_S=4$}\label{tab:lexicode8-4-4}
\begin{tabular}{|c|c|c|c|}
\hline \multicolumn{1}{|c|}{$i$} &
 \multicolumn{1}{|c|}{id.vector $v_i$} &
\multicolumn{1}{c|}{size of $\C^{lex}_i$} &
\multicolumn{1}{c|}{size of $\C^{ML}_i$} \tabularnewline
\hline\hline 1&11110000 & 4096  &4096 \tabularnewline\hline 2&
11001100 &256  & 256 \tabularnewline\hline 3& 10101010 & 64 & 64
\tabularnewline\hline 4& 10011010 & 16 & -- \tabularnewline\hline
5& 10100110 & 16 & -- \tabularnewline\hline 6& 00111100 & 16 & 16
\tabularnewline\hline 7& 01011010 & 16 & 16 \tabularnewline\hline
8& 01100110 & 16 & 16 \tabularnewline\hline 9& 10010110 & 16 & 16
\tabularnewline\hline 10&01101001 & 32 & 32 \tabularnewline\hline
11&10011001 & 16 & 16 \tabularnewline\hline 12&10100101 & 16 & 16
\tabularnewline\hline 13&11000011 & 16 & 16 \tabularnewline\hline
14&01010101 & 8 & 8 \tabularnewline\hline 15&00110011 & 4 & 4
\tabularnewline\hline 16&00001111 & 1 & 1 \tabularnewline\hline
\end{tabular}
\end{table}

We can see that these two codes have the same identifying vectors,
except for two vectors $10011010$ and $10100110$  in the lexicode
$\C^{lex}$ which form the difference in the size of these two
codes. In  addition, there are several  sub-codes of $\C^{lex}$
for which the corresponding unlifted codes  are nonlinear:
$\C^{lex}_4$, $\C^{lex}_5$, $\C^{lex}_7$, $\C^{lex}_8$,
$\C^{lex}_{11}$, and $\C^{lex}_{12}$. However, all these  unlifted
codes are  cosets of linear codes.
\end{example}
\vspace{0.3cm}

In general, not all unlifted codes of lexicodes based on the
Ferrers tableaux form representation are linear or cosets of some
linear codes. However, if  we construct a binary constant
dimension lexicode with only one identifying vector, the unlifted
code is always linear. This phenomenon can be explained as an
immediate consequence from the main theorem in~\cite{Zan97}.
However, it does not explain why some of unlifted codes in
Example~\ref{exm:8-4-4} are cosets of linear codes, and why
$\C_9^{lex}$ is linear ($d_H(v_5,v_9)<4$)?


Based on Theorem~\ref{thm:distance}, Lemma~
\ref{lm:first_id}, and Lemma~\ref{lm:second_id}, we suggest  an improved search of
a constant dimension  $(n,M,d,k)_q$ code,  which
will be called a \textit{lexicode with a seed}.

In the first step we construct a  maximal sub-code $\C_1$  which
corresponds to the identifying vector
$\underset{k}{\underbrace{11\ldots 1}}
\underset{n-k}{\underbrace{00\ldots 0}}$. This sub-code
corresponds to the largest Ferrers diagram. In this step we can
take any known $[k\times(n-k), (n-k)(k-\frac{d}{2}+1),
\frac{d}{2}]$ MRD code (e.g.~\cite{Gab85}) and consider its
codewords as the unlifted codewords  of
$\C_1$.

In the second step we construct a  sub-code $\C_2$  which
corresponds to the  identifying vector $\underset{k-\delta}
{\underbrace{11\ldots 1}}\underset{\delta}{\underbrace{00\ldots
0}} \underset{\delta}{\underbrace{11\ldots 1}}
\underset{n-k-\delta} {\underbrace{00\ldots 0}}$. According to
Lemma~\ref{lm:first_id}, we cannot use identifying vectors with
larger Ferrers diagrams (except for the identifying vector
$\underset{k}{\underbrace{11\ldots 1}}
\underset{n-k}{\underbrace{00\ldots 0}}$ already used). If there
exists a Ferrers diagram MRD code with the corresponding
parameters, we can take any known construction of such code (see
in Subsection~\ref{subsec:FD rank-metric}) and build from it the corresponding sub-code. If
a code which attains the bound of Theorem~\ref{thm:upper_rank} is
not known, we take the largest known Ferrers diagram rank-metric
code with the required parameters.

In the third step we construct the other sub-codes, according to the
lexicographic order based on the Ferrers tableaux form representation.
We first calculate the Hamming distance between the identifying vectors and
examine the subspace distance only of subspaces which are not pruned out by
Lemmas \ref{lm:first_id} and \ref{lm:second_id}.

\begin{example}
\label{exm:10-5-6}

Let $n=10$, $k=5$, $d=6$, and $q=2$.
By the construction of a lexicode with a seed we obtain  a constant
dimension code of size $32890$. (A code of
size $32841$ was obtained by the multilevel construction).
\end{example}

\begin{example}
Let  $n=7$, $k=3$, $d=4$, and $q=3$. By the construction a lexicode
with a seed we obtain  a constant
dimension code of size $6691$. This code attains the upper bound
of Theorem~\ref{trm:upper bound from Steiner Structure}.
\end{example}

We introduce now  a variant of the construction of a lexicode with
a seed. As a seed we take a constant dimension code obtained by
the multilevel construction of Chapter~\ref{ch:bounds and constr}
and try to add some more
codewords using the lexicode construction. Similarly, we can take
as a seed any subset of codewords obtained by any given
construction and to continue by applying the lexicode with a seed
construction.


\begin{example}
\label{ex:4-4-9} Let $n=9$, $k=d=4$, and $q=2$. Let $\C$ be a
$(9,2^{15}+2^{11}+2^7,4,4)_2$ code obtained as follows. We take
three codes of sizes $2^{15}$, $2^{11}$, and $2^7$, corresponding
to identifying vectors $111100000$, $110011000$, and $110000110$,
respectively, and then continue by applying the lexicode with a
seed construction. For the identifying vector $111100000$ we can
take as the unlifted code, any code which attains the bound of
Theorem~\ref{thm:upper_rank}. To generate the codes for the last
two identifying vectors with the corresponding unlifted codes
(which attains the bound of Theorem~\ref{thm:upper_rank}), we
permute the order of entries in the Ferrers diagrams and apply the
lexicode construction.
 The Ferrers diagrams which correspond to the
identifying vector $110011000$  and $110000110$ are
$$
\begin{array}{ccccc}
 \bullet & \bullet & \bullet & \bullet & \bullet\\
 \bullet & \bullet & \bullet & \bullet & \bullet \\
 & & \bullet & \bullet  & \bullet  \\
 &  & \bullet & \bullet & \bullet  \\
\end{array},\;\;\;
\begin{array}{ccccc}
 \bullet & \bullet & \bullet & \bullet & \bullet\\
 \bullet & \bullet & \bullet & \bullet & \bullet \\
 & & & & \bullet  \\
 &  & & & \bullet  \\
\end{array},
$$
respectively. The coordinates' order of their entries (defined in Subsection~\ref{subsec:FD order}) is:
$$
\begin{array}{ccccc}
15 & 13 & 9 & 5& 1\\
16 & 14 & 10 & 6& 2\\
 & & 11 & 7 & 3  \\
 &  & 12 & 8 & 4 \\
\end{array},\;\;\;
\begin{array}{ccccc}
11 & 9 & 7 & 5& 1\\
12 & 10 & 8 & 6& 2\\
 & &  &  & 3  \\
 &  &  &  & 4 \\
\end{array},
$$
respectively. The order of the coordinates that we use to form an MRD code (lexicode) is
$$\begin{array}{ccccc}
11 & 7  & 5 & 3 & 1\\
15 & 12 & 8 & 2 & 4\\
 & &     13 & 9 & 6 \\
 & &     16 &14 & 10 \\
\end{array},\;\;
\begin{array}{ccccc}
9 & 7  & 5 & 3 & 1\\
11 & 10 & 8 & 2 & 4\\
 & &      &  & 6 \\
 & &      & & 12 \\
\end{array}.
$$
As a result, we obtain a code of  size $37649$ which is the
largest known constant dimension code with these parameters.
\end{example}

\begin{remark}
The decoding of a code $\C$ constructed by the search method
depends on the nature of the seed code ($\C_s$) and the size of
rest of the code ($\C_r = \C \setminus \C_s$) produced by the
greedy search. For example, if the identifying vectors of $\C_s$
form a constant weight code with minimum distance $d$, the related
rank-metric codes have an efficient decoding algorithm, and $\C_r$
is relatively of small size then we can use the decoding algorithm
mentioned in Chapter~\ref{ch:bounds and constr} to decode $\C_s$.
For the decoding of
the small code $\C_r$ we will use a look-up table.
\end{remark}
\begin{remark}
It should be noted that the improvements yielded by the search
method are not dramatic. Nevertheless, it is interesting to
realize that simple greedy algorithm can be effective in enlarging
a code obtained by a mathematical method.
\end{remark}

\chapter{Conclusion and Open Problems}
\label{ch:conclusion}

The main purpose of this work was to investigate  codes in the
Grassmannian and in the projective space,
to present new bounds and constrictions for such codes and
to provide efficient coding techniques.

 Different representations
of subspaces in $\Ps$ were shown in Chapter~\ref{ch:representation and distance}.
The representations of a subspace by its Ferrers tableaux form, and by its
identifying vector and the matrix in reduced row echelon form play
an essential role in our constructions of error-correcting codes in $\Ps$.
These representations are also important for distance computation between  two subspaces,
and for an enumerative coding in $\Gr$.

Lifted MRD codes were considered in Chapter~\ref{ch:designs}. It was proved
that the codewords of such codes form the blocks of transversal designs in sets, and also
blocks of  subspace transversal designs.
This work is the first to present the connections between
codes in the Grassmannian space and codes in the Hamming space:
first, by showing the relationship between the Hamming distance
of identifying vectors and subspace distance of related subspaces;
and second, by using an incidence matrix of a transversal design
obtained from a lifted MRD code as a parity-check matrix for
a linear code in the Hamming space.

New bounds and constructions for error-correcting codes in the
Grassmannian and in the projective space
were given in Chapter~\ref{ch:bounds and constr}. A multilevel
coding approach to construct codes  was presented. The method
makes usage of four tools, an appropriate constant weight code,
the reduced row echelon form of a linear subspace, the Ferrers
diagram related to this reduced row echelon form, and rank-metric codes
related to the Ferrers diagram. The constructed codes by this method
are usually the best known today for most parameters (except for $k=3$;  $k=4$ with $n=8$;
and the parameters of several constant dimension lexicodes).
The structure of the transversal designs obtained from lifted MRD codes
is used to obtain upper
bounds on the sizes of constant dimension codes which contain the
lifted MRD code. Codes which attain these bounds
are constructed for $k=3$, and $k=4$ with $n=8$.
These codes are the largest known codes with their parameters.
 The puncturing
operation on codes in the projective space was defined. This
operation was applied to obtain punctured codes from our constant dimension
codes. These punctured codes are considerably larger than codes
constructed by the multilevel method.

Three methods of enumerative coding for the Grassmannian were
presented in Chapter~\ref{ch:Enum and Lexi}. The first is based on
the representation of subspaces
by their identifying vector and their reduced row echelon form.
The second is based on the Ferrers tableaux form representation of
subspaces. The complexity of the first method is superior on the
complexity of the second one. The third method is a combination of
the first two. On average it reduces the constant in the first
term of the complexity compared to the complexity of the first
method.
The enumerative coding is based on a lexicographic order for the Grassmannian
related to a specific representation. This order can be used to
form constant dimension lexicodes. Several ideas to make the search
for such lexicodes more efficient were described.
Some of the codes obtained by
this search are the largest known
error-correcting codes in the Grassmannian with their parameters.

\vspace{0.5cm}

We conclude with a list of open problems for future research.
\begin{enumerate}

\item  Find new connections between constant dimension codes and
($q$-analogs of) combinatorial designs.

\item Which other properties have LDPC codes obtained from lifted
MRD codes? We would like to see improvements on the bounds given
in Section~\ref{sec:Linear codes}. In addition, we would like to
know the performance of these codes with various decoding
algorithms~\cite{DPTRU02,Ric03}.

\item Is there a specification for the best constant weight code
which should be taken for our multilevel approach? Our discussion
on the Hamming code and lexicodes is a first step in this
direction.

\item Is the upper bound of Theorem~\ref{thm:upper_rank} attained
for all parameters? Our constructions for optimal Ferrers diagram
rank-metric codes suggest that the answer is positive.

\item What are the general upper bounds on a size of
$(n,M,2\delta,k)_q$ code which contains a lifted MRD code?

\item Are the upper bounds of Theorems~\ref{trm:upper bound from
Steiner Structure} and~\ref{trm:bound 2k-k} and related general
bounds for other parameters (as we just asked) are attained for
all parameters?

\item Can the codes constructed in Constructions I, II, and III be
used, in a recursive method, to obtain new bounds on
$\cA_q(n,d,k)$ for larger $n$?


\item The upper bound on $\cA_2
(7,4,3)$ is 381. If a code which attains the bound exists, it
contains only 128 codewords of a rank-metric code which should be
lifted (compared to 256 codewords of an MRD code). Assume a
maximal linear rank-metric code of size 128 is taken and lifted.
To which size we can extend the obtained constant dimension code?
Similarly, the upper bound on $\cA_2 (8,4,4)$ is $17 \cdot 381 =
6477$. If a code which attains the bound exists it contains $17
\cdot 128$ codewords of a rank-metric code which should be lifted.
Clearly, such a rank-metric code cannot be linear.

\item One of the main research problems is to improve the lower
bounds on $\cA_q(n,d,k)$, with codes which do not contain the
lifted MRD codes. Only such codes can close the gap between the
lower and the upper bounds on $\cA_q(n,d,k)$ for small $q$ and
small $d$ (e.g. the cyclic codes for $k=3$~\cite{EV08,KoKu08}).
\end{enumerate}

\addcontentsline{toc}{chapter}{Bibliography}
\bibliographystyle{plain}

\begin{thebibliography}{10}

\bibitem{AAK01}
R.\ Ahlswede, H.\,K.\ Aydinian, and L.\,H.\ Khachatrian,
``On perfect codes and related concepts,'' \emph{Designs, Codes, Crypt.}, vol.
22, pp. 221--237, 2001.

\bibitem{ACLY00}
R. Ahlswede, N. Cai, S.-Y. R. Li and R. W. Yeung,
``Network Information Flow'',
\emph{IEEE Trans. on Inform. Theory}, vol. 46, no. 4, pp. 1204-1216, July 2000.


\bibitem{AHKXL04}
B. Ammar, B. Honary, Y. Kou, J. Xu, and S. Lin, ''Construction
of low-density parity-check codes based on balanced
incomplete block designs,'' \emph{IEEE Trans. Inform. Theory},
vol. 50, no. 6, pp. 1257-1568, Jun. 2004.

\bibitem{And97}
I. Anderson, {\emph Combinatorial Designs and Tournaments},
Clarendon press, Oxford, 1997.

\bibitem{And84}
G. E. Andrews, {\em The Theory of Partitions}, Cambridge
University Press, 1984.

\bibitem{Ber80}
E. R. Berlekamp, ``The technology of error-correcting codes'',
\emph{Proc. IEEE}, vol. 68, pp. 564--593, May 1980

\bibitem{Beu74}
A. Beutelspacher, ``On parallelisms in finite projective spaces,''
\emph{Geometriae Dedicata}, vol. 3, pp. 35--45, 1974.



\bibitem{BrIm00}
V. Braun and K. A. S. Immink, ``An Enumerative coding technique
for DC-free runlength-limited sequences,'' \emph{IEEE Trans. on
Commun.}, vol.\,IT-48, no.\,1, pp.\,2024--2031, Dec. 2000.


\bibitem{Bra05}
M. Braun, A. Kerber, and R. Laue ``Systematic construction
of $q$-analogs of $t-(v,k,\lambda)$-designs,''
\emph{Designs, codes and Cryptography}, vol. 34, pp. 55--70, 2005.


\bibitem{BrEi92}
A. E. Brouwer and C. A. van Eijl, ``On the $p$-rank of strongly regular graphs,''
\emph{Algebra and Combinatorics}, vol. 1,  pp.\ 329--346, 1992.



\bibitem{YeCa06}
R. W. Yeung and N. Cai, ``Network error correction, part I:
Basic concepts and upper bounds,'' \emph{Commun. Information. Syst.},
vol.~6, pp. 19-36, May 2006.

\bibitem{CaYe06}
N. Cai and R. W. Yeung, ``Network error correction, part II: Lower
bounds,'' \emph{Commun. Information. Syst.}, vol.~6, pp. 37-54,
May 2006.

\bibitem{HandCD}
C. J. Colbourn and J. H. Dinitz,
\emph{Handbook of Combinatorial Designs}, Chapman and Hall/CRC, 2007 (Second edition).

\bibitem{CoSl86}
J. H. Conway and N. J. A. Sloane, ``Lexicographic codes:
error-correcting codes from game theory,'' \emph{IEEE Trans.\
Inform. Theory}, vol.~IT-32, pp. 337-348, May 1986.

\bibitem{Cover}
T.\,M.\ Cover, ``Enumerative source encoding,'' \emph{IEEE Trans.
Inform. Theory}, vol.\,IT-19, no.\,1, pp.\,73--77, Jan. 1973.


\bibitem{Del78}
P. Delsarte, ``Bilinear forms over a finite field, with
applications to coding theory,''\! \emph{Journal of Combinatorial
Theory, Series A}, vol. 25, pp. 226-241,~1978.

 \bibitem{DPTRU02}
  C. Di, D. Proietti, I. E. Telatar, T. J. Richardson, and R. L. Urbanke,
  "Finite-length analysis of low-density parity-check codes on the binary erasure channel",
   \emph{IEEE Trans. Inform. Theory},
 48, no. 6, 1570-1579, 2002.

\bibitem{Dou93}
S. Dougherty, ``Nets and their codes,'' \emph{Designs, Codes,
Crypt.}, vol. 3, pp. 315--331, 1993.

 \bibitem{EtSi08}
T. Etzion and N. Silberstein ``Construction of error-correcting codes for
random network coding '', in proceedings of \emph{IEEE 25th Convention
of Electrical and Electronics Engineers in Israel}, pp. 070--074,
Eilat, Israel, December 2008.


 \bibitem{EtSi09}
T. Etzion and N. Silberstein, ''Error-correcting codes in
projective space via rank-metric codes and Ferrers diagrams'',
\emph{IEEE Trans.\ Inform. Theory}, vol. 55, no.7, pp. 2909--2919,
July 2009.



\bibitem{EV}
T. Etzion and A. Vardy, ``Error-correcting codes in projective
space'', in proceedings of \emph{International Symposium on
Information Theory}, pp. 871--875, Toronto, Ontario, Canada,  July 2008.

\bibitem{EV08}
T. Etzion and A. Vardy, ``Error-correcting codes in projective
space'', \emph{IEEE Trans.\ Inform. Theory}, vol.\,57, no.\,2,
pp.\,1165--1173, February 2011.

\bibitem{EV10}
T. Etzion and A. Vardy, ``On $q$-Analogs for Steiner Systems and Covering Designs'',
 \emph{Advances in Mathematics of Communications}, vol. 5, no. 2, pp. 161--176, 2011.



\bibitem{Ful97}
W. Fulton, {\em Young Tableaux}, Cambridge University Press, 1997.


\bibitem{Gab85}
E. M. Gabidulin, ``Theory of codes with maximum rank distance,''
\emph{Problems of Information Transmission}, vol.~21, pp. 1-12,
July 1985.

\bibitem{GaYa10a}
M. Gadouleau and Z. Yan, ``Packing and covering properties of subspace
codes for error control in random linear network coding,''
 \emph{IEEE Trans. Inform. Theory},  vol.~56, no. 5, pp. 2097--2108, May 2010.

\bibitem{GaYa10}
M. Gadouleau and Z. Yan, ``Constant-rank codes and their connection to constant-dimension codes,''
 \emph{IEEE Trans. Inform. Theory},  vol.~56, no. 7, pp. 3207--3216, July 2010.

\bibitem{Gal62}
R. G. Gallager, ''Low density parity check codes'',
\emph{IRE Trans.\ Inform. Theory}, vol. IT-8, pp. 21--28,
January 1962.

\bibitem{HSS99}
A. S. Hedayat, N. J. A. Sloane, and J. Stufken, {\emph Orthogonal
arrays. Theory and applications}, Springer, 1999.

\bibitem{HKMKE}
T. Ho, R. Koetter, M. M\'{e}dard, D. R. Karger, and M. Effros, ``The benefits of coding
over routing in a randomized setting'', proceedings of \emph{International Symposium
on Information Theory}, Yokohoma, p. 442, June-July 2003.


\bibitem{HMKKESL}
T. Ho, M. M\'{e}dard, R. Koetter, D. R. Karger, and M. Effros, J. Shi, and B. Leong,
``A random linear network coding approach to multicast",
\emph{IEEE Trans.\
Inform. Theory}, vol.~IT-52, pp. 4413-4430, Oct. 2006.




\bibitem{HoJo85}
R. A. Horn, C. R. Johnson, {\emph Matrix analysis},
Cambridge university press, 1985.

\bibitem{HDLA10}
Q. Huang, Q. Diao, S. Lin, K. Abdel-Ghaffar ``Cyclic and Quasi-Cyclic LDPC
Codes on Row and Column Constrained Parity-Check Matrices and Their Trapping Sets'',
\emph{arxiv.org/abs/1012.3201}.

\bibitem{HDLA11}
Q. Huang, Q. Diao, S. Lin, K. Abdel-Ghaffar ``Trapping sets of structured LDPC codes'',
in proceedings of \emph{International Symposium on
Information Theory}, pp. 366--370, Saint Petersburg, Russia, August 2011.


\bibitem{Imm99}
K. A. S. Immink, {\em Codes for Mass Data Storage Systems},
Shannon Foundation Publishers, 1999.


\bibitem{JoWe01}
S. Johnson and S. R. Weller, ''Regular low-density parity-check codes
from combinatorial designs,'' \emph{Proc. 2001 IEEE
Inform. Theory Workshop}, Cairns, Australia, pp. 90-92, Sept. 2-7, 2001.

\bibitem{JoWe04}
S. J. Johnson and S. R. Weller, ``Codes for iterative decoding
from partial geometries,'' \emph{IEEE Trans. on comm.}, vol. 52, pp. 236--243, 2004.

\bibitem{KV03}
N. Kashyap and A. Vardy, ''Stopping sets in codes from designs'',
Available: http://www.mast.queensu.ca/$\sim$nkashyap/Papers/stopsets.pdf,
preprint, 2003.

\bibitem{KhKs09}
A. Khaleghi and F. R. Kschischang, ``Projective space codes for the injection metric'',
in proceedings of \emph{11th Canadian Workshop on Information Theory}, pp. 9--12, Ottawa, ON,  2009

\bibitem{KhSK09}
A. Khaleghi, D. Silva, and F. R. Kschischang, ``Subspace codes'',
in proceedings of \emph{12th IMA International Conference on Cryptography and Coding}, pp. 1--21,  2009



\bibitem{Knu71}
D. E. Knuth, ``Subspaces, subsets. and partitions ,'' \emph{J.\
Combin.\ Theory}, vol.\,10, pp.\,178--180, 1971.


\bibitem{Knu99}
    D. E. Knuth,
    {\em The Art of Computer Programming, Vol.2, Seminumerical Algorithms},
    Third Ed.,
    Addison-Wesley, 1997.


\bibitem{KK}
R.\ Koetter and F.\,R.\ Kschischang, ``Coding for errors and
erasures~in~random network coding,'' \emph{IEEE Trans.\ Inform.
Theory}, vol.\,54, no.\,8, pp.\,3579--3591, August 2008.

\bibitem{KoKu08}
A.\,Kohnert and S.\,Kurz, ``Construction of large
constant-dimension~codes with a prescribed minimum distance,''
\emph{Lecture Notes in Computer Science}, vol.\,5393,
pp.\,31--42, December 2008.

\bibitem{KLF01}
Y. Kou, S. Lin, amd M. P. C. Fossorier, ''Low density parity check codes
based on finite geometries: a rediscovery and new results'',
\emph{IEEE Trans. Inform. Theory}, vol. 47, no.7, pp. 2711--2736, 2001.

\bibitem{Kur02}
O. F. Kurmaev, ``Enumerative coding for constant-weight binary
sequences with constrained run-length of zeros,'' \emph{Problems
of Inform. Tran.}, vol. 38, no.\,1, pp.\,249--254, 2002.

\bibitem{LM05}
S. Landner and O. Milenkovic, ``Algorithmic and combinatorial analysis of
trapping sets in structured LDPC codes,'' \emph{Int. Conf. Wireless Networks,
Communications and Mobile Computing},
Maui, HI, pp.630--635, June 2005.

\bibitem{LaMi07}
S. Laendner and O. Milenkovic, ''LDPC codes based on Latin squares:
cyclic structure, stopping set, and trapping set analysis,'' \emph{IEEE Trans.
Commun.}, vol. 55, no. 2, pp. 303--312, Feb. 2007.

\bibitem{LTLMH08}
L. Lan, Y. Y. Tai, S. Lin, B. Memari and B. Honary,
''New construction of quasi-cyclic LDPC codes based on special
classes of BIBDs for the AWGN and binary erasure channels,''
\emph{IEEE Trans. Commun.}, vol 56, no.1, pp.39-48, Jam. 2008.

\bibitem{Lev60}
V. L. Levenshtein, ``A class of systematic codes ,'' \emph{Soviet
Math. Dokl. 1},  pp.\,368--371, 1960.



\bibitem{vLWi92}
J. H. van Lint and R. M. Wilson,
    {\em A course in Combinatorics},
    Cambridge University Press, 2001 (second edition).

\bibitem{MWSl78}
F. J. MacWilliams and N. J. A. Sloane, {\emph The theory of error-correcting codes},
North-Holland, 1978.


\bibitem{MGR08}
F. Manganiello, E. Gorla, and J. Rosenthal, ``Spread codes and spread decoding in network
coding'', in proceedings of \emph{International Symposium on
Information Theory}, pp. 881--885, Toronto, Ontario, Canada,  July 2008.

\bibitem{Milne82}
S. Milne, ``Mappings of subspaces into subsets ,'' \emph{J.\
Combin.\ Theory}, \,Series A, vol.\,33, pp.\,36--47, 1982.


\bibitem{NMS71}
T. V. Narayana, R. M. Mathsen, and J. Sarangi, ``An algorithm for
generating partitions and its applications,'' \emph{J.
Combinatorial Theory}, vol. 11, pp. 54--61,~1971.


\bibitem{Ric03}
T. Richardson, ``Error floors of LDPC codes,'' \emph{Proc. of the
41st Annual Allerton Conf. Commun., Constrol and Comp.},
Monticello, IL, pp.1426--1435, October 2003.

\bibitem{Rot91}
R. M. Roth, ``Maximum-rank array codes and their application to
crisscross error correction,'' \emph{IEEE Trans.\ Inform. Theory},
vol.~37, pp. 328-336, March 1991.

\bibitem{Rus}
F. Ruskey, {\em Combinatorial Generation}, Working Version,
University of Victoria, Victoria, Canada, 2001.


\bibitem{ScEt02}
M.\ Schwartz and T.\ Etzion, ``Codes and anticodes in the
Grassman graph'', \emph{J.\ Combin.\ Theory, Ser.\,A}, vol. 97, pp. 27--42, 2002.


\bibitem{Sha48}
C. E. Shannon, ''A mathematical theory of communication'',
\emph{Bell Systems tech. Journal}, no.27, pp. 623--656, 1948.


\bibitem{SiEt09c}
N. Silberstein and T. Etzion,
``Enumerative Encoding in the Grassmannian Space'',
in \emph{IEEE Information Theory Workshop}, pp. 544--548, Taormina, Sicily, October 2009.

\bibitem{SiEt09}
N. Silberstein and T. Etzion, ``Enumerative Coding for
Grassmannian Space'',
\emph{IEEE Trans.\ Inform. Theory}, vol. 57, no.1, January 2011.


\bibitem{SiEt10}
N. Silberstein and T. Etzion, ``Large constant dimension codes and
lexicodes,''  \emph{Advances in Mathematics of Communications}, vol. 5,
no. 2, pp. 177--189, 2011.


\bibitem{SiEt11is}
N. Silberstein and T. Etzion, ``Codes and Designs Related to Lifted MRD Codes'',
in proceedings of \emph{International Symposium on
Information Theory}, pp. 2199--2203, Saint Petersburg, Russia, August 2011.



\bibitem{SKK08}
D. Silva, F.\,R.\ Kschischang, and R.\ Koetter, ``A rank-metric
approach to error control in random network coding,'' \emph{IEEE
Trans. Inform. Theory},  vol.~54, pp. 3951--3967, September
2008.

\bibitem{SK09}
D. Silva and  F.\,R.\ Kschischang, ``On metrics for error correction in network coding,'' \emph{IEEE
Trans. Inform. Theory},  vol.~55, no. 12 pp. 5479--5490, December 2009.


\bibitem{Ska10}
V. Skachek, ``Recursive code construction for random networks,''
 \emph{IEEE Trans. Inform. Theory},  vol.~56, no. 3, pp. 1378--1382, March 2010.

\bibitem{Sta86}
R. P. Stanley, {\em Enumerative Combinatorics}, Monterey, CA:
Wadsworth, 1986, vol. 1.


\bibitem{Tan01}
  R. M. Tanner, "Minimum Distance Bounds by Graph Analysis", \emph{IEEE Trans. Inform. Theory},
  47, 808-821, 2001.


\bibitem{TXLA-G05}
  H. Tang, J. Xu, S. Lin, K. A. S. Abdel-Ghaffar, "Codes on Finite Geometries", \emph{IEEE Trans. Inform. Theory},
  51, no. 2, 572--596, 2005.



\bibitem{Tho87}
S. Thomas, ``Designs over finite fields,''
\emph{Geometriae Dedicata}, vol. 21, pp. 237--242, 1987.

\bibitem{ToZh10}
S. Topalova and S. Zhelezova, ``2-spreads and transitive and
orthogonal 2-parallelisms of PG(5,2),'' \emph{Graphs and Comb.},
vol.26, pp. 727--735, 2010.


\bibitem{TrRo10}
A.-L. Trautmann  and J. Rosenthal, ``New improvements on the
echelon-Ferrers construction'', in proc. of \emph{Int. Symp. on
Math. Theory of Networks and Systems}, pp. 405--408, July 2010.

\bibitem{VKK02}
B. Vasic, E. M. Kurtas, and A. Kuznetsov, ''LDPC code based on mutually
orthogonal Latin rectangles and their applications in perpendicular
magnetic recording,'' \emph{IEEE Trans. Magn.}, vol. 38, no. 5, pp. 2346-2348,
Sep. 2002.

\bibitem{VaMi04}
B. Vasic and O. Milenkovic, ''Combinatorial construction of low-density
parity-check codes for iterative decoding,'' \emph{IEEE Trans. Inf. Theory}, vol.
50, no. 6, pp. 1156--1176, June 2004.

\bibitem{WXS-N03}
H. Wang, and C. Xing and R. Safavi-Naini,
``Linear authentication codes: bounds and constructions'',
\emph{IEEE Trans. on Inform. Theory,}
vol. 49, no. 4 pp. 866--872 , 2003.

\bibitem{Xia08}
S.-T. Xia,
``A Graham-Sloane type contruction of constant-dimension codes'',
in proccedings of \emph{Fourth Workshop on Network Coding,
Theory and Applications}, pp. 1--5, Jan. 2008.


\bibitem{XiFu09}
S.-T.Xia  and F.-W. Fu,
``Johnson type bounds on constant dimension codes'',
\emph{Designs, Codes, Crypto.},
vol. 50 no. 2, pp. 163 - 172, February 2009



\bibitem{ZZS71}
G. Zaicev, V. Zinoviev, and N. Semakov, ``Interrelation of Preparata and Hamming codes
and extensions of Hamming codes to new double error-correcting codes ,''
\emph{The 2nd Int. Symp. Inform. Theory},  pp. 257--263, 1971.

\bibitem{Zan97}
A. J. van Zanten, ``Lexicographic order and linearity,'' \emph{Designs, Codes, and
Cryptography}, vol. 10, 85--97, 1997.


\bibitem{ZC05}
    M. Zhu and K. M. Chugg,
     "Lower bounds on stopping distance of linear codes and their applications",
    \emph{Proc. 43rd Allerton Conf. on Communications, Control, and Computing},
Monticello, Sep. 2005.

























%
%
%
%
%
%
%


%



%
%













\end{thebibliography}

\end{document}